\documentclass[useAMS,usenatbib]{mn2e}
\usepackage[dvips]{graphicx}
\usepackage{amsmath}
\usepackage{natbib}
\usepackage{pifont}
\usepackage{txfonts}
\usepackage{mathrsfs}
\usepackage{hyperref}
\usepackage{epsfig}
\usepackage{slashbox}
\usepackage{threeparttable}

\title[Planetary host star model atmospheres]{Planetary host stars:\\ Evaluating uncertainties in cool model atmospheres}

\author[]{
      I. Bozhinova, Ch. Helling, A. Scholz \\
      SUPA, School of Physics and Astronomy, University of St Andrews, North Haugh, St Andrews,  KY16 9SS, UK
    }

\begin{document}    
    
\maketitle

  \begin{abstract}
         
M-dwarfs are emerging in the literature as promising targets for
detecting low-mass, Earth-like planets.  An important step in
this process is to determine the stellar parameters of the
M-dwarf host star as accurately as possible. Different
well-tested stellar model atmosphere simulations from different
  groups are widely applied to undertake this task. This paper
  provides a comparison of different model atmosphere families to
  allow a better estimate of systematic errors on host-star stellar
  parameter introduced by the use of one specific model atmosphere
  family only. We present a comparison of the {\sc ATLAS9}, {\sc
  MARCS}, {\sc Phoenix} and {\sc Drift-Phoenix} model atmosphere
families including the M-dwarf parameter space (T$_{\rm
    eff}=2500$K$\,\ldots\,$4000K, log(g)=3.0$\,\ldots\,$5.0,
  [M/H]=$-2.5\,\ldots\,0.5$). We examine the differences in the (T$_{\rm gas}$, p$_{\rm
  gas}$)-structures, in synthetic photometric fluxes and in colour
indices. Model atmospheres results for higher log(g) deviate
  considerably less between different models families than those for
  lower log(g) for all T$_{\rm eff}=2500$K$\,\ldots\,$4000K examined.
  We compiled the broad-band synthetic photometric fluxes for all
  available model atmospheres (incl. M-dwarfs and brown dwarfs) for the UKIRT WFCAM
  ZYJHK, 2MASS JHKs and Johnson UBVRI filters, and calculated related
  colour indices. Synthetic colours in the IR wavelengths diverge by
  no more than 0.15 dex amongst all model families. For all spectral 
  bands considered, model discrepancies in colour
diminish for higher T$_{\rm eff}$ atmosphere simulations. We notice
differences in synthetic colours between all model families and
observed example data (incl. Kepler 42 and GJ1214). 

  \end{abstract}
   

\section{Introduction}
	
	Ever since the discoveries of the first extra-solar planets
        (\citealt{wol}, \citealt{mayor}, \citealt{charb}),
        exoplanetary science has been one of the hot topics in
        astronomy in the past two decades. High-precision instruments
        and missions such as {\sc HARPS} \citep{mayor2}, {\sc
          CoRoT} \citep{auvergne}, {\sc Kepler} 
        \citep{batalha} and the future {\sc PLATO}\footnote{http://sci.esa.int/plato/}-mission have allowed the number of 		known exoplanets to grow rapidly. Up to date, the Exoplanet
        Encyclopaedia ({\it exoplanet.eu}) lists a total of 1822  
        planets in 1137 planetary systems. Better instruments and
        enhanced observational techniques are pushing the boundaries
        of detectable planets down to the Super-Earth group. In order
        to achieve this goal, target host
        stars decrease in mass. M-dwarfs, and also brown dwarfs \citep{triaud13}, are suggested as they have smaller radii,
        masses and are less luminous, presenting opportunities for
        detecting smaller planets orbiting around them, possibly even
        within their respective habitable zones. The solar
        neighbourhood has been photometrically, spectroscopically and
        astrometrically studied by the RECONS team \citep{henry}
        in order to understand the distribution of stellar types
        nearby. Their latest finding \citep{dieterich} indicate
        that M and later type stars 
        account for 60-70\% of the stellar population
        within 10\,pc of the Sun. The fact that they are so numerous
        additionally increases the chances of planet detections,
        making M-dwarfs and brown dwarfs even more desirable survey targets. 
		 On the other hand, habitability on planets around these stars will be 
		limited by their magnetic activity (see \citealt{vidotto} for details).

	The Exoplanet Encyclopaedia list a total of 36 confirmed
        planets around M-dwarfs with about 2/3 of them with masses
        under 0.2M$_{\rm J}$. There are no detections of planets around brown dwarfs so far.
         Data from Kepler suggests that early
        M-dwarfs have an occurrence rate of, on average,
        0.90$^{+0.04}_{-0.03}$ planets per star with planet parameters
        in range 0.5-4R$_{\rm Earth}$ and P $\rm < $ 50days
        \citep{dressing}. \cite{montet} combine radial velocity
        and adaptive optics direct imaging observations for a sample
        of 111 M stars. They report that 6.5 $\pm$ 3.0\% of the
        M-dwarfs host a gas giant with mass between 1-13M$_{\rm J}$
        and semi-major axes of less than 20AU, corresponding to 0.083
        $\pm$ 0.019 planets per star in that parameter space.  These
        results suggest that planets around M dwarfs are abundant,
        motivating future studies to characterize them in detail (e.g. \citealt{oneh, mann, raj,newt}).
	
	    The evaluation of planet parameters is tightly
        correlated with the host star's parameters
        (e.g. \citealt{torr12, griff13}). Therefore our
        knowledge about extrasolar planets is limited by how
        well we can characterize the host stars.  The challenge of
        determining fundamental stellar parameters is not new (see
        \citealt{rojas13}) and not restricted to planetary host stars
        (e.g. \citealt{casa13} and references
        therein). \cite{burrows2011} generate evolutionary tracks for
        brown dwarfs and very low mass stars (VLMs) for different
        atmospheric metallicities with and without clouds. By
        comparing observational data to these tracks, their study
        demonstrates a variety of plausible stellar radii, and
        narrowing down this range for a given mass depends on precise
        estimates of stellar age and metallicity.  \cite{lee} use
        inverse modelling of directly-imaged data for HR 8799B. Their
        results indicate that reasonable fits to the data can be
        obtained for both cloudy and cloud-free atmospheres but with
        different values for metallicities and element abundances.
	    Such studies  indicate the difficulty of inferring
        precise values for stellar parameters based on
        atmospheric models.  Both, variations in underlying physical
        assumptions between models and different parameter values,
        within the same model can lead to a spread in estimates for
        stellar mass and radii. It is therefore important  to be aware of the limitations of
        model atmospheres and how they
        compare to each other.
	
	This paper focuses on the comparison of different model atmosphere families with some focus on the M-dwarf parameter space:
        effective temperature T$_{\rm eff}=4000\,\ldots\,2500$K, surface gravity
        spans log(g)$=3.0\,\ldots\,5$ (included young M-dwarfs, log(g)$<$4.0, and Brown Dwarfs, log(g)=5.0), and  metallicity  [M/H]$=-2.5\,\ldots\,0.5$ (Appendix~\ref{appA}). 
        Not all parameter combinations are available for all model
        families.  Section~\ref{ss:atm} gives a brief overview
        of the atmosphere models used in this study. In Sect.~\ref{ss:com} we explore the
        similarities and differences in the atmospheric structure of
        the model families.  In Sect.~\ref{ss:phot} we present the results for the
        synthetic photometry comparisons. Section~\ref{ss:dis}
         contains our discussion.

\section{Atmosphere model families in comparison}\label{ss:atm}

 The following model atmosphere families are included in the comparison study presented in this paper:\\*[-0.4cm]
\begin{itemize}
\item {\sc ATLAS9}\footnote{http://user.oat.ts.astro.it/castelli/grids.html} (\citealt{kurucz}, \citealt{castelli}),
\item  {\sc MARCS}\footnote{http://marcs.astro.uu.se}  \citep{gustafsson},
\item  (cloud-free) {\sc PHOENIX-ACES-AGSS-COND-2011}\footnote{http://phoenix.astro.physik.uni-goettingen.de}, version 16.01.00B (hereafter {\sc Phoenix})  \citep{husser},
\item {\sc Drift-Phoenix}  (\citealt{dehn}, \citealt{hellingb},\citealt{witte}, \citealt{witte2011}).
\end{itemize}
  All models assume LTE, hydrostatic and chemical equilibrium
  and obey radiative and convective flux conservation. They
  model a homogeneous, 1D oxygen-rich atmosphere in plane-parallel
  geometry. {\sc Phoenix} models were available in spherical symmetry.

 The grid of {\sc ATLAS}
models utilised spans the  following range of stellar parameter: T$_{\rm eff}$ = 3500 K$\,\ldots\,$4000 K, log(g) = 3.0$\,\ldots\,$5.0,  [M/H] = $0.5\,\ldots-2.5$.  All {\sc ATLAS} models were
calculated with the convection option switched on but with the
overshooting option switched off. The mixing length parameter 
$\alpha=l/H_{\rm p}= 1.25$, ($H_{\rm p} = {\rm p} / ({\rm {g  \rho}})$, $H_{\rm p}$ -- local pressure scale height, ${\rm p}$ - local gas pressure, $\rho$ - local gas density, $g$ - local gravitational acceleration, where $ d{\rm p} / dr = g\rho$), 
the micro-turbulence velocity v$_{\rm turb}$ = 2.0 km/s, and solar element abundances from \citealt{grevesse98} are used in all  {\sc ATLAS} models considered here.

 The {\sc MARCS} models used span a grid of T$_{\rm eff}$ = 2500K$\,\ldots\,$4000K, log(g) =
3.0$\,\ldots\,$5.5,  [M/H] = $0.5\,\ldots-2.5$. For all {\sc MARCS} models, v$_{\rm
  turb}$= 2 km/s, mixing length parameter with
$l/H_{\rm p}= 1.5$ and solar element abundances \citep{grevesse07}.
	
The {\sc Phoenix} models considered are for T$_{\rm
eff}$ = 2500K$\,\ldots\,$4000K, log(g) = 3.0$\,\ldots\,$5.5, [M/H] =
  0.0 and $\alpha$-element abundance of [$\alpha$/M] = 0.0.
   Values for the mixing length parameter $l/H_{\rm p} \sim
    3.0\,\ldots\,\sim1.8$ depending on the stellar parameters as
    depicted in Fig. 2 in
    \cite{husser}. Figures~\ref{tp2500}\,--\,\ref{tp4000} provide the
    detailed information regarding the model atmospheres compared. The
    micro-turbulence velocities v$_{\rm turb}<1.5$ km/s according to
    Fig. 3 in \cite{husser}. The element abundances are solar (\citealt{asplund}).

The {\sc Drift-Phoenix} models are aimed specifically at late-type
stars (M-dwarfs, brown dwarfs) and giant planet atmospheres as they
also model dust cloud formation.  The Drift module deals with dust
treatment, calculating a consistent cloud structure and passing it to
the main radiative transfer code ({\sc Phoenix}).  The subset of
models used is for the solar metallicity models with 2500K $<$ T$_{\rm
  eff} <$ 3000K and 3.0 $<$ log(g) $<$ 5.5.  Mixing length is set
  to 2.0 scale heights and micro-turbulence velocity is 2.0 km/s.
Solar elements abundances are those from \citep{grevesse07}

The different model atmosphere families of models cover different
parts of the M-dwarf regime, with {\sc ATLAS} barely touching
early-type M stars, {\sc Drift-Phoenix} covering the late end of this
spectral type and {\sc MARCS} and {\sc Phoenix} spanning the entire
M-dwarf range. The different sets of element abundances applied
  for different model families are summarized in Table
  \ref{el_abund}. All non-solar metalicities are derived from scaled
  solar values. More detailed information about the models,
  e.g. regarding the used opacity sources, are provided in the
  discussion Sect~\ref{ss:diffm}.  

The model atmospheres under investigation do not contain one M-dwarf
parameter set that is common to all of them. Therefore, we 
  compare subsets of model families: the {\sc ATLAS}$+${\sc MARCS}
models for T$_{\rm eff}$ = 3500K and T$_{\rm eff}$ = 4000K and varying
log(g) and [M/H] values, {\sc ATLAS}$+${\sc MARCS}$+${\sc Phoenix} for
for T$_{\rm eff}$ = 3500K and T$_{\rm eff}$ = 4000K, [M/H] = 0.0 and
varying log(g), as well as the {\sc MARCS}+{\sc Drift-Phoenix}, {\sc
  MARCS}+{\sc Phoenix} and {\sc Phoenix}+{\sc Drift-Phoenix} models
for solar metallicity and varying T$_{\rm eff}$ and log(g) values. A
total of 141 models were examined.  Appendix~~\ref{appA} summarize the
parameter values of all models used.

\begin{table}
    \centering
      \caption {  Element abundances used in the model atmosphere families.
      			}
      \label{el_abund}
      
      \begin{tabular}{|c|c|c|c|}
      
	    \hline    
	    & \hspace{-0.5cm}Grevesse \& Sauval 1998 & \hspace{-0.2cm}Grevesse et al. 2007 & \hspace{-0.2cm}Asplund et al.  2009 \\
            & ({\sc ATLAS})           &  ({\sc MARCS},        &  {\sc Phoenix}\\
            &                         &  {\sc Drift-Phoenix}) & \\ \hline
	    
		C & 8.52	 $\pm$ 0.06 & 8.39 $\pm$ 0.05 & 8.43 $\pm$ 0.05 \\ 
		N & 7.92 $\pm$ 0.06 & 7.78 $\pm$ 0.06 & 7.83 $\pm$ 0.05 \\  
		O & 8.83 $\pm$ 0.06 & 8.66 $\pm$ 0.05 & 8.69 $\pm$ 0.05 \\ 
		Na & 6.33 $\pm$ 0.03 & 6.17 $\pm$ 0.04 & 6.24 $\pm$ 0.04 \\
		Mg & 7.58 $\pm$ 0.05 & 7.53 $\pm$ 0.09 & 7.60 $\pm$ 0.04 \\
		Al & 6.47 $\pm$ 0.07 & 6.37 $\pm$ 0.06 & 6.45 $\pm$ 0.03 \\
		Si & 7.55 $\pm$ 0.05 & 7.51 $\pm$ 0.04 & 7.51 $\pm$ 0.03 \\
		S & 7.33 $\pm$ 0.11 & 7.14 $\pm$ 0.05 & 7.12 $\pm$ 0.03 \\
		K & 5.12 $\pm$ 0.13 & 5.08 $\pm$ 0.07 & 5.03 $\pm$ 0.09 \\
		Ca & 6.36 $\pm$ 0.02 & 6.31 $\pm$ 0.04 & 6.34 $\pm$ 0.04 \\
		Ti & 5.02 $\pm$ 0.06 & 4.90 $\pm$ 0.06 & 4.95 $\pm$ 0.05 \\
		Fe & 7.50 $\pm$ 0.05 & 7.45 $\pm$ 0.05 & 7.50 $\pm$ 0.04 \\
		V & 4.00 $\pm$ 0.02 & 4.00 $\pm$ 0.02 & 3.93 $\pm$ 0.08 \\
		Cr & 5.67 $\pm$ 0.03 & 5.64 $\pm$ 0.10 & 5.64 $\pm$ 0.04 \\
	    
	     \hline

      \end{tabular}

  \end{table}

\section{Comparing the atmosphere structures}\label{ss:com}
	
Model atmosphere simulations provide the numerical
solution to energy transfer by radiation and convection, hydrostatic
equilibrium and gas-phase equilibrium chemistry. The radiative energy
transfer is  likely to carry  inconsistencies between the model families as it depends on element abundances, gas-phase number densities and line lists for those species taken into account as opacity sources in each of the atmosphere models.  Other differences between model atmosphere results from different codes are caused by different numerical schemes used, difference in convergence criteria applied, maybe by differences in the machines where the code is run, or also by different hidden parameters like e.g. the outer integration boundary. This paper can only present the effect of the sum of all these factors on the results from different model families and showcase how and if the results differ. Without a dedicated benchmark study, like e.g. \cite{hellinga}, a more detailed assessment of the differences between the model families is not possible.    

The local gas temperatures and gas pressures affect number
densities of chemical species, which in turn affects opacities and, hence, 
 result in differences in the emergent spectral energy
distribution (SED). We therefore start our investigation by examining the local (T$_{\rm gas}$, p$_{\rm gas}$) structures of model atmospheres for a given set of T$_{\rm eff}$, log(g) and [M/H] values.

	\begin{figure*}
	    \begin{tabular}{cc}
	    	  \vspace{3mm}
	      \includegraphics[width=0.30\linewidth, keepaspectratio, angle=90]{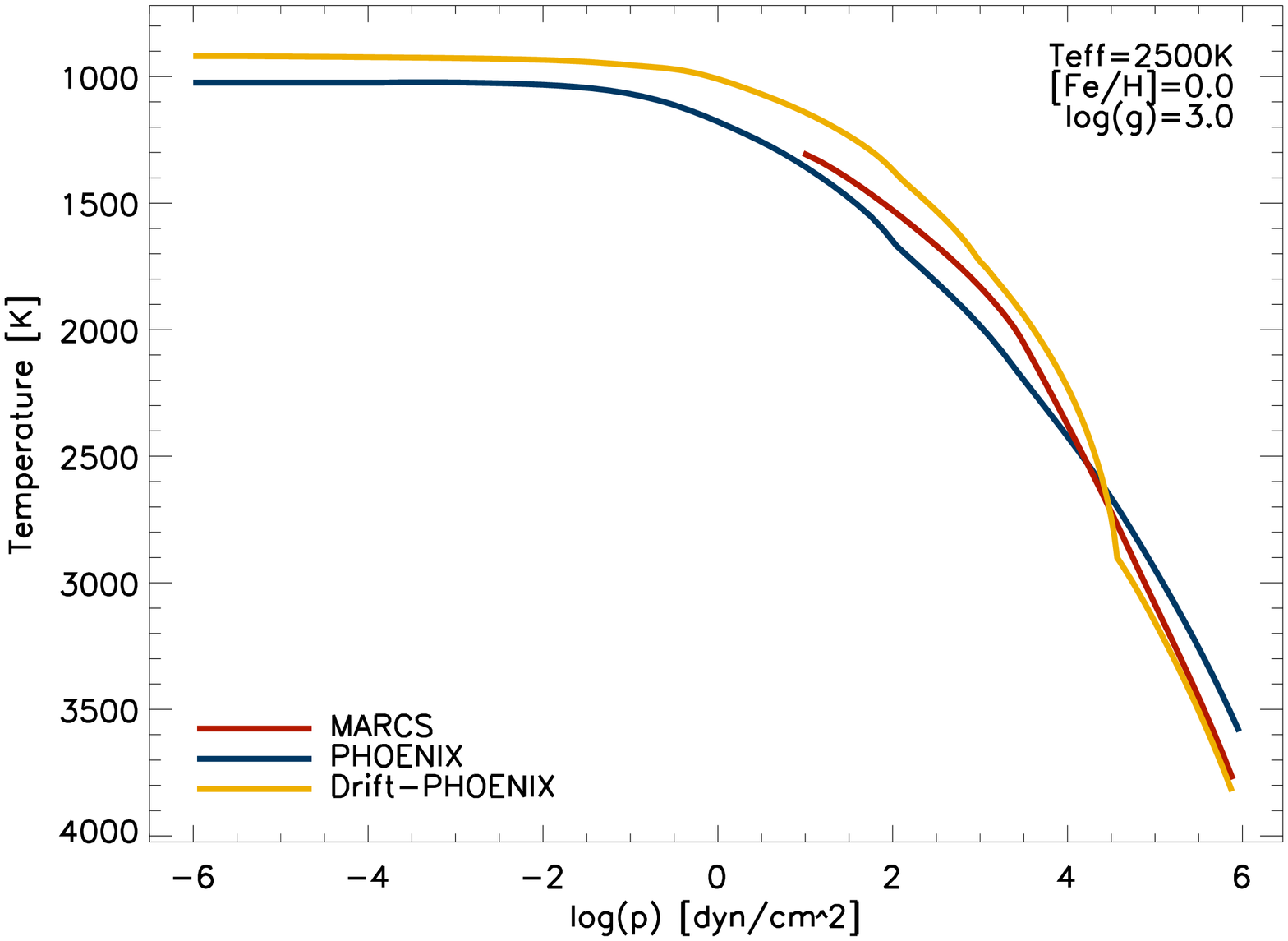}\hspace{5mm}
	      \includegraphics[width=0.30\linewidth, keepaspectratio, angle=90]{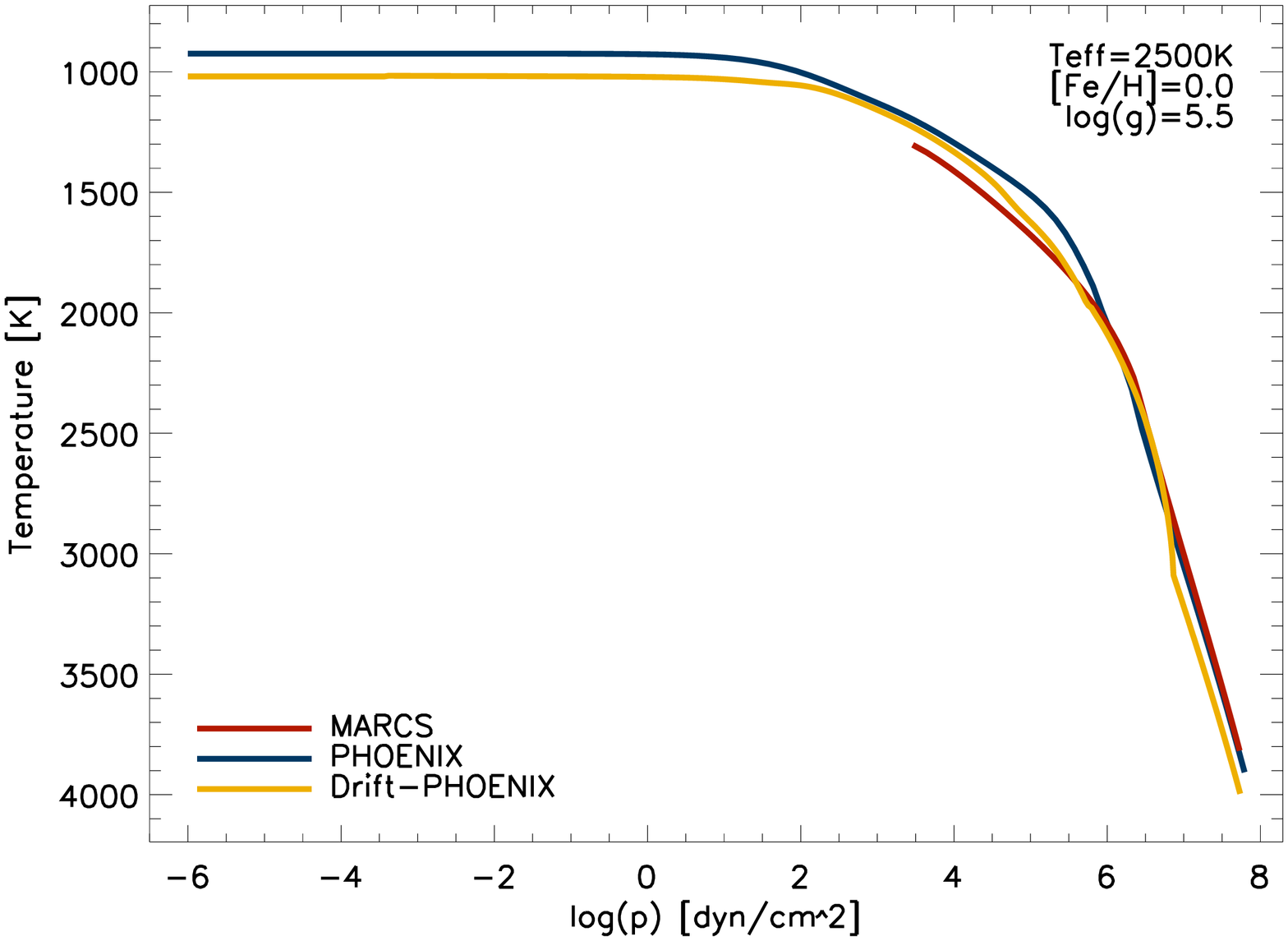}\\
	      \vspace{3mm}
	      \includegraphics[width=0.31\linewidth, keepaspectratio, angle=90]{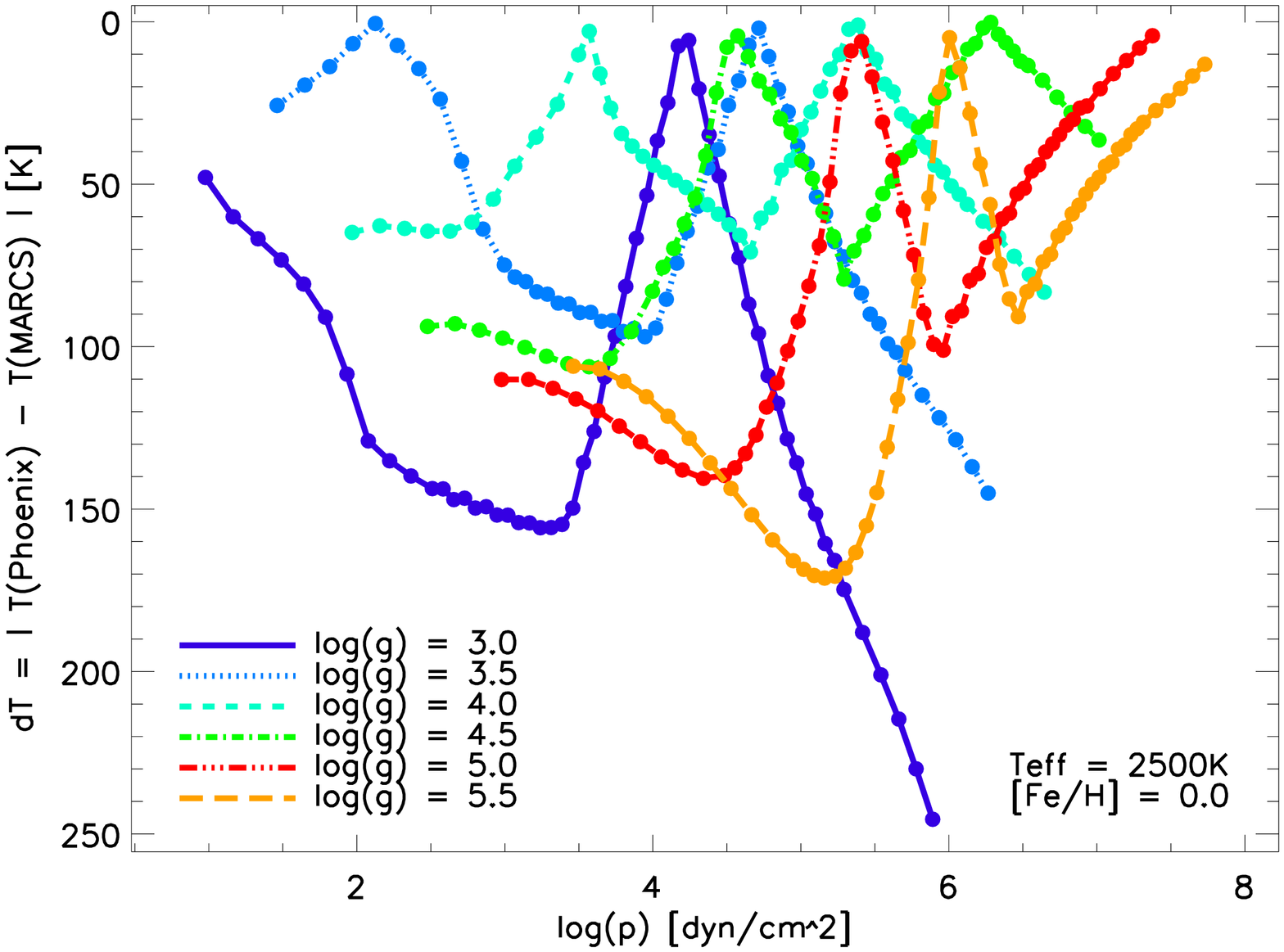}\hspace{5mm}
	      \includegraphics[width=0.31\linewidth, keepaspectratio, angle=90]{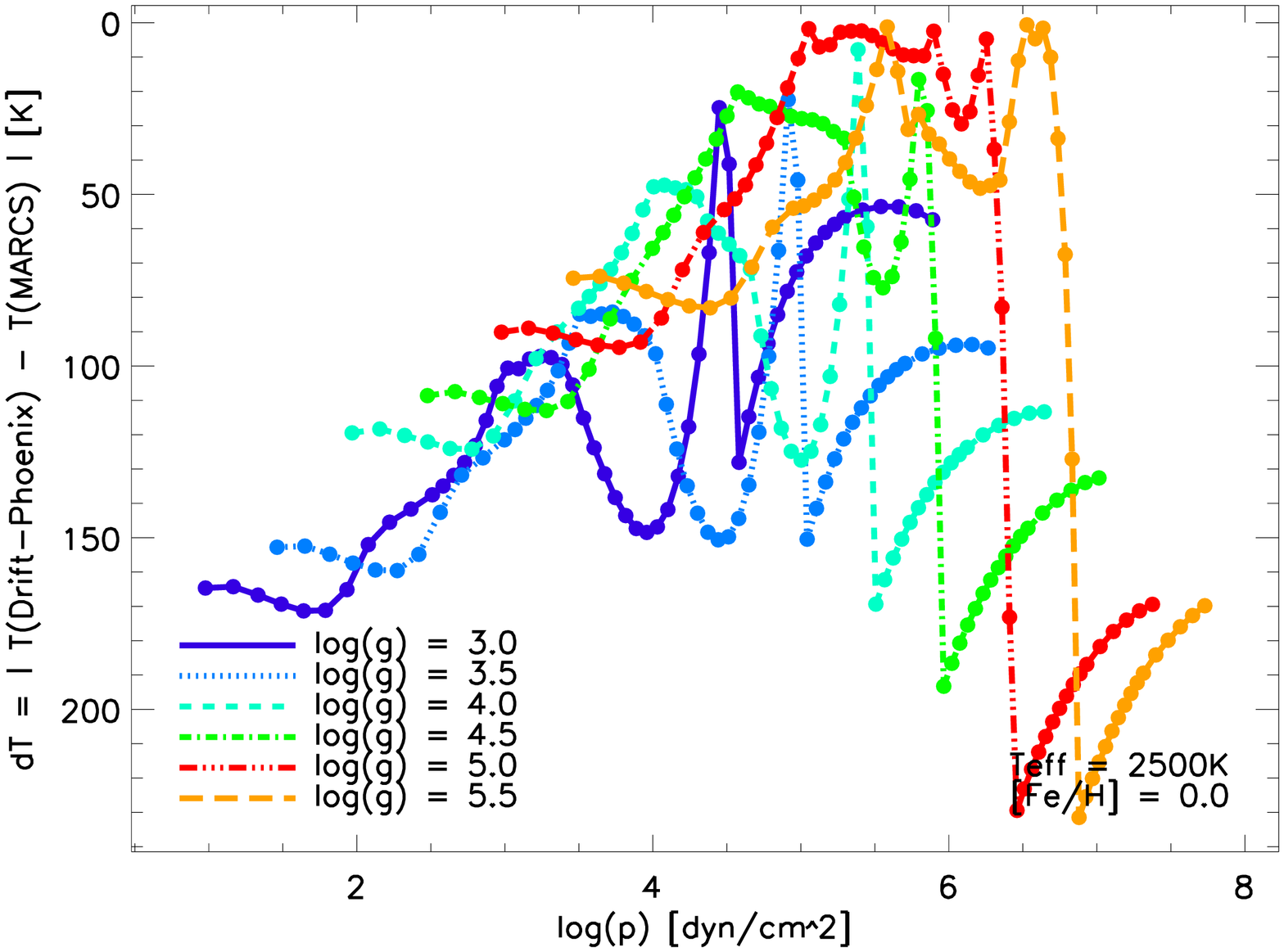}\\
	      \includegraphics[width=0.31\linewidth, keepaspectratio, angle=90]{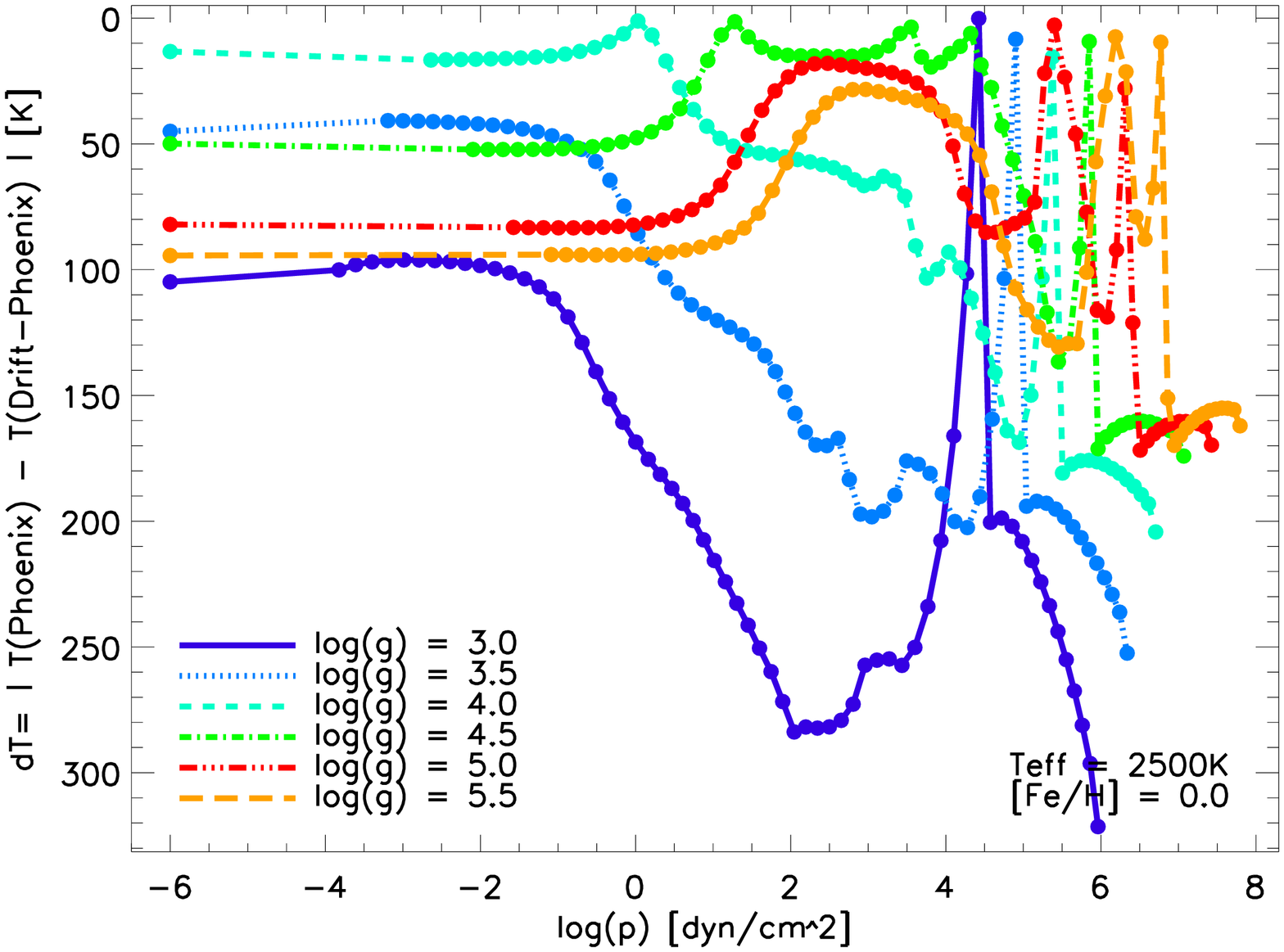}
          \end{tabular}
	  \caption{
          {\bf Top row:} local gas temperature-pressure structures for T$_{\rm eff}$ = 2500K (log(g) = 3.0: top left;  log(g) = 5.5: top right) for {\sc MARCS}, {\sc Phoenix} and 
		    {\sc Drift-Phoenix}.  {\bf Middle row:} left - residual 
		    temperature values  $d T_{\rm gas}$ between {\sc Phoenix} and {\sc MARCS}; right -  $d T_{\rm gas}$ between {\sc Drift-Phoenix}
		    and {\sc MARCS}. {\bf Bottom row:} $d T_{\rm gas}$  between {\sc Drift-Phoenix}
		    and {\sc Phoenix} model atmosphere results. Mixing length parameter: {\sc Phoenix} - $l/H_{\rm p}$=2.79 for log(g) = 3.0 and $l/H_{\rm p}$=3.5 for log(g) = 5.5; {\sc MARCS} - $l/H_{\rm p}$=1.5 for all models; {\sc Drift-Phoenix} - $l/H_{\rm p}$=2.0 for all models.
		    }
	  \label{tp2500}	
	\end{figure*}   
   
	\begin{figure*}   
	    \begin{tabular}{cc}	
	    	  \vspace{3mm}      
	      \includegraphics[width=0.30\linewidth, keepaspectratio, angle=90]{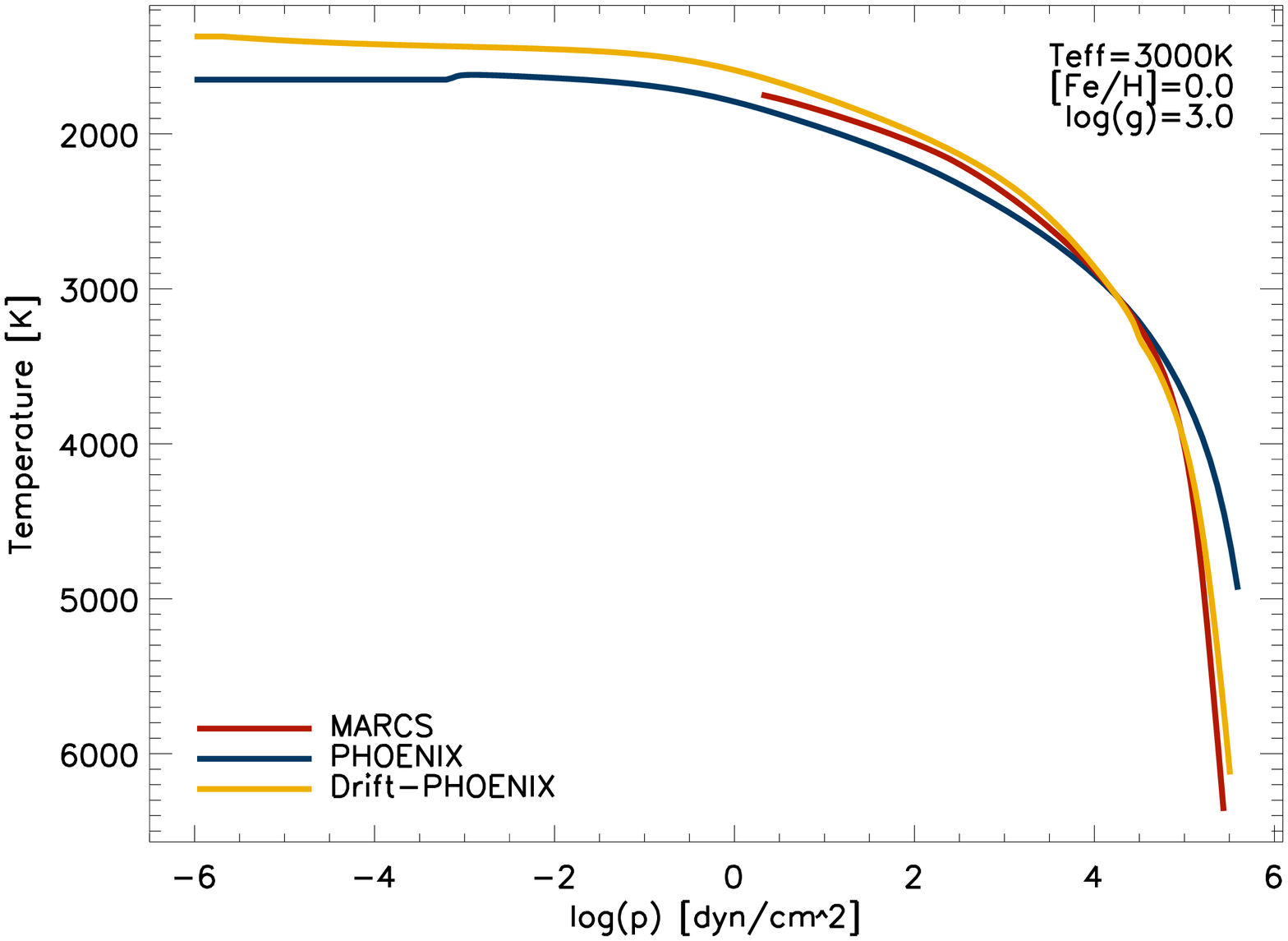}\hspace{5mm}
	      \includegraphics[width=0.30\linewidth, keepaspectratio, angle=90]{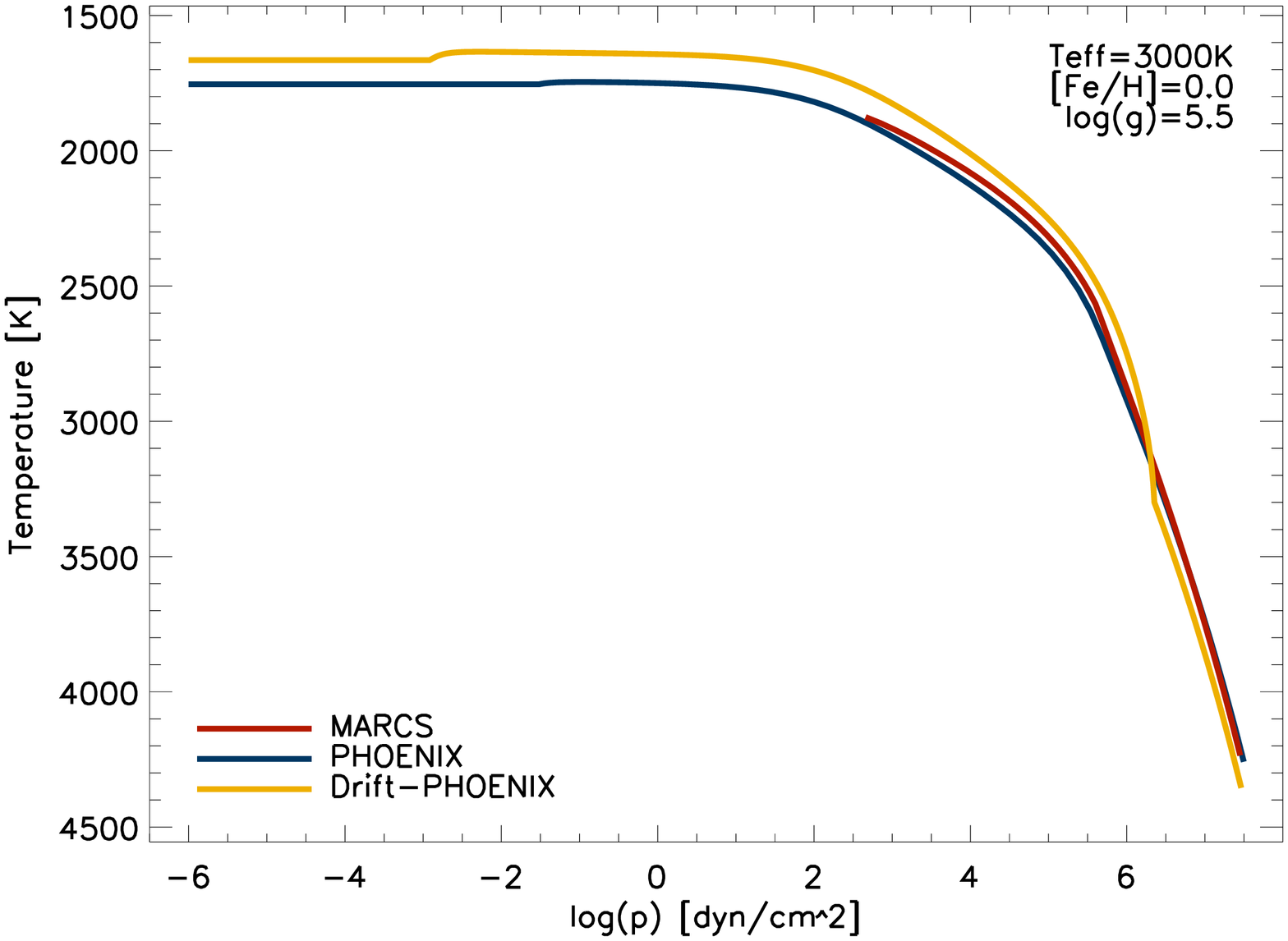}\\
	      \vspace{3mm}
	      \includegraphics[width=0.31\linewidth, keepaspectratio, angle=90]{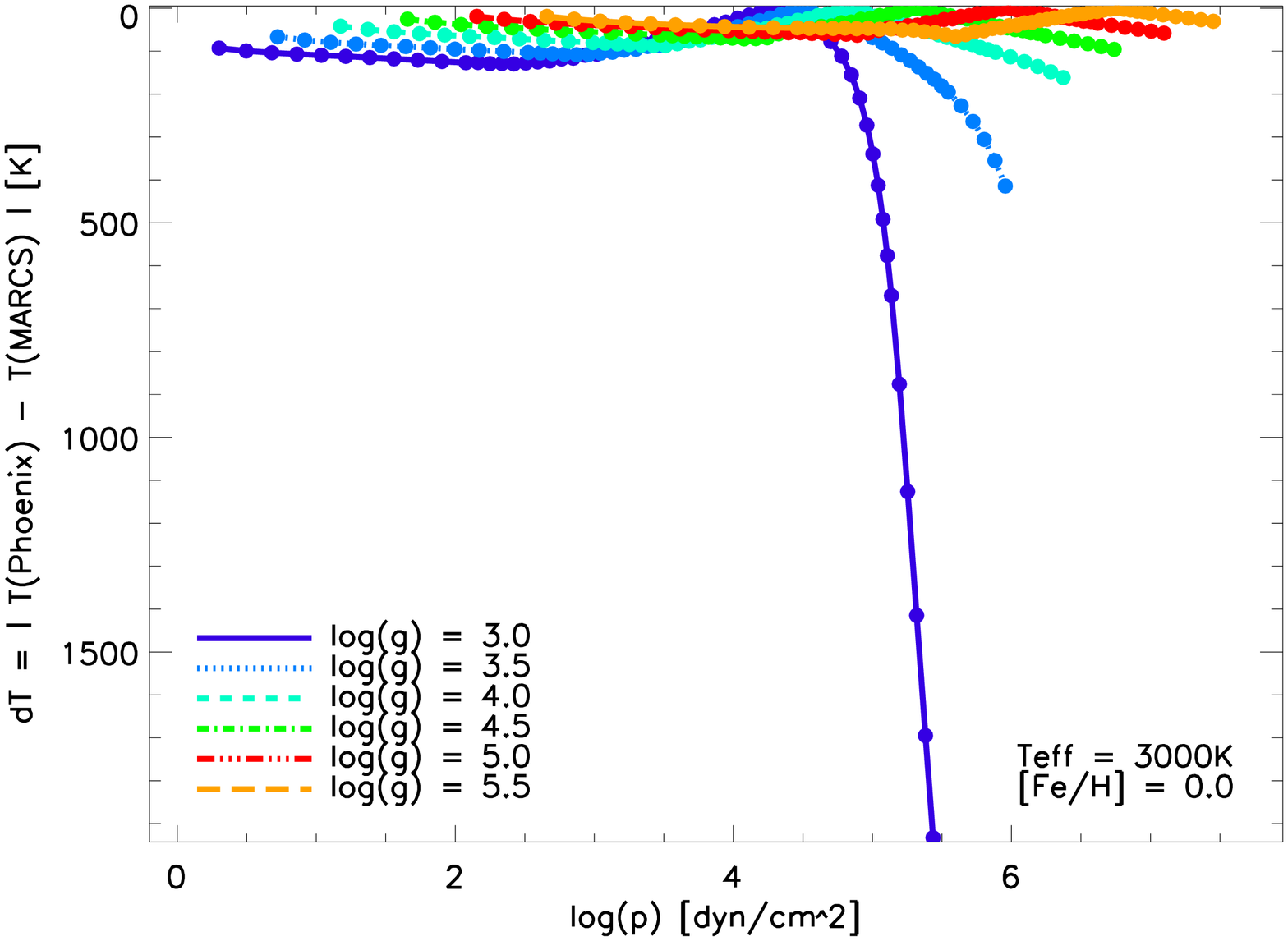}\hspace{5mm}
	      \includegraphics[width=0.31\linewidth, keepaspectratio, angle=90]{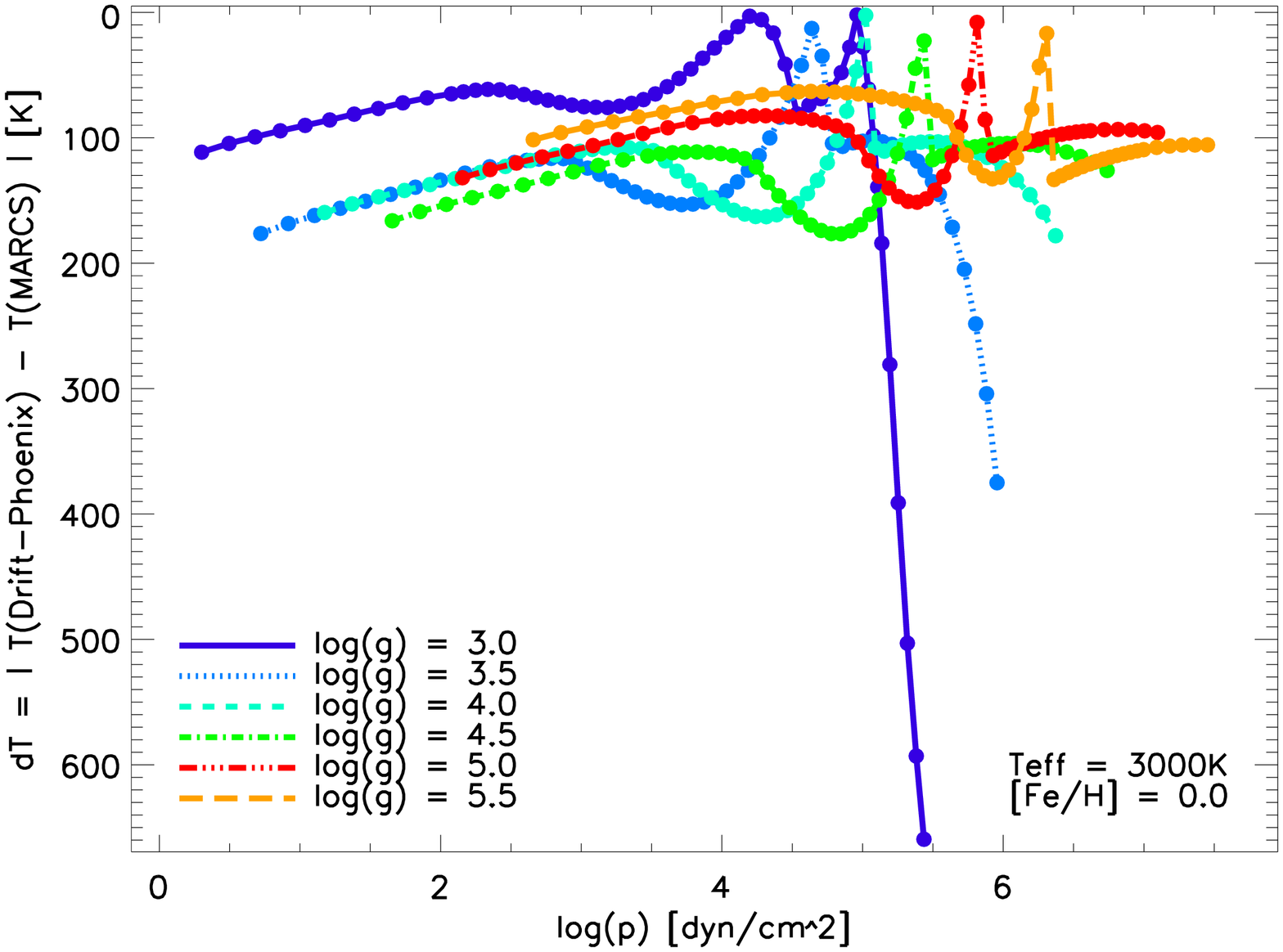}\\
	      \includegraphics[width=0.31\linewidth, keepaspectratio, angle=90]{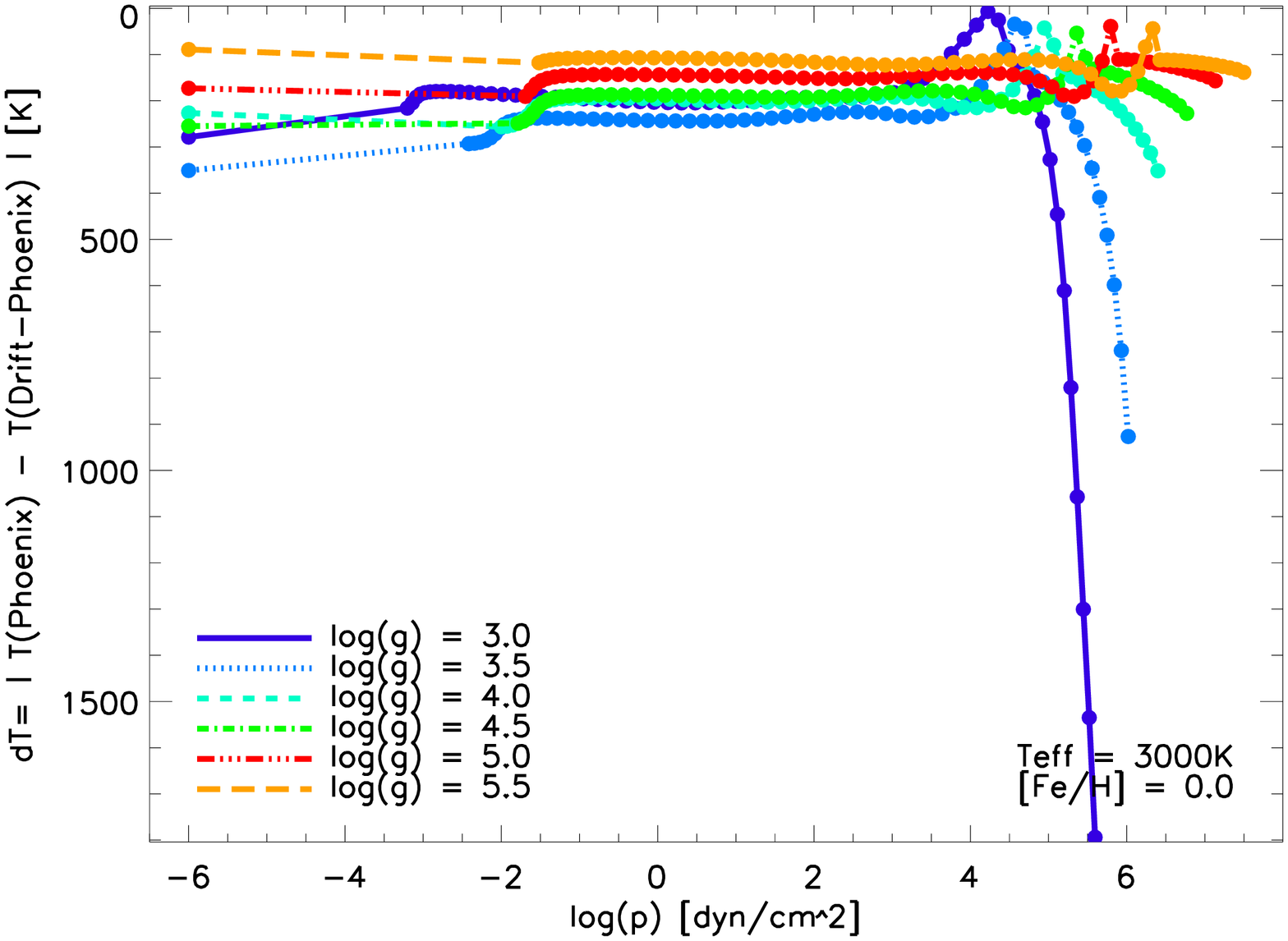}
	      
	    \end{tabular}	  
	  \caption{
		    {\bf Top row:} local gas temperature-pressure structures for T$_{\rm eff}$ = 3000K (log(g) = 3.0: top left;  log(g) = 5.5: top right) for {\sc MARCS}, {\sc Phoenix} and 
		    {\sc Drift-Phoenix}.  {\bf Middle row:} left - residual 
		    temperature values $d T_{\rm gas}$ between {\sc Phoenix} and {\sc MARCS}; right - $d T_{\rm gas}$  between {\sc Drift-Phoenix}
		    and {\sc MARCS}. {\bf Bottom row:} $d T_{\rm gas}$  between {\sc Drift-Phoenix}
		    and {\sc Phoenix} model atmosphere results. Mixing length parameter: {\sc Phoenix} - $l/H_{\rm p}$=2.3 for log(g) = 3.0 and $l/H_{\rm p}$=3.25 for log(g) = 5.5; {\sc MARCS} - $l/H_{\rm p}$=1.5 for all models; {\sc Drift-Phoenix} - $l/H_{\rm p}$=2.0 for all models.
		    }
	  \label{tp3000}
	  
	\end{figure*}

	\begin{figure*}
	  
	    \begin{tabular}{cc}
	      \vspace{3mm}
	      \includegraphics[width=0.30\linewidth, keepaspectratio, angle=90]{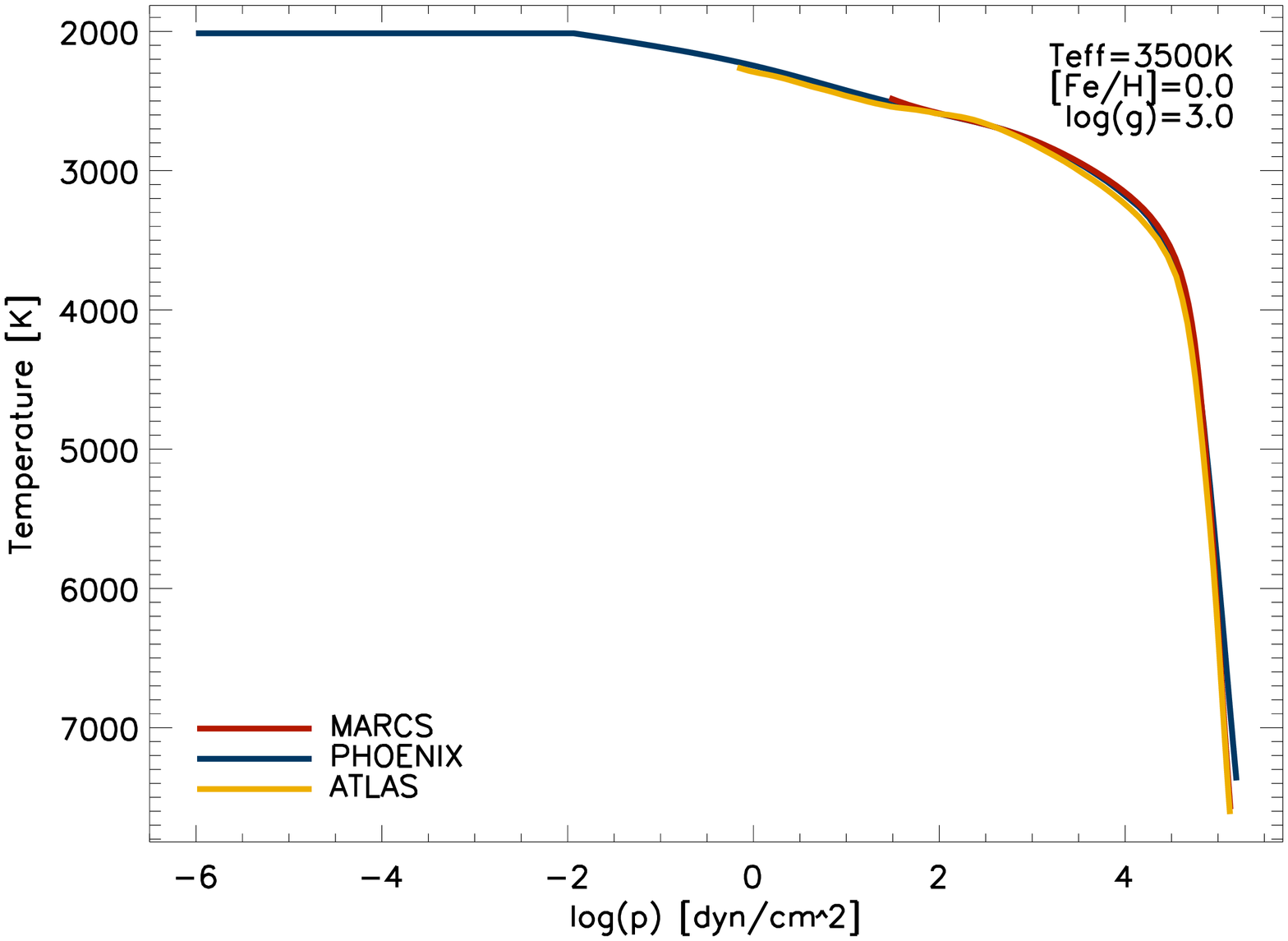}\hspace{5mm}
	      \includegraphics[width=0.30\linewidth, keepaspectratio, angle=90]{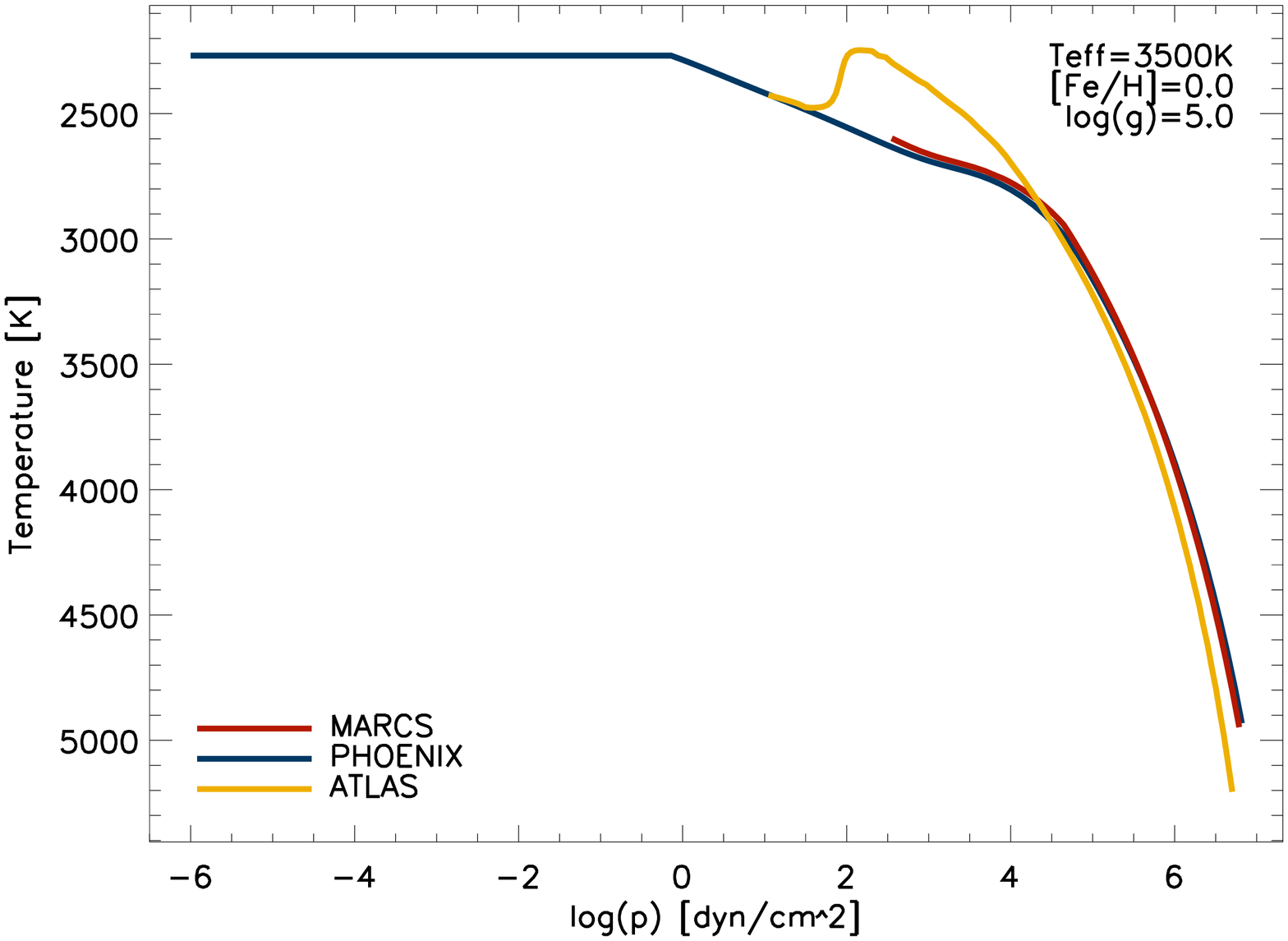}\\
	      \vspace{3mm}
	      \includegraphics[width=0.31\linewidth, keepaspectratio, angle=90]{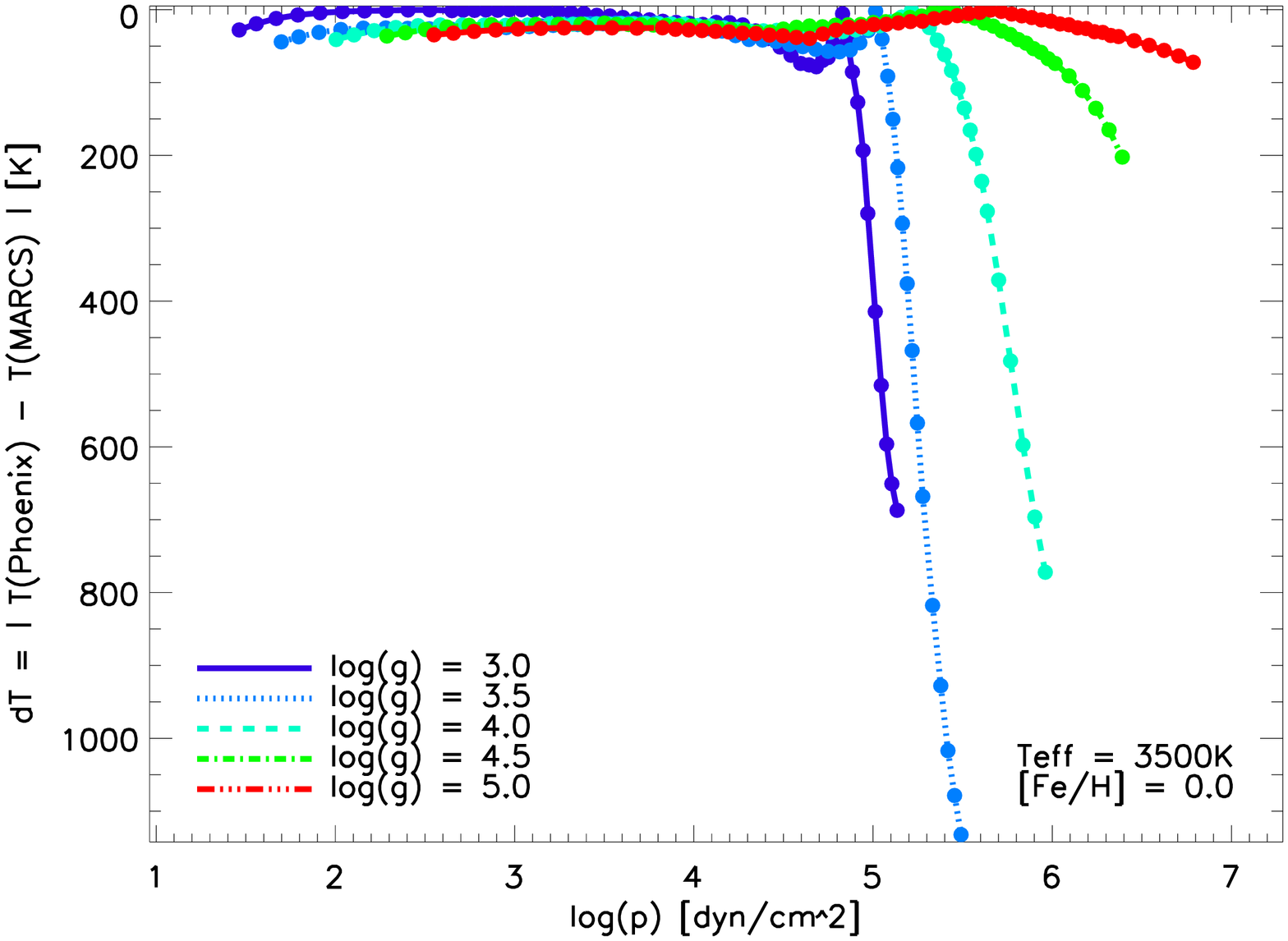}\hspace{5mm}
	      \includegraphics[width=0.31\linewidth, keepaspectratio, angle=90]{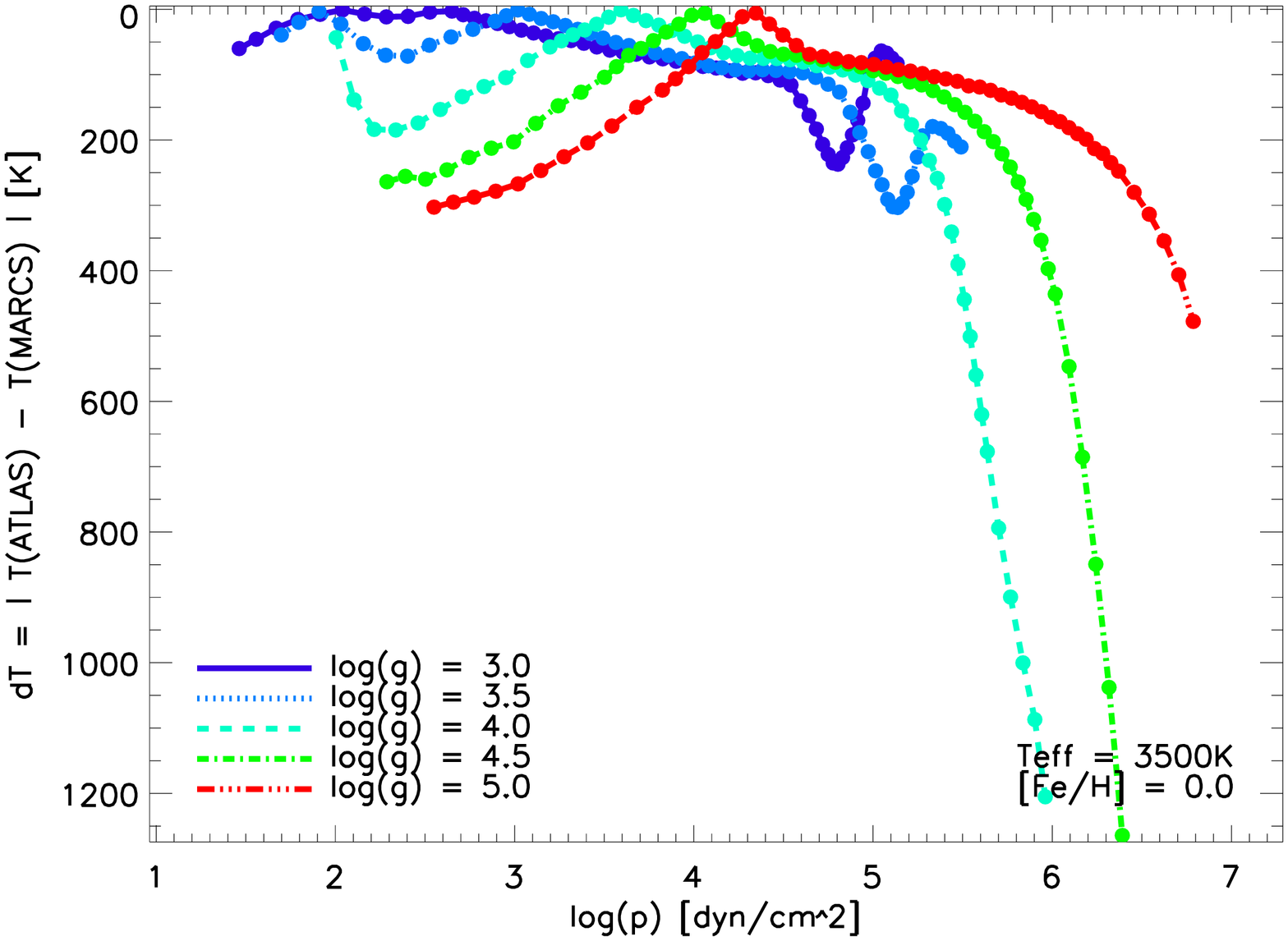} \\
	      \includegraphics[width=0.31\linewidth, keepaspectratio, angle=90]{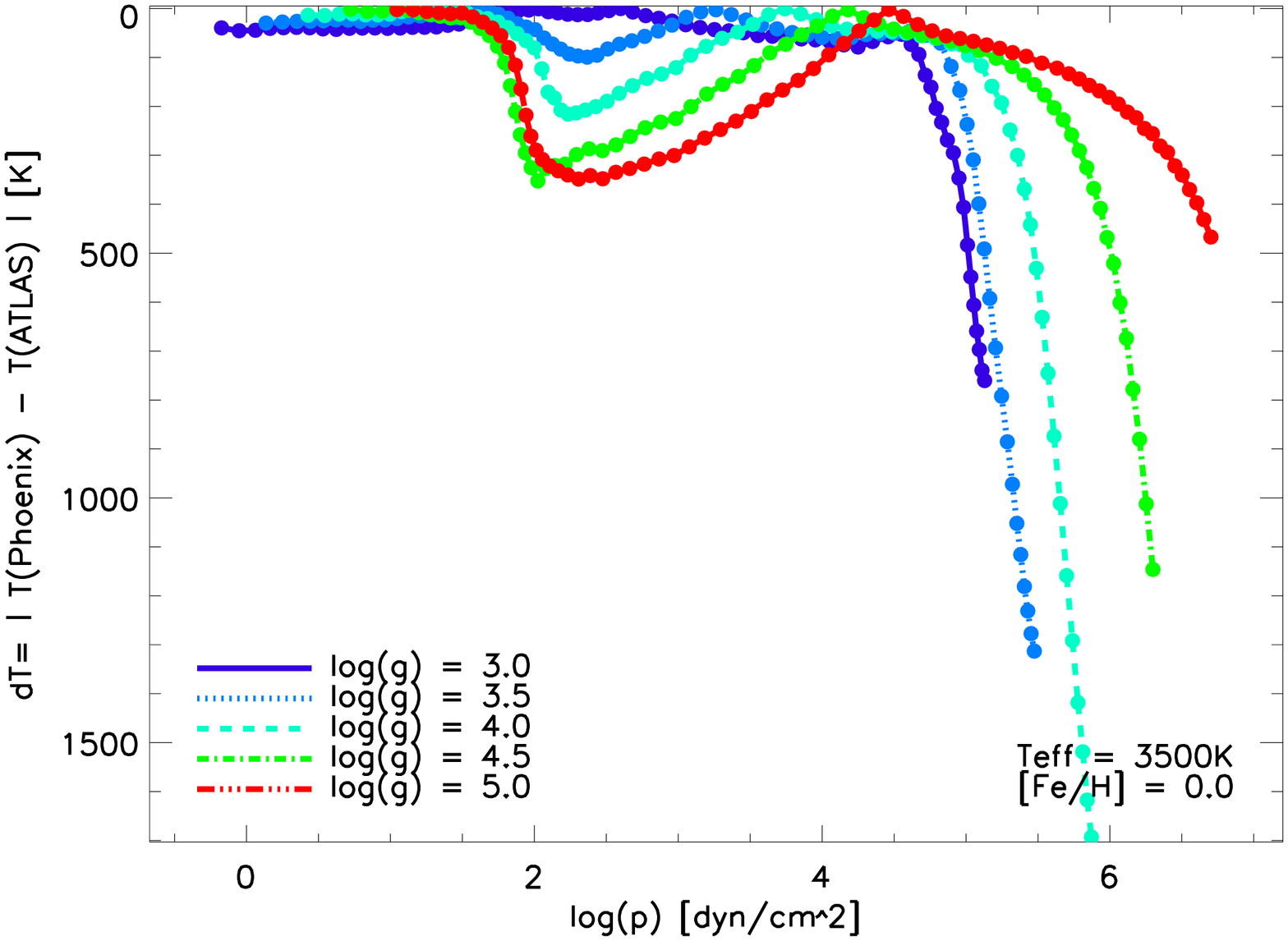}
	      
	    \end{tabular}

	  \caption{
		    {\bf Top row:} local gas temperature-pressure structures for T$_{\rm eff}$ = 3500K  (log(g) = 3.0: top left;  			log(g) = 5.0: top right) for {\sc MARCS}, {\sc Phoenix} and {\sc ATLAS}. {\bf Middle row:} left - residual 
		    temperature values $d T_{\rm gas}$ between {\sc Phoenix} and {\sc MARCS}; right - $d T_{\rm gas}$  between 				{\sc ATLAS} and {\sc MARCS}. {\bf Bottom row:} $d T_{\rm gas}$  between {\sc Phoenix} and {\sc ATLAS} model 				atmosphere results. Mixing length parameter: {\sc Phoenix} - $l/H_{\rm p}$=1.99 for log(g) = 3.0 and $l/H_{\rm p}$=2.39 for log(g) = 5.0; {\sc ATLAS} - $l/H_{\rm p}$=1.25 for all                                                                                              			models; {\sc MARCS} - $l/H_{\rm p}$=1.5 for all models. 
		    }
		    
	  \label{tp3500}
	  
	\end{figure*}
	
	\begin{figure*}
	  
	    \begin{tabular}{cc}
	      \vspace{3mm}
	      \includegraphics[width=0.30\linewidth, keepaspectratio, angle=90]{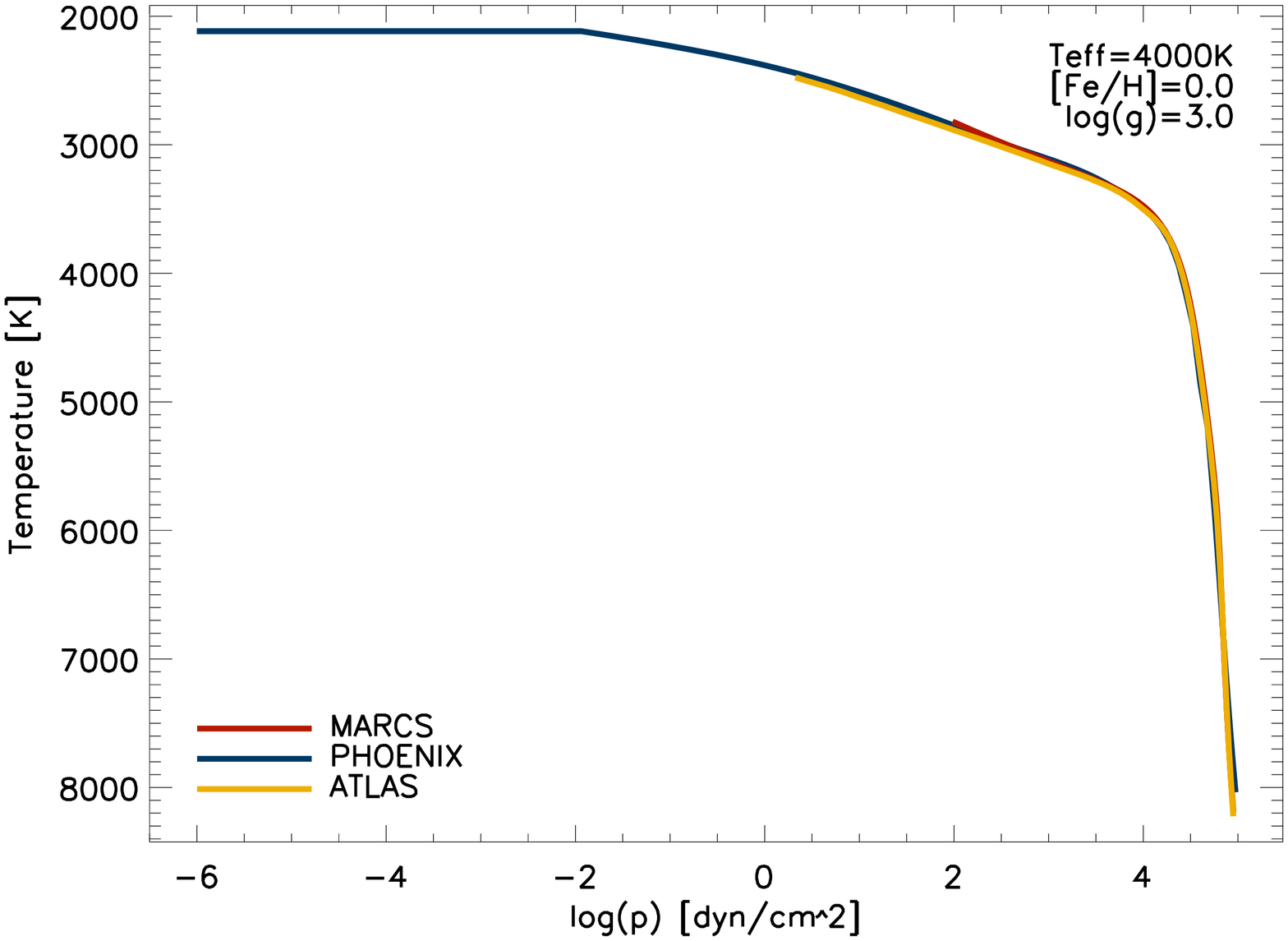}\hspace{5mm}
	      \includegraphics[width=0.30\linewidth, keepaspectratio, angle=90]{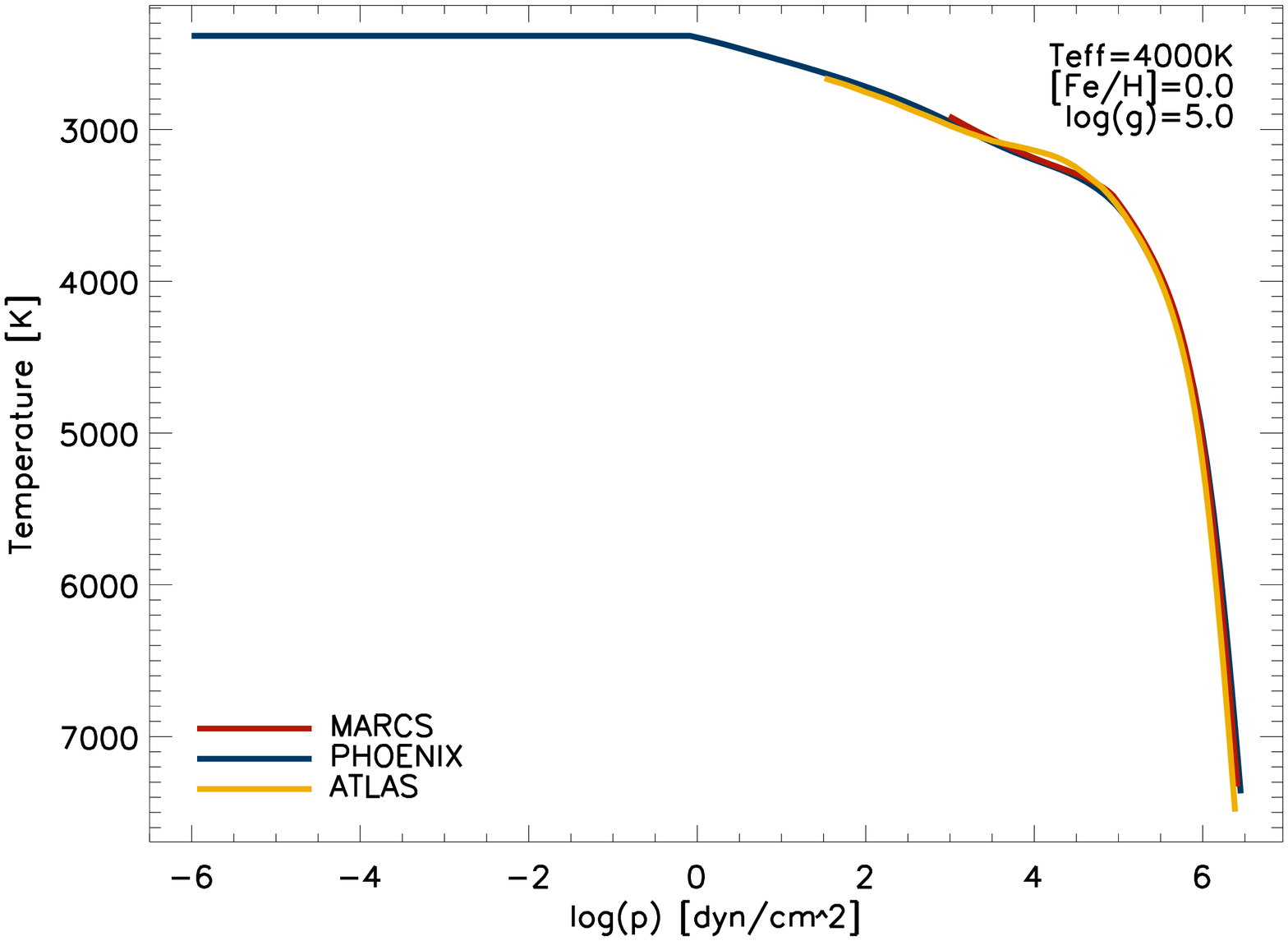}\\
	      \vspace{3mm}
	      \includegraphics[width=0.31\linewidth, keepaspectratio, angle=90]{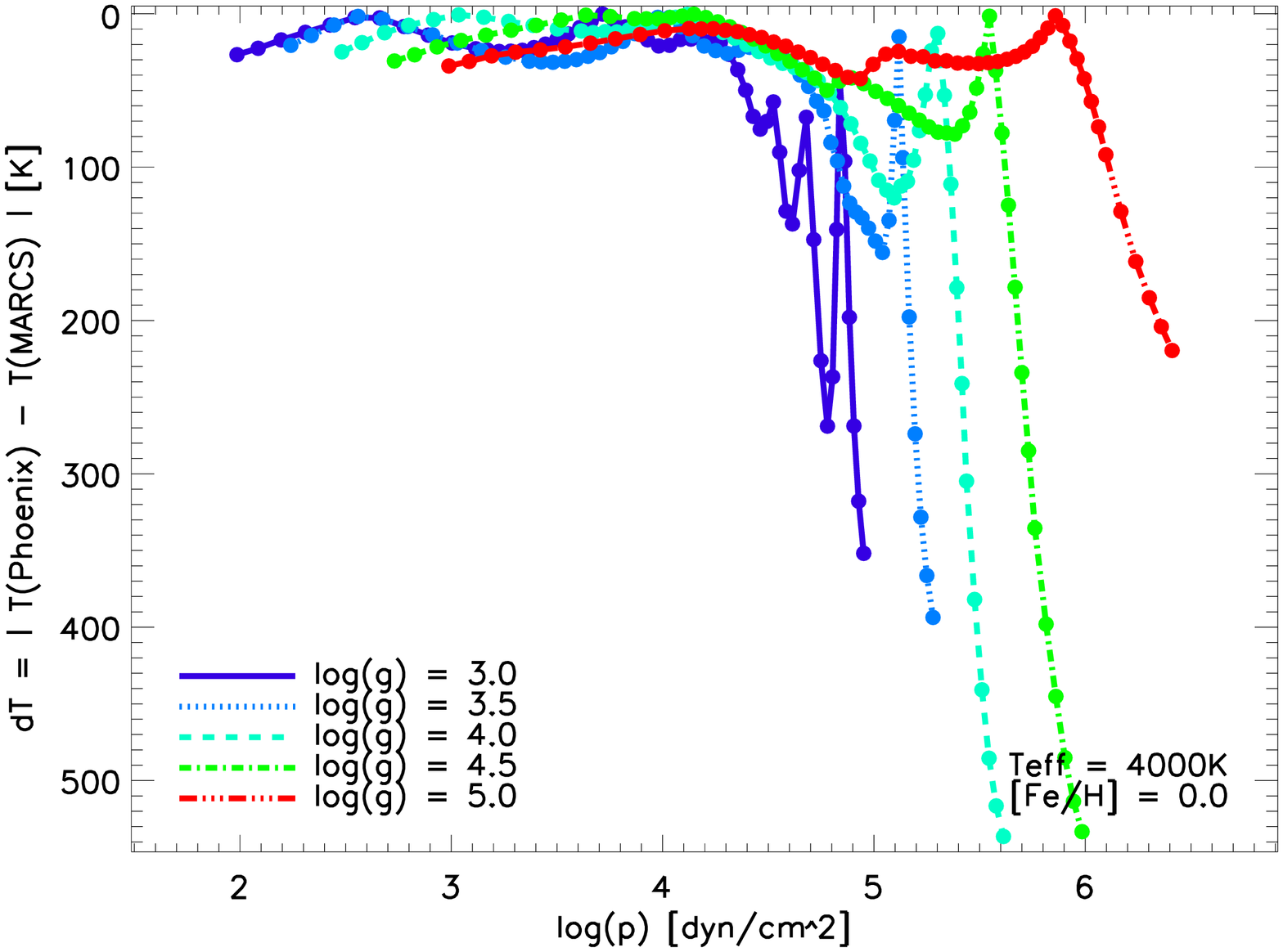}\hspace{5mm}
	      \includegraphics[width=0.31\linewidth, keepaspectratio, angle=90]{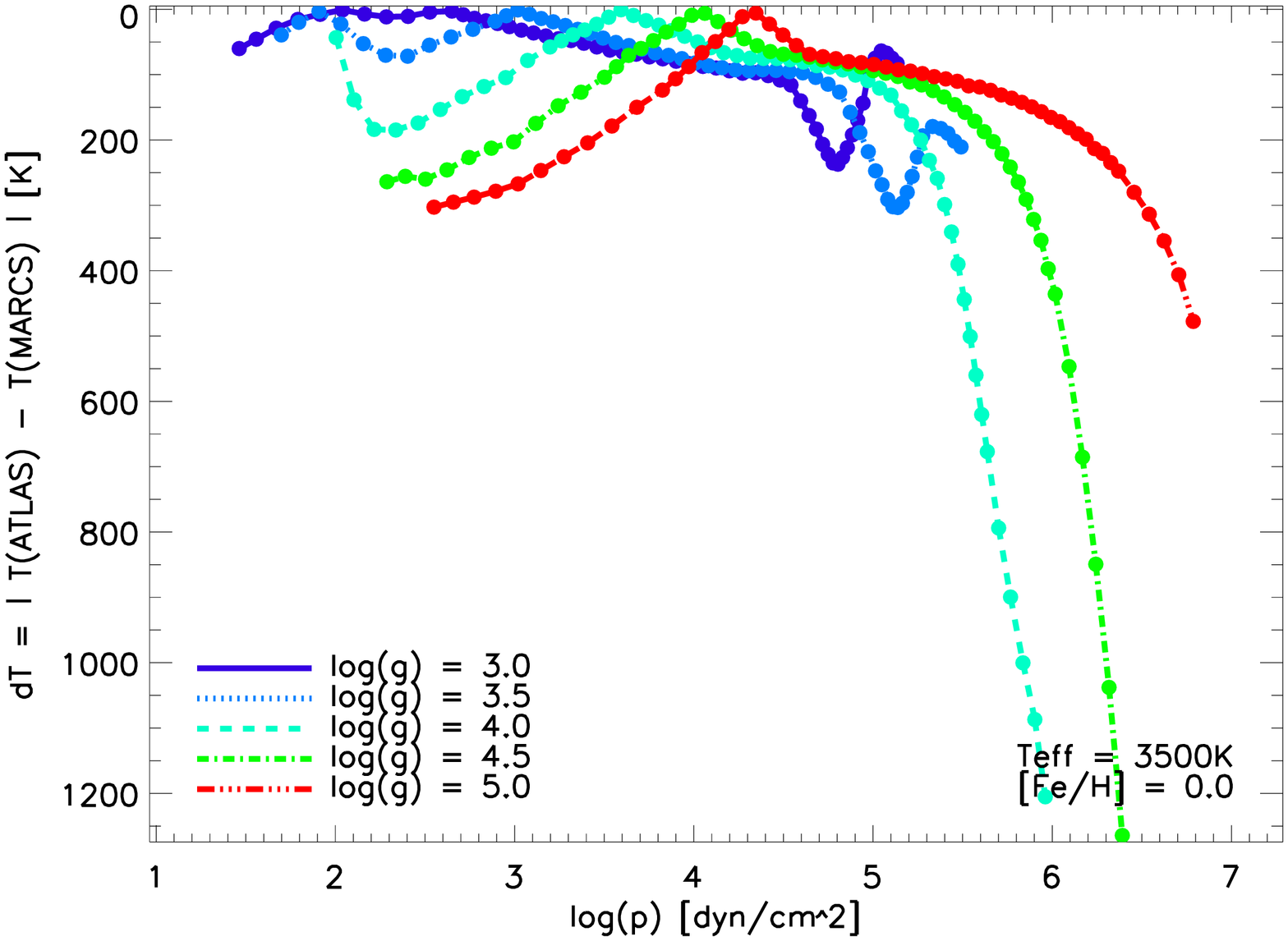} \\
	      \includegraphics[width=0.31\linewidth, keepaspectratio, angle=90]{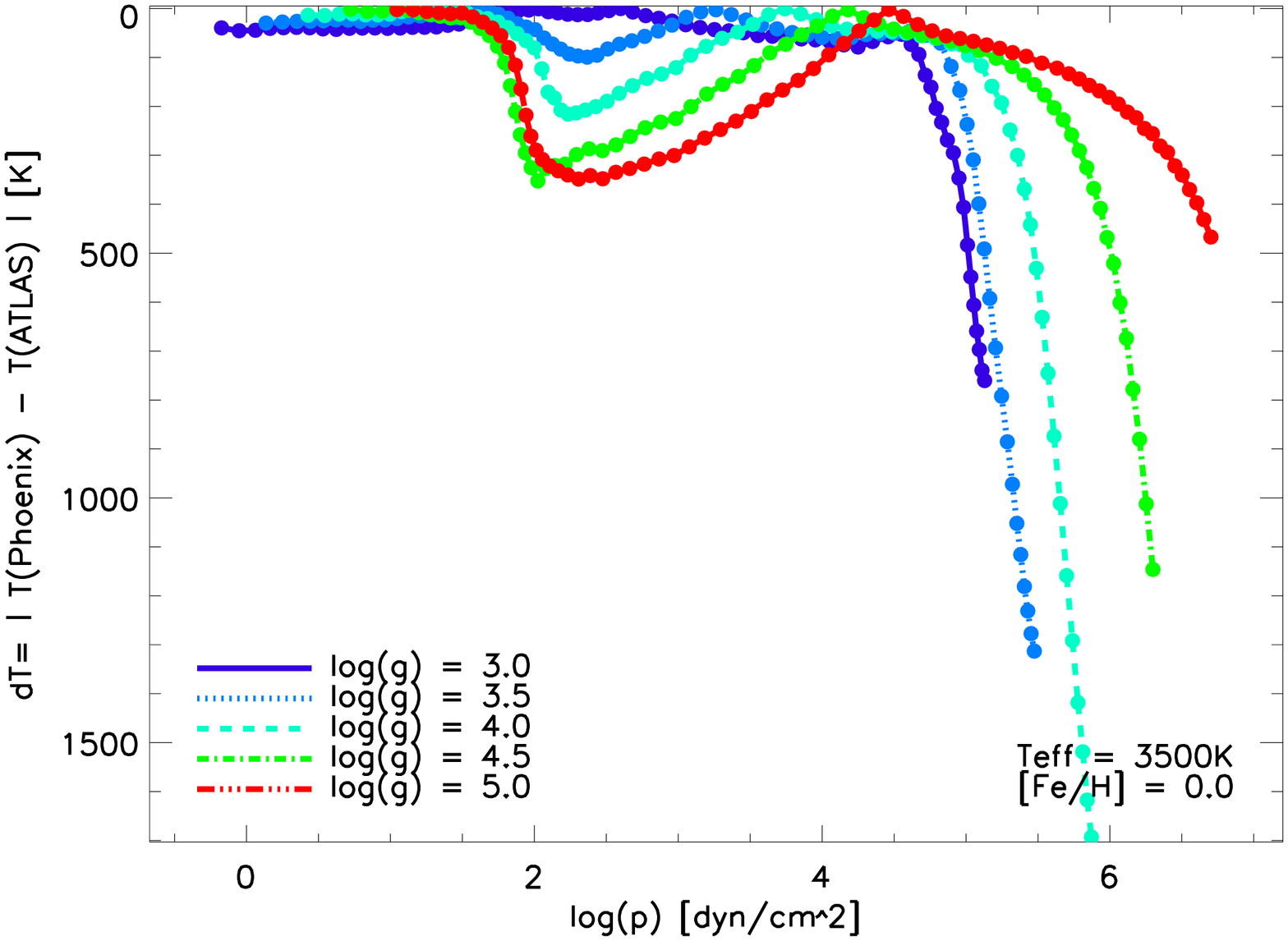}
	      
	    \end{tabular}

	  \caption{
		    {\bf Top row:} local gas temperature-pressure structures for T$_{\rm eff}$ = 4000K  (log(g) = 3.0: top left;  			log(g) = 5.0: top right) for {\sc MARCS}, {\sc Phoenix} and {\sc ATLAS}. {\bf Middle row:} left - residual 
		    temperature values $d T_{\rm gas}$ between {\sc Phoenix} and {\sc MARCS}; right - $d T_{\rm gas}$  between 				{\sc ATLAS} and {\sc MARCS}. {\bf Bottom row:} $d T_{\rm gas}$  between {\sc Phoenix} and {\sc ATLAS} model 				atmosphere results. Mixing length parameter: {\sc Phoenix} - $l/H_{\rm p}$=1.82 for log(g) = 3.0 and $l/H_{\rm p}$=1.96 for log(g) = 5.0; {\sc ATLAS} - $l/H_{\rm p}$=1.25 for 	all models; {\sc MARCS} - $l/H_{\rm p}$=1.5 for all models. 
		    }
		    
	  \label{tp4000}
	  
	\end{figure*}
  
 Figures \ref{tp2500} -- \ref{tp4000} present  the comparison of the (T$_{\rm
  gas}$, p$_{\rm gas}$)-structures  of {\sc MARCS}, {\sc
  Phoenix}, {\sc Drift-Phoenix} and {\sc ATLAS}\footnote{The 'kink' in the {\sc ATLAS} local temperature-pressure profile in Fig \ref{tp3500}, top row, right panel, is visible in other models with T$_{\rm eff} = 3000K$ and solar metallicity.} for solar metallicity and T$_{\rm
  eff}$ = 2500, 3000, 3500 and 4000K, respectively.  We observe better
agreement between {\sc MARCS}, {\sc Phoenix} and {\sc Drift-Phoenix} for T$_{\rm eff}$ = 2500K than
 for higher T$_{\rm eff}$ models. The 2500K sets of models do not vary by more than 300K (except for high pressure values). For all effective
temperatures (T$_{\rm eff}$=2500-4000K), the model atmospheres with higher surface gravity  (brown dwarfs)  agree better between different
model families than those with lower surface gravity (giant gas planets, young brown dwarfs). Note these differences are hard to see in the top rows of Figs. \ref{tp2500} -- \ref{tp4000} due to the scale of the plots. For this reason we provide plots of the calculated residuals in rows 2 and 3 of the figures.    
  
  We compare the hot {\sc ATLAS} and {\sc MARCS}
models for T$_{\rm eff}$ = 3500K and T$_{\rm eff}$ = 4000K.  While the
T$_{\rm eff}$ = 3500K models compare better in the low
metallicity range -1.5 $<$ [M/H] $<$ -2.5, the T$_{\rm eff}$ = 4000K
models display better agreement for higher metalicities [M/H] = +0.5 and [M/H] = 0.0. For
both T$_{\rm eff}$, the biggest discrepancies lie within the [M/H] =
-1.0 models, with local gas temperature differences $d T_{\rm gas}>1500$K for the T$_{\rm eff}$ = 3500K and  $d T_{\rm gas}>1200$K for the T$_{\rm eff}$ = 4000K case.  All  model families diverge with increasing local
pressure, i.e. deeper in the atmosphere, regardless of  T$_{\rm eff}$, log(g) or metallicity [M/H]. The detailed plots for T$_{\rm eff}$ = 4000K, log(g) = 5.0 can be found in Appendix \ref{s:appB}, Figure \ref{B1}.

In summary, we find that for the higher effective temperature values
(3500K, 4000K) the {\sc ATLAS}, {\sc Phoenix} and {\sc MARCS}  (T$_{\rm
  gas}$, p$_{\rm gas}$)-structures diverge from each other with an average of $\sim$600K in local temperature and in extreme cases well over 1000K. The
{\sc MARCS}, {\sc Phoenix} and {\sc Drift-Phoenix}  differ by an average of
$\sim$300K for 2500K $<$ T$_{\rm eff}$ $<$ 4000K, with some extreme cases of over 1000K. Agreement
improves as the surface gravity increases.

\section{Comparing synthetic photometry}\label{ss:phot}

 The (T$_{\rm gas}$, p$_{\rm gas}$)-structure determines the emergent
 spectral energy distribution for stars.  In order to compare the SEDs
 of the different model atmosphere families, we perform synthetic photometry for all models considered.  We convolve the model SEDs to the (UKIDSS) UKIRT  WFCAM ZYHJK \citep{hewett2006}, 2MASS JHKs \citep{cohen2003} and Johnson UBVRI \citep{johnson} filter systems, based on  codes used in \cite{sinclair}. The wavelength ranges for these used  filters are summarized in Table \ref{filters}.

  The convolved broad-band flux F$_{\rm R}$ is given by \citep{straizys}
    \begin{equation}
	    F_R(\lambda) = \frac {\int_{\lambda_1}^{\lambda_2}F(\lambda)R(\lambda)d\lambda} {\int_{\lambda_1}^{\lambda_2}R(\lambda)d\lambda}
    \end{equation}
   where R($\lambda$) is the throughput function (only filter transmission for the optical (UBVRI) bands, but filter transmission plus detector throughput for 2MASS/UKIDSS); and 
  $\lambda_{\rm 1}$ and $\lambda_{\rm 2}$ are the limits of the filter
  wavelength range.  Zero-point calibration is performed using the HST
  spectrum of Vega \citep{vega}.
  
  We proceed to calculate synthetic colour indices for each model atmosphere family. The colour indices are defines as
  \begin{equation}
      m_1 - m_2 = -2.5(\rm{ log(\frac{F_{\rm R1}}{F_{\rm R1,Vega}}) + log(\frac{F_{\rm R2}}{F_{\rm R2,Vega}}) }).
  \end{equation}
A complete set of the synthetic photometry results is provided in Appendix~\ref{s:appC}. In the following,  we compare the ratios between the synthetic broad-band fluxes for all pairs of corresponding models in each filter. The closer the value to 1.0, the more similar the model atmosphere results are.
  
  \subsection{Optical bands (Johnson UBVRI filters)}

      The broad-band fluxes of {\sc ATLAS} and {\sc MARCS} model atmospheres in the
      optical differ significantly more than
      those in the IR range.  The flux ratios for T$_{\rm eff}$ =
      3500K are deviating from 1.0 significantly more (as high as
      $\sim$1.8) than those for T$_{\rm eff}$=4000K (less than
      $\sim$1.3).  The {\sc ATLAS} models predict more flux than {\sc MARCS} in the U-band for metalicities above [M/H] = -1.0 and then drop to as low as about 20\% less flux for [M/H] $>$ -1.0 for both effective temperatures. In the V-band, the {\sc ATLAS} model predict systematically higher fluxes than {\sc MARCS}. Corresponding plots are provided in Figures \ref{B2} and \ref{B3}.

  \subsection{IR bands}\label{ssIR}
\begin{figure*}
    \begin{tabular}{cc}
    \vspace{6mm}
	\includegraphics[width=0.30\linewidth, keepaspectratio, angle=90]{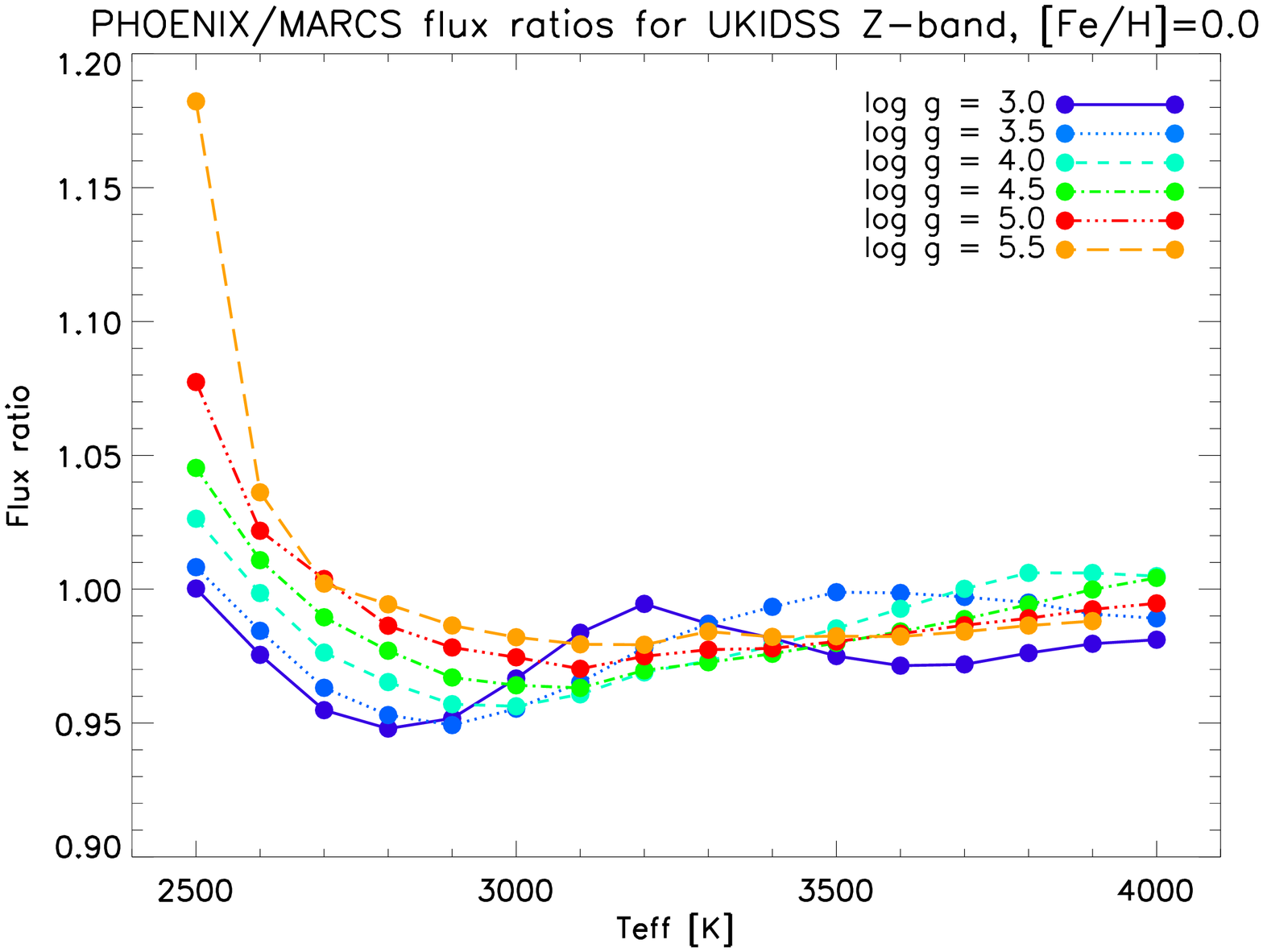}\hspace{10mm}
	\includegraphics[width=0.30\linewidth, keepaspectratio, angle=90]{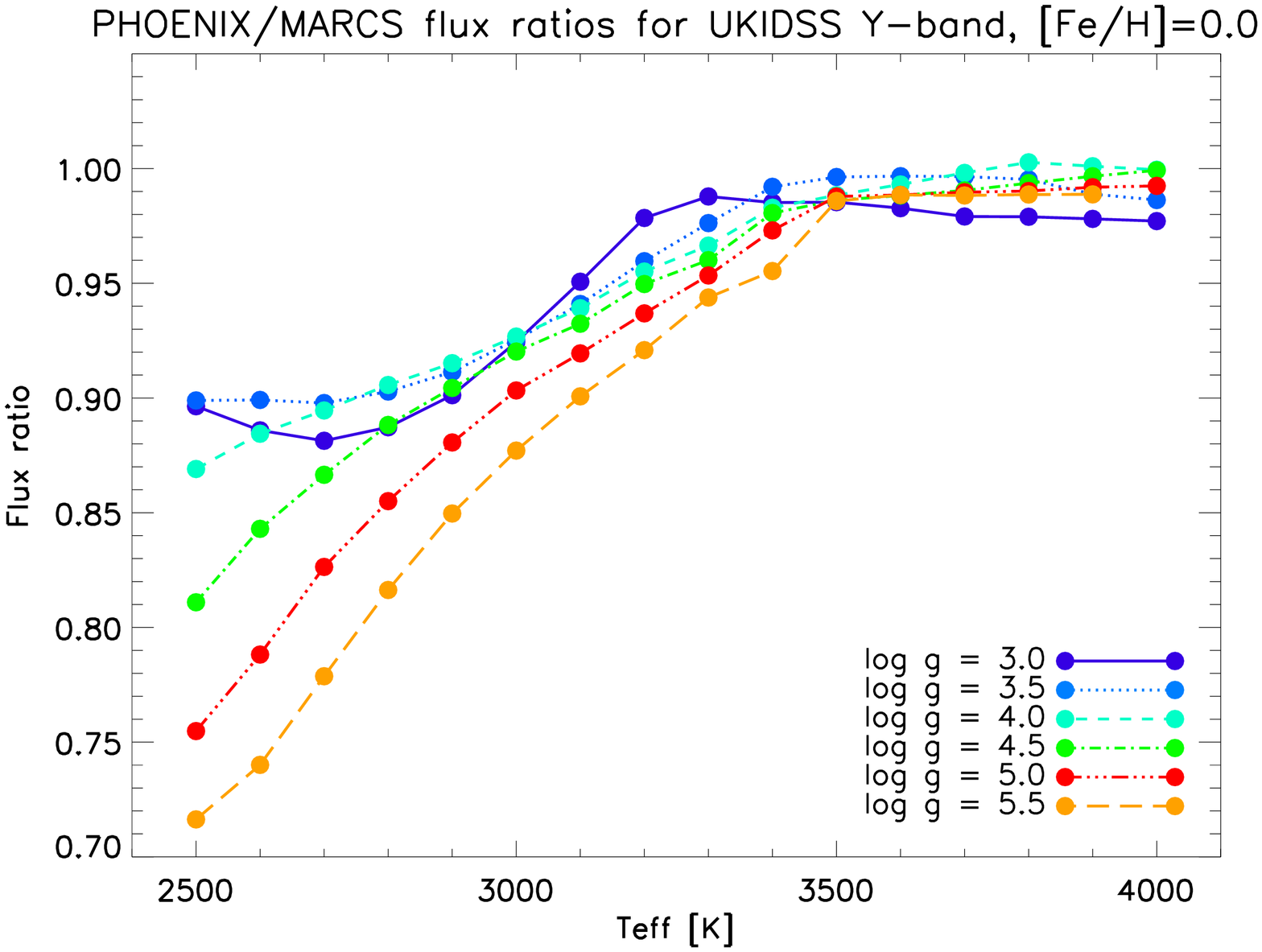}\\
	\vspace{6mm}
	\includegraphics[width=0.30\linewidth, keepaspectratio, angle=90]{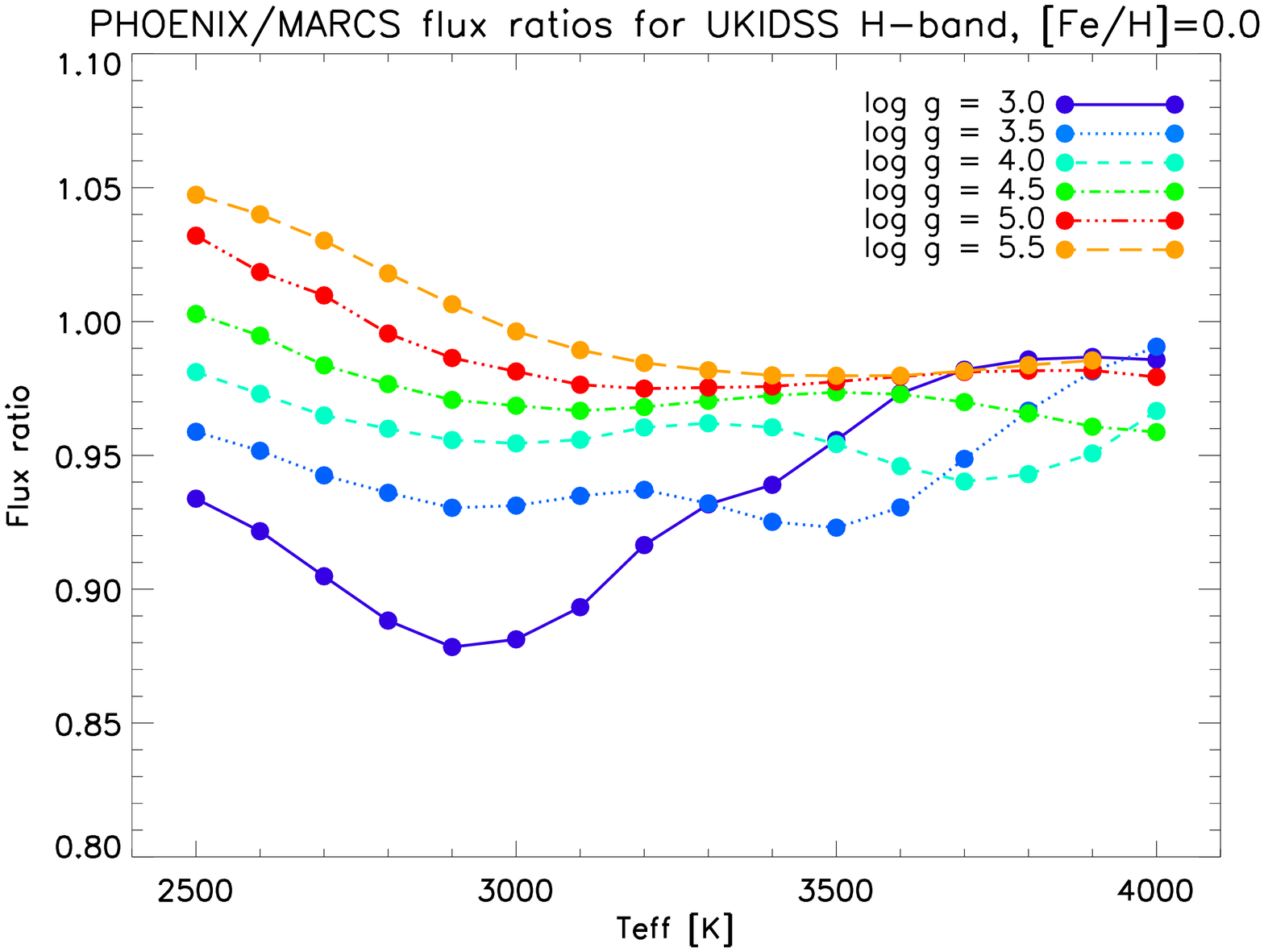}\hspace{10mm}
	\includegraphics[width=0.30\linewidth, keepaspectratio, angle=90]{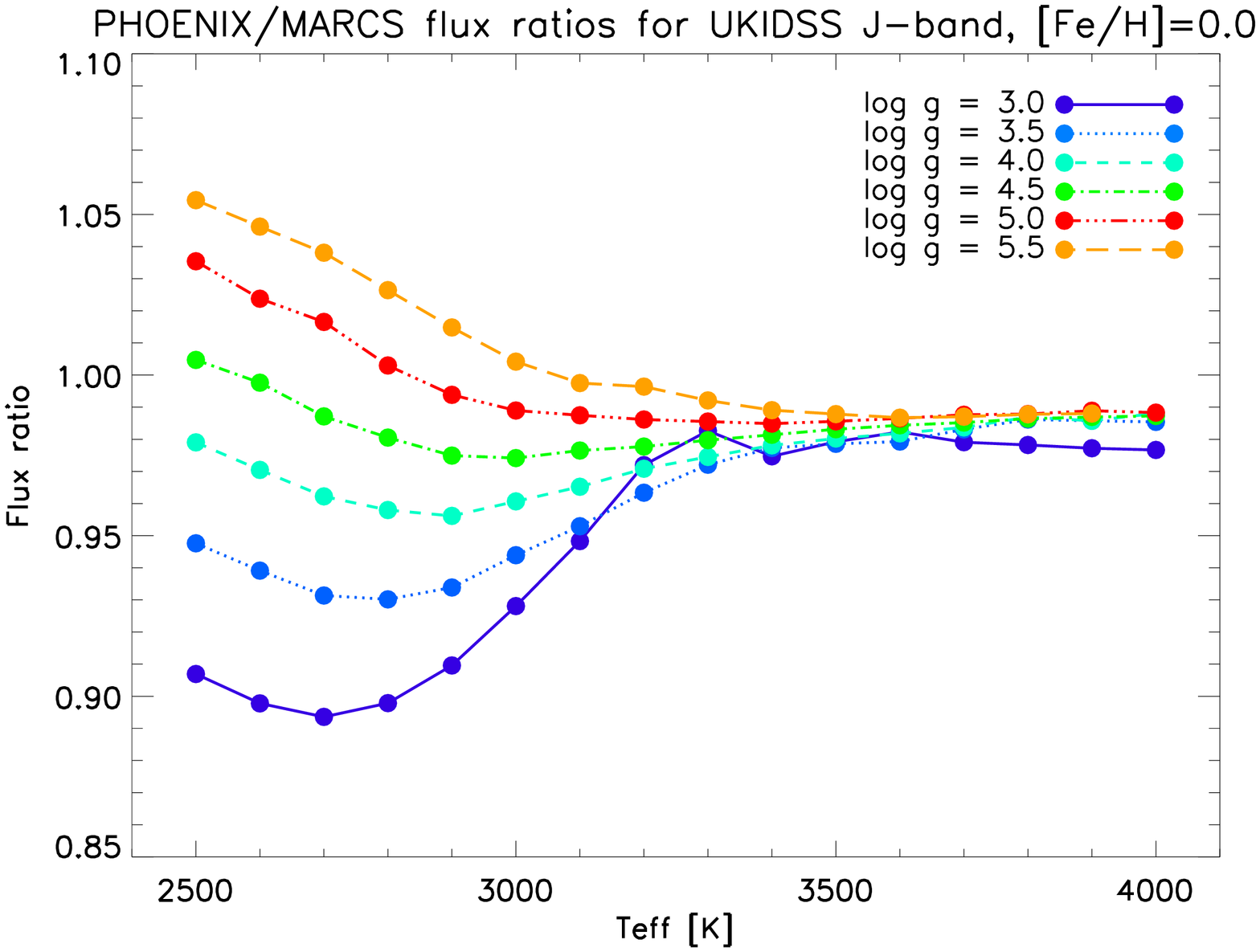}\\
	\includegraphics[width=0.30\linewidth, keepaspectratio, angle=90]{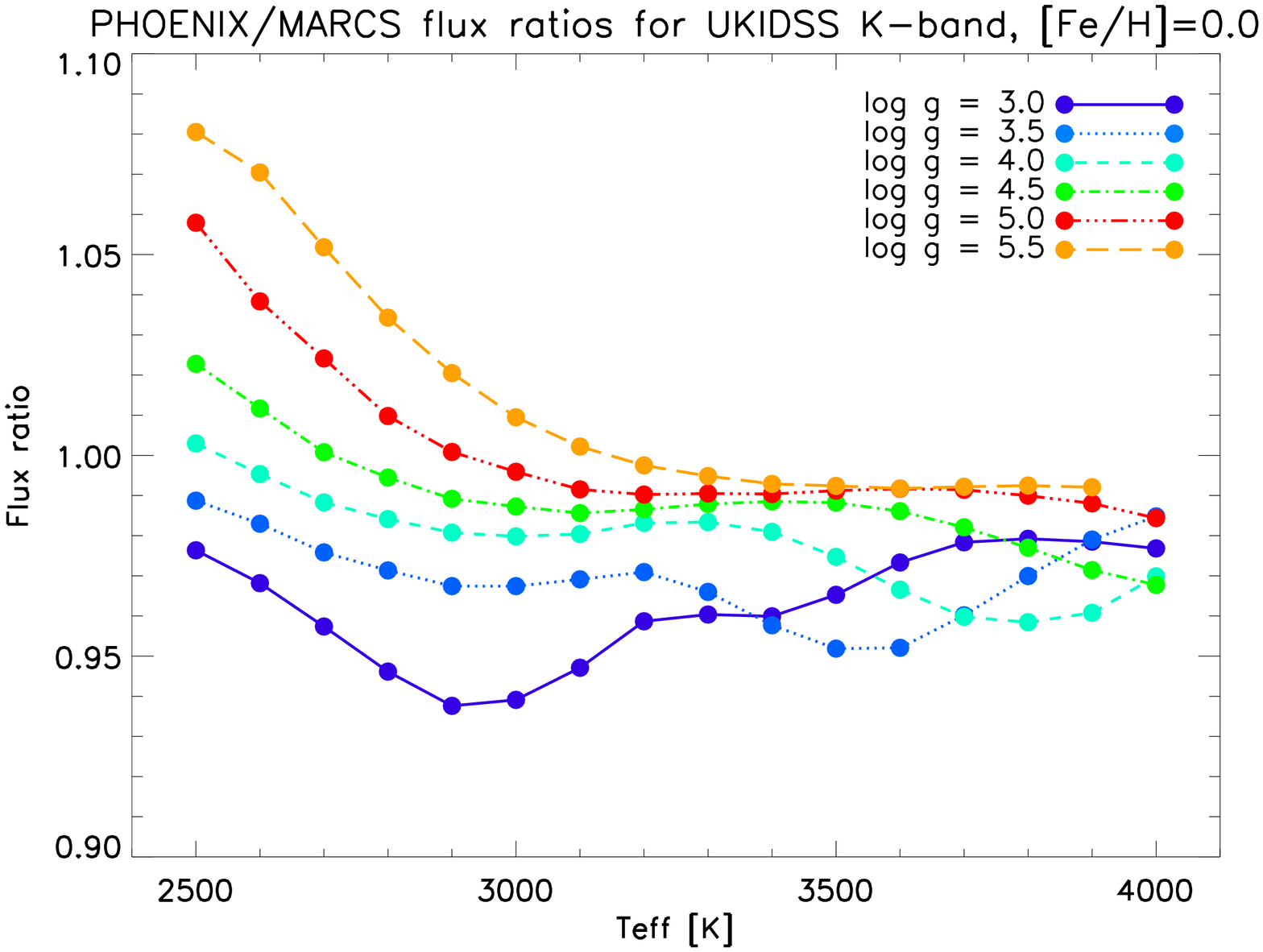}
    \end{tabular}
     \caption{Convolved flux ratios for  {\sc Phoenix} and {\sc
         MARCS} atmosphere models for the UKIDSS filter system.
       Curves are colour coded with respect to log(g) value: orange -
       5.5, red - 5.0, green - 4.5, cyan - 4.0, blue - 3.5, dark blue
       - 3.0. All models here are of solar metallicity.  }
    \label{fluxRatios1}
  
  \end{figure*}

\begin{figure*}
    \begin{tabular}{cc}
    \vspace{6mm}
	\includegraphics[width=0.30\linewidth, keepaspectratio, angle=90]{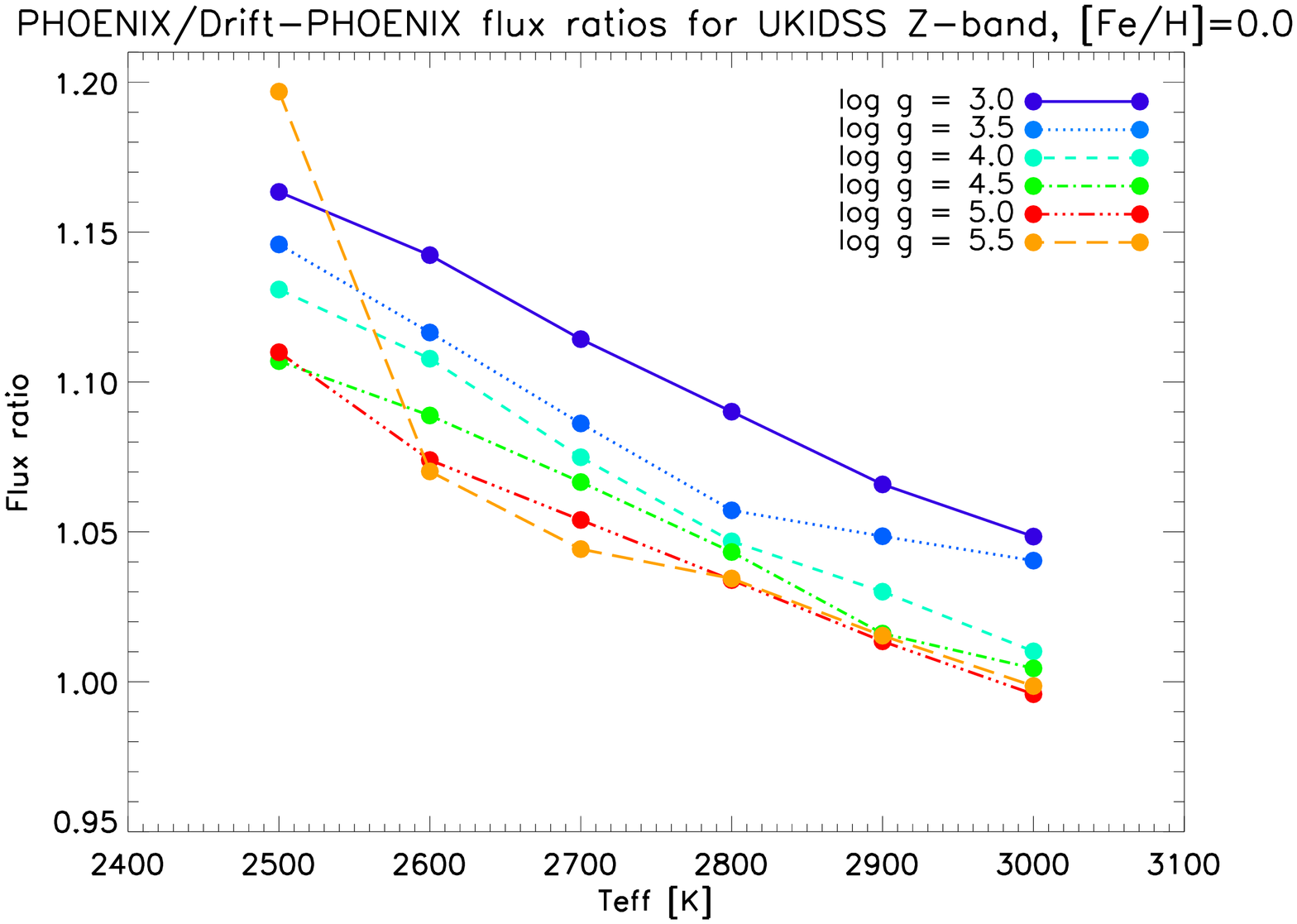}\hspace{10mm}
	\includegraphics[width=0.30\linewidth, keepaspectratio, angle=90]{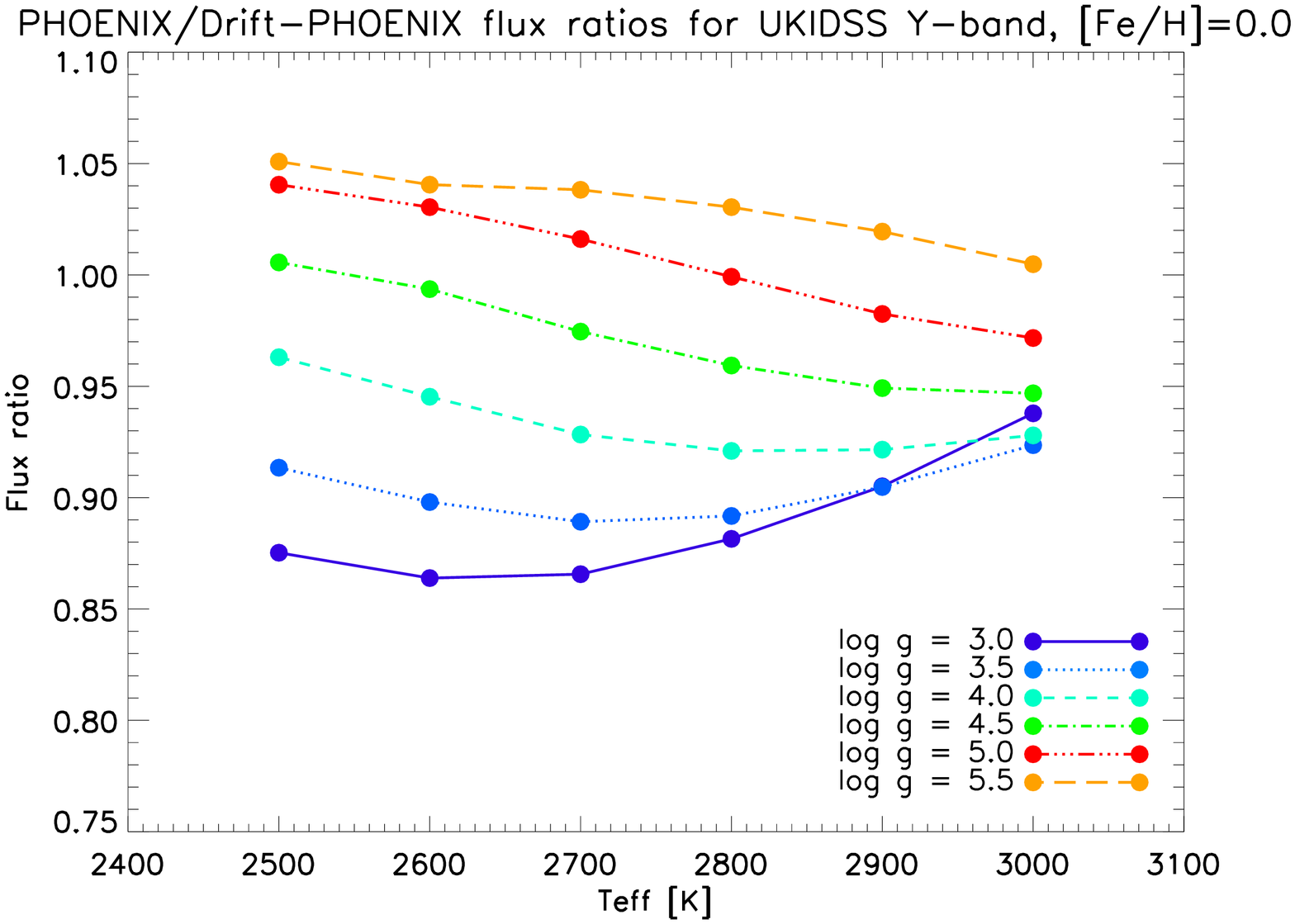}\\
	\vspace{6mm}
	\includegraphics[width=0.30\linewidth, keepaspectratio, angle=90]{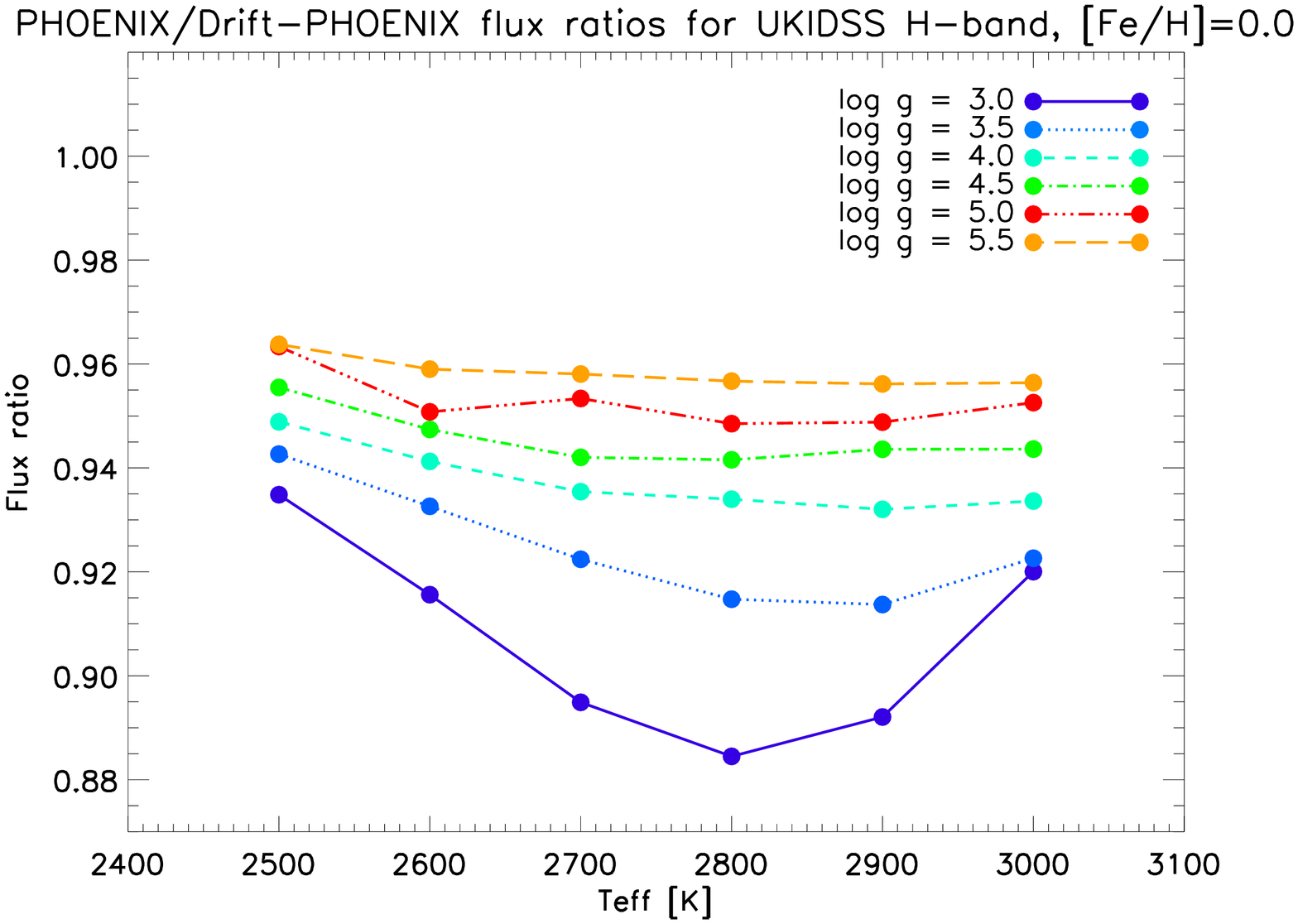}\hspace{10mm}
	\includegraphics[width=0.30\linewidth, keepaspectratio, angle=90]{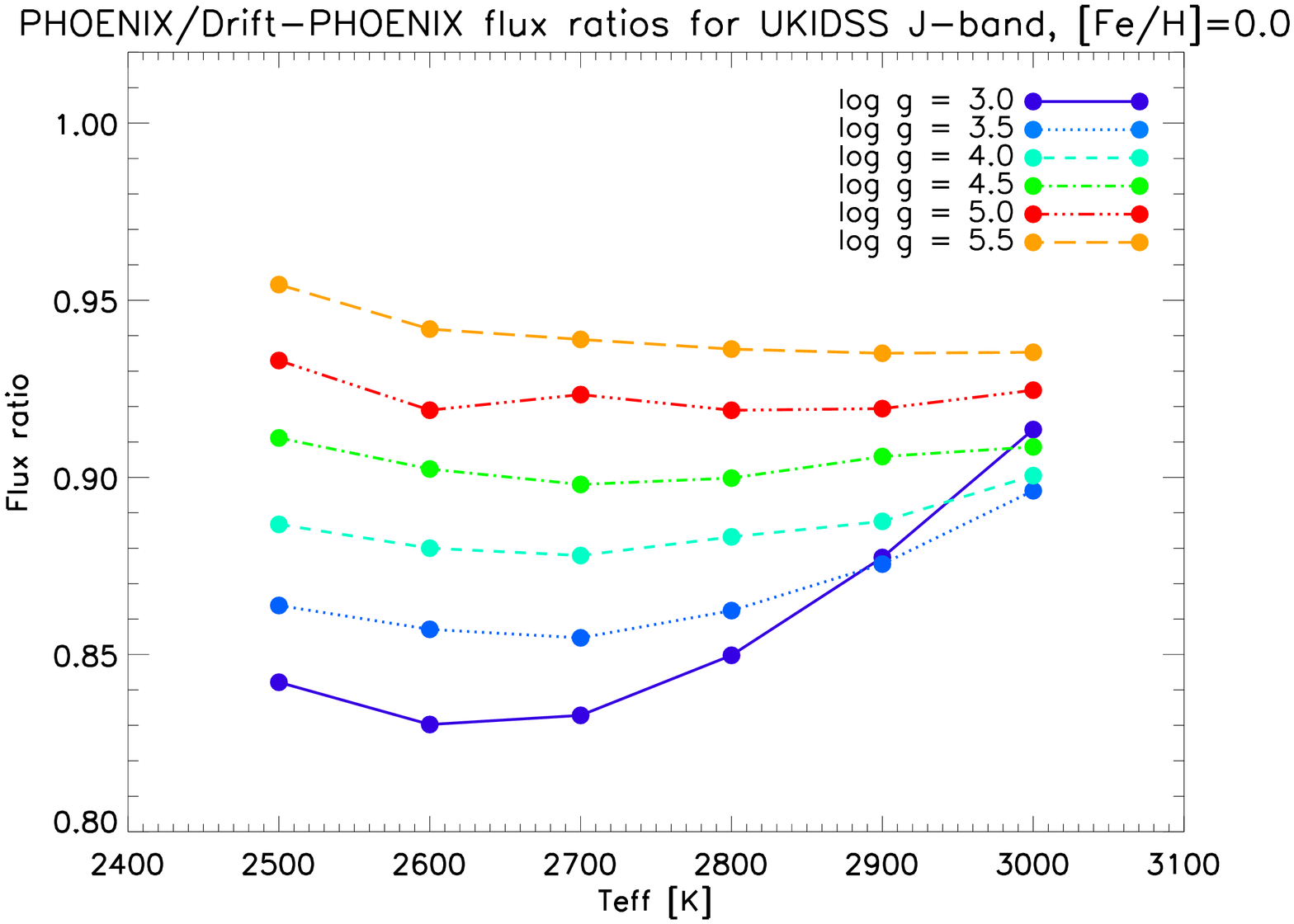}\\
	\includegraphics[width=0.30\linewidth, keepaspectratio, angle=90]{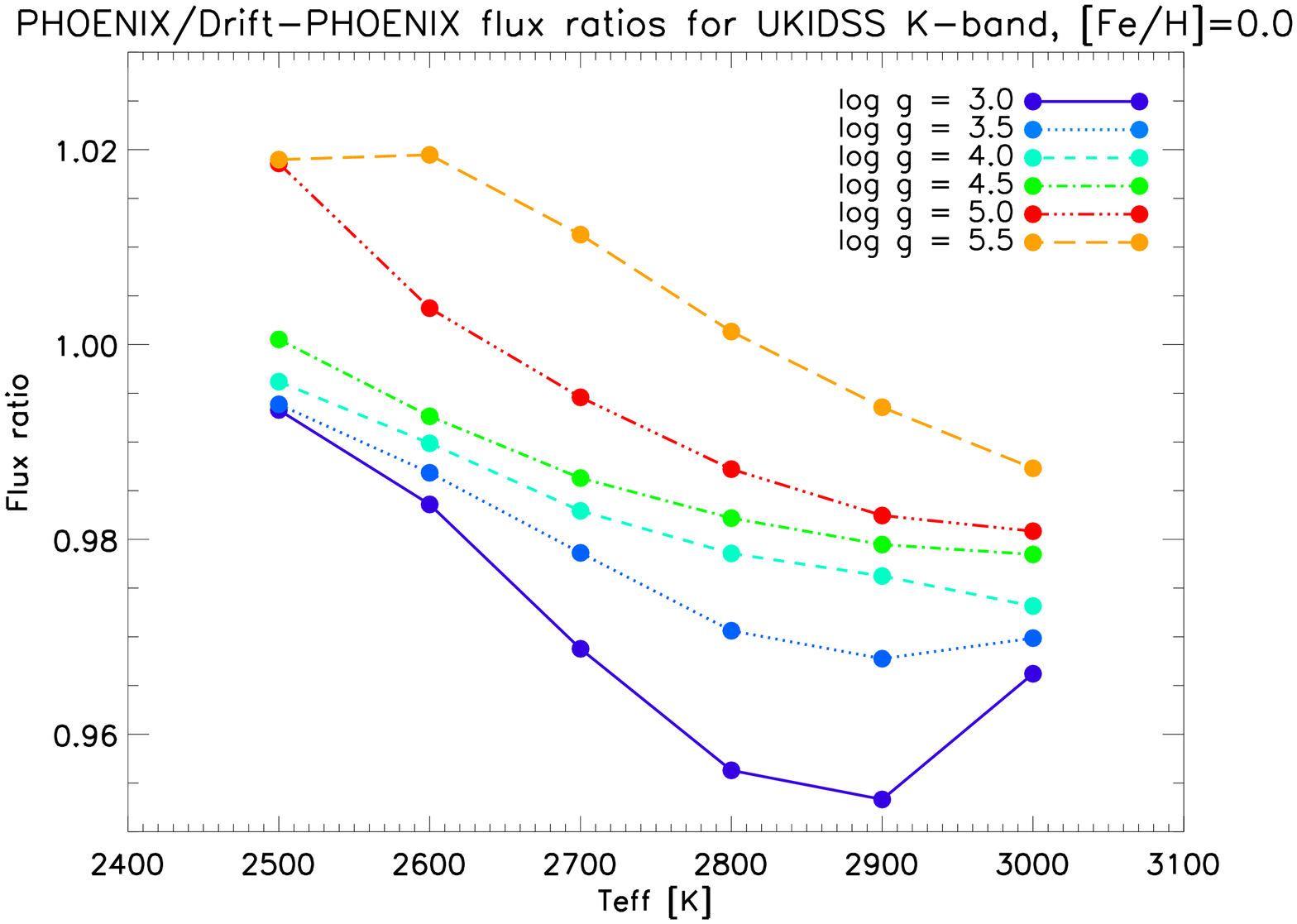}	    \end{tabular}    
     \caption{Convolved flux ratios for  {\sc Phoenix} and {\sc
         Drift-Phoenix} atmosphere models for the UKIDSS filter
       system. }
    \label{fluxRatios2}
  \end{figure*}

Figures \ref{fluxRatios1} and \ref{fluxRatios2} represent the UKIDSS
photometric flux ratios for  {\sc Phoenix} to {\sc MARCS} and {\sc
  Phoenix} to {\sc Drift-Phoenix}.  {\sc Phoenix} and {\sc MARCS}
display a good agreement in the range T$_{\rm eff}$ = 3500$-$4000K. A
possible explanation for the decreasing discrepancies in this T$_{\rm
  eff}$ range relative to T$_{\rm eff} < $3000K is the lack of dust as
effective temperature rises. Dust should not have an impact on the
atmospheric structures in models with T$_{\rm eff} > $3000K \citep{witte}.

For decreasing effective temperature, the {\sc Phoenix} models
systematically predict more flux in the IR bands than the {\sc
  MARCS} model atmospheres for higher log(g) values and less flux than {\sc MARCS} for low log(g). The Y-band band is an exception to this trend,
where, for T$_{\rm eff} < $ 3500K all {\sc Phoenix} models predict
less flux than {\sc MARCS} with the flux ratio dropping as low as 0.7. Both model families do not include cloud formation in the model atmospheres considered here, hence, the differences in fluxes may point to differences in the molecular opacities (line lists and/or gas-phase chemistry data).      
  
We compare the {\sc Phoenix} and {\sc Drift-Phoenix} model atmospheres (2500K $\leq$ T$_{\rm eff}$ $\leq$ 3000K) in order to check if the
difference in dust treatment is sufficient to explain differences in synthetic fluxes (Fig.
\ref{fluxRatios2}).  Differences between these two
families are generally smaller than in comparisons with the {\sc MARCS} model atmospheres. 
For the H and J bands, all {\sc Phoenix} atmosphere models predict less flux than
{\sc Drift-Phoenix}. This result is unexpected as these bands are heavily affected by dust and the {\sc Phoenix} models are dust-free. Therefore, while still an important factor, the dust treatment alone cannot explain the observed trends in the comparison of the synthetic fluxes. All models produce very similar fluxes in the K band. For all bands, except in the Z band, the flux ratio is highest for the higher surface gravity  values. This trend is reversed in the Z band, which also
appears to vary the most with change in T$_{\rm eff}$.
  
\begin{figure*}
      \begin{tabular}{cc}
      \vspace{5mm}
	\includegraphics[width=0.312\linewidth, keepaspectratio, angle=90]{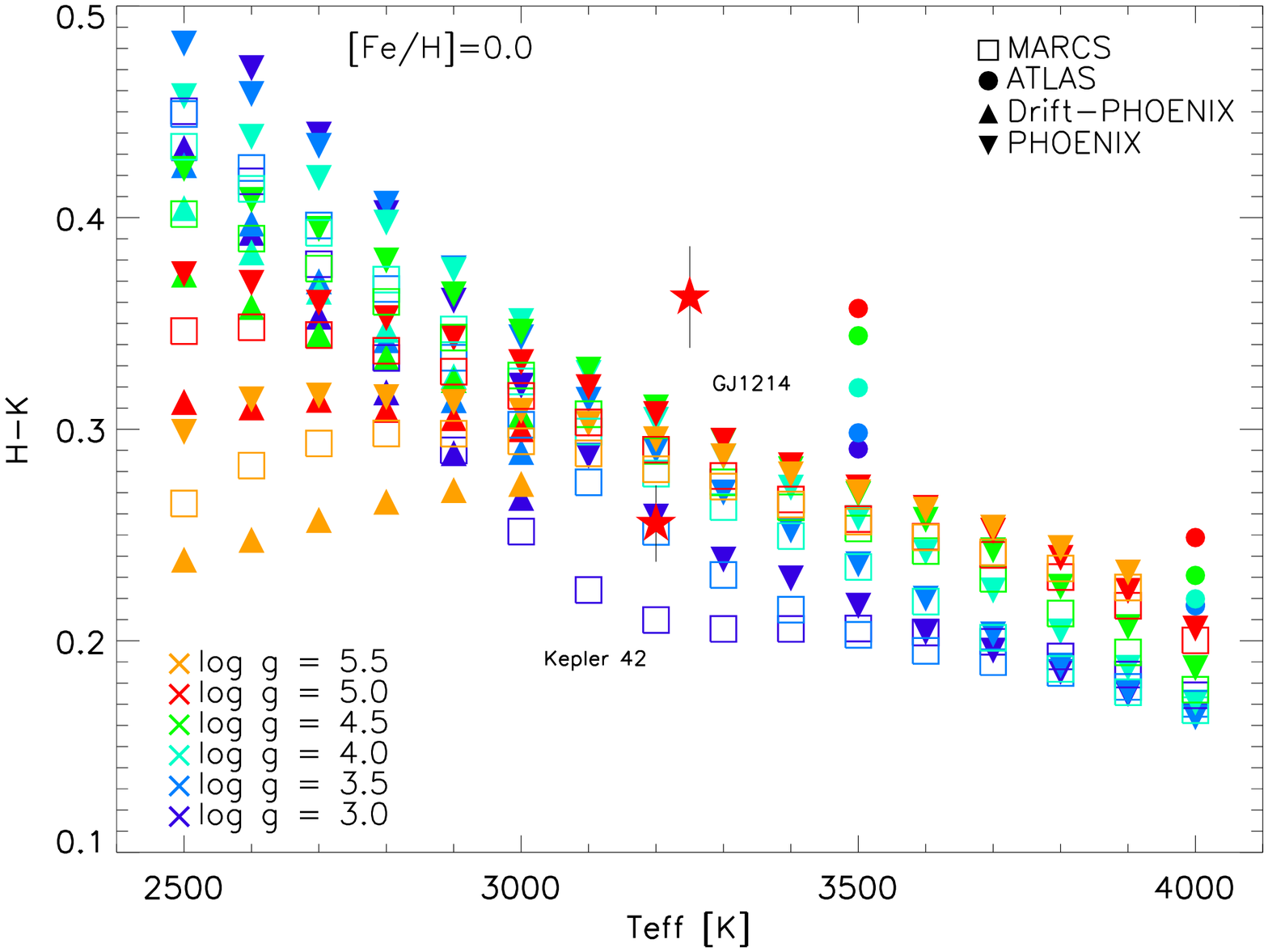}\hspace{5mm}
	\includegraphics[width=0.312\linewidth, keepaspectratio, angle=90]{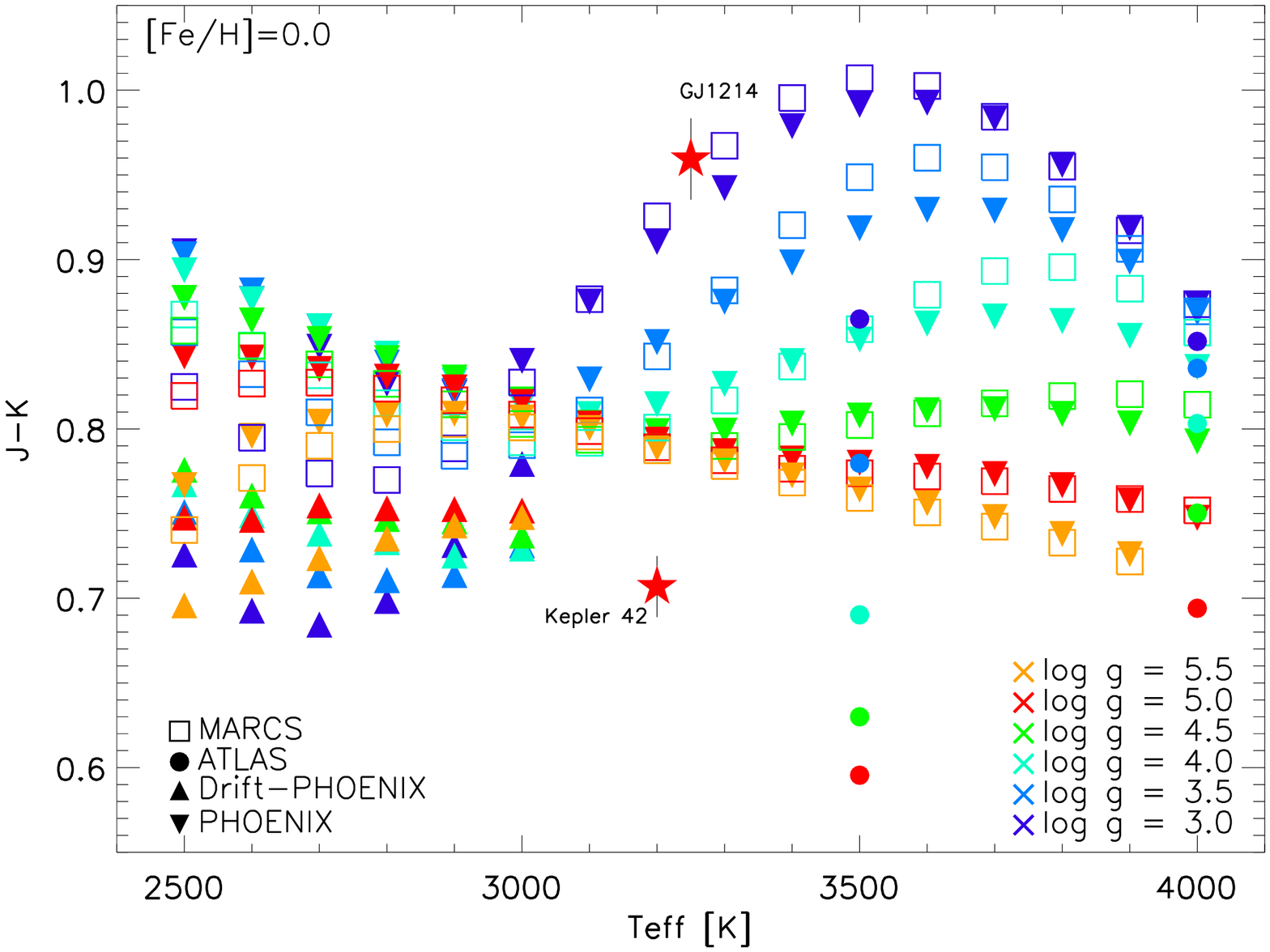}\\
	\includegraphics[width=0.30\linewidth, keepaspectratio, angle=90]{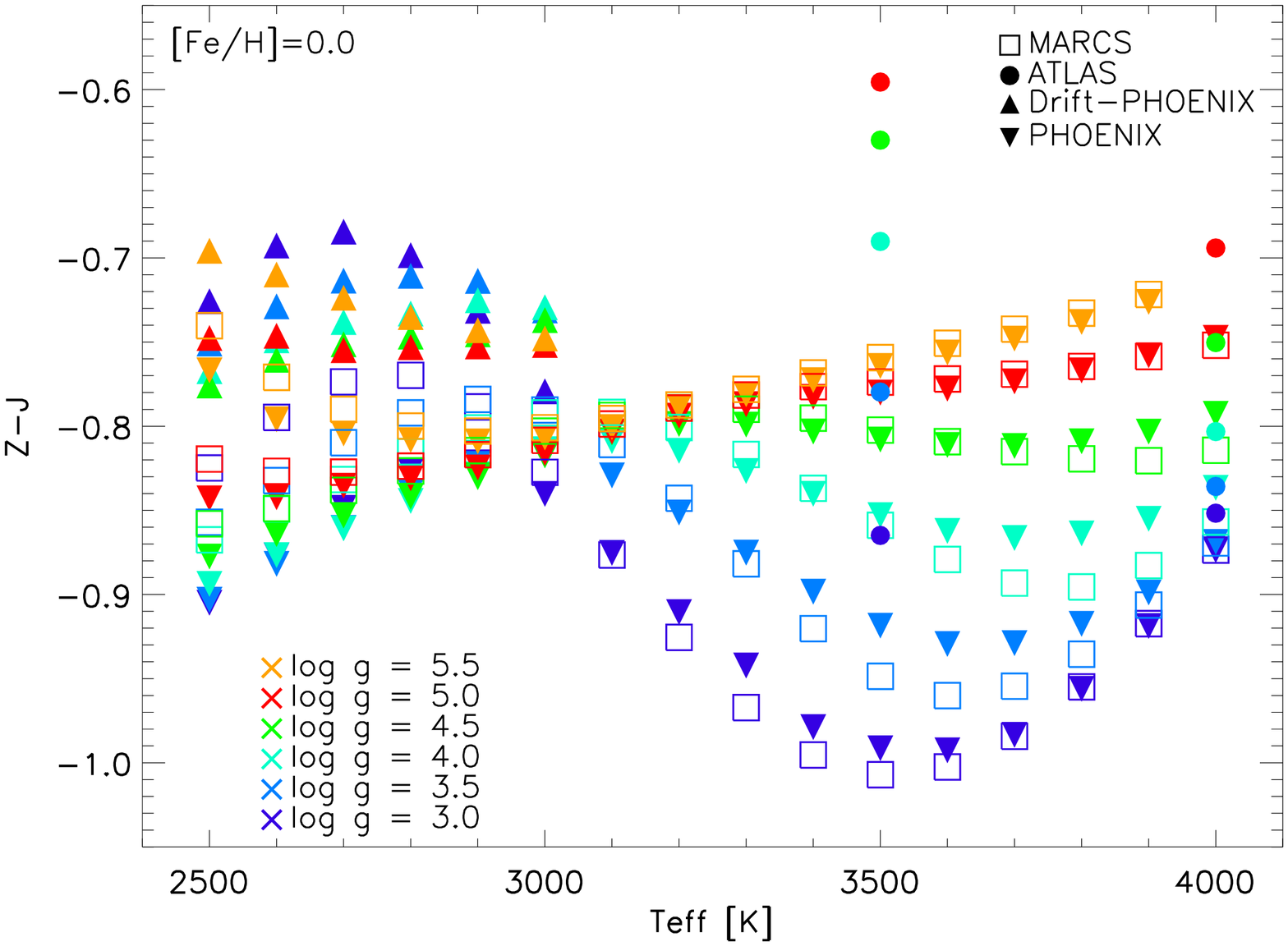}\hspace{5mm}
	\includegraphics[width=0.30\linewidth, keepaspectratio, angle=90]{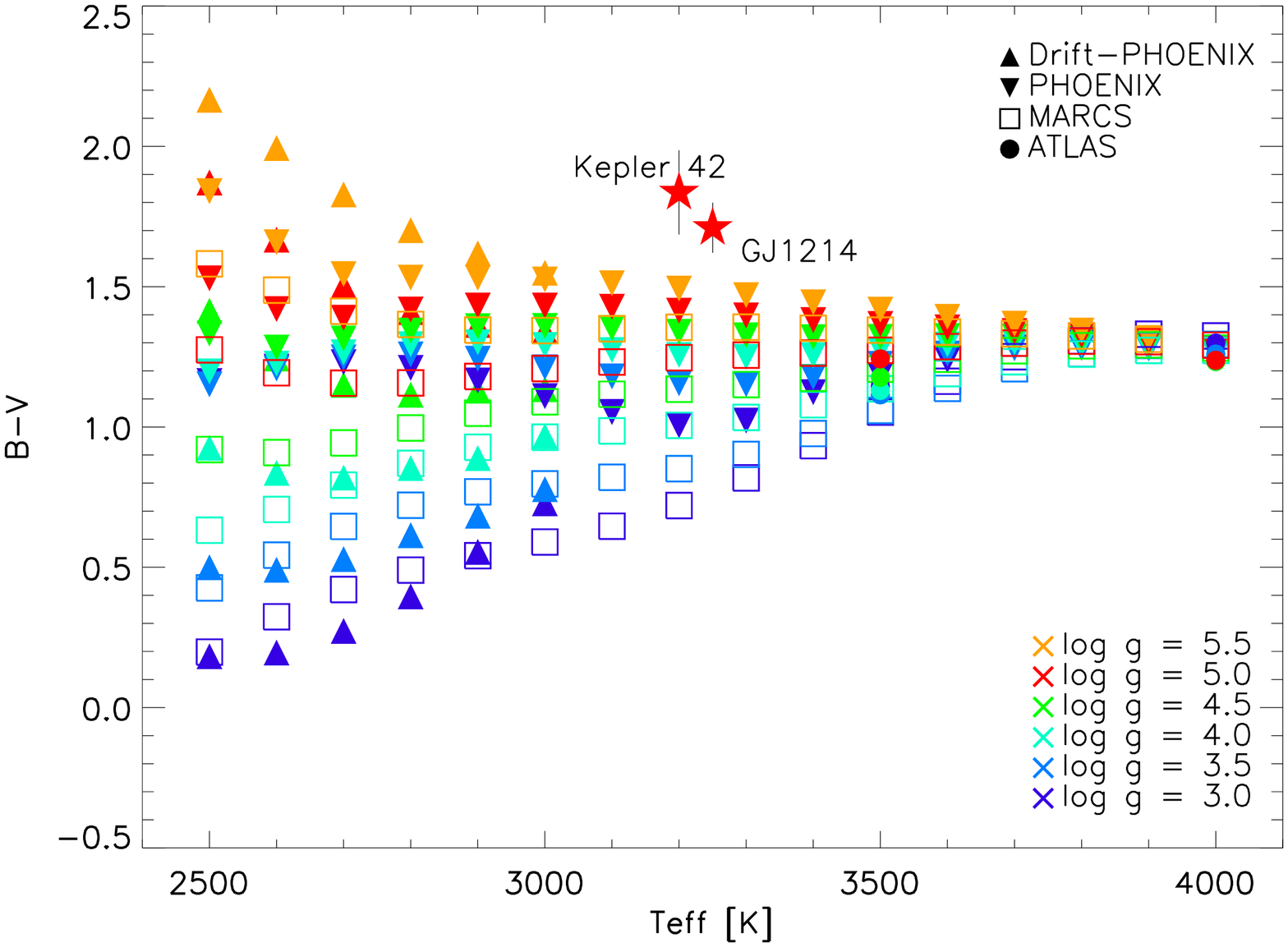}
        \end{tabular}
     \caption{ Colour indices versus effective temperature for all
       model atmosphere families (circles - {\sc ATLAS}, squares -
       {\sc MARCS}, lower triangles - {\sc Phoenix}, upper triangles -
       {\sc Drift-Phoenix}) of solar metallicity.  The first three plots refer to the UKIDSS filter
       system. The two red stars are observed data for GJ 1214 and Kepler 42 (\citealt{GJ1214, Kepler42, cutri}). The 			   2MASS measurements for these two stars have been shifted into the UKIDSS system using the transformations given 		   in \citealt{hewett2006}. Colour coding is used for different values for log(g) with a step of 0.5
       dex. \bf {\bf Top left:} For lower T$_{\rm eff}$, models with lower log(g) have higher H-K values than models with high log(g) . The trend inverses at higher temperatures. {\bf Top right:} No clear trend with respect to log(g) is visible for lower T$_{\rm eff}$. At higher temperatures models with lower log(g) show higher J-K values. {\bf Bottom left:} No trend for lower T$_{\rm eff}$. At higher temperatures models with higher log(g) have higher Z-J values. {\bf Bottom right:} For low temperatures, models with high log(g) show a higher B-V value. The B-V difference with respect to log(g) diminishes for higher T$_{\rm eff}$.}
    \label{colours}
\end{figure*}

  We also present the colour indices calculated for each model
  atmosphere family considered here (Figure \ref{colours} and Appendix~\ref{s:appC}).  All colours show considerable differences for model atmospheres T$_{\rm eff}< 3000$K. 
  Dust  starts to form in small amounts at T$_{\rm
    eff}\approx2700$K and the resulting element depletion of the
  gas-phase may contribute to the increasing differences with
  decreasing T$_{\rm eff}$ below 2700K \citep{witte}.
  In particular, the B-V magnitudes differ by up to half a magnitude
  between the {\sc Drift-Phoenix} and {\sc MARCS} models in the low
  temperature half of the plot. The {\sc ATLAS} models appear to
  differ significantly from all other model families considered here.

\section{Discussion}\label{ss:dis}
	
  \subsection{Different model assumption}\label{ss:diffm}
  
 The differences in the atmospheric (T$_{\rm gas}$, p$_{\rm
   gas}$)-structures and the resulting SEDs arises from differences in
 input data (element abundances, opacity sources), physical
 assumptions (mixing length, overshooting, dust/ no dust), the
 choice of material values (equilibrium constants, line lists), but also from more technical details like convergence criteria and/or inner/outer boundary choices. It is outside the scope of this paper to identify in more detail why the
 model atmosphere results differ as this would require a dedicated
 benchmark study.
    
The {\sc ATLAS} atmosphere models were developed 
for hotter stars and cover a wide range of metallicities, surface
gravities and effective temperatures, from hot O and B down to early
type M stars. The latest models use improved opacity distribution
functions (ODFs) as described in \cite{castelli}. The atomic and molecular line lists
 for the new ODFs are from the old \cite{kurucz90} ODFs with some changes. 
 A new TiO list from \cite{schwenke98} is used. Additionally
 H$_{\rm 2}$O lines are adopted from \cite{partridge97}. Furthermore extra bands have been added for some 
 molecules such as CN, OH and SiO.
 Linelest and ODFs can be found on Kurucz\footnote{http://kurucz.harvard.edu/} 
and Castelli's\footnote{http://wwwuser.oats.inaf.it/castelli/} webpages. 
Element abundances are solar and adopted from \cite{grevesse98}. The
models are calculated assuming mixing-length theory  without
overshooting, with a mixing length parameter l$ \slash H_{\rm p}$ =
1.25 and line-broadening by a micro-turbulent velocity of v$_{\rm turb}$ = 2.0 km$\slash$s.
    
The {\sc MARCS} models \citep{gustafsson} have focused on
F, G and K stars  extending into the M-dwarf regime. The
opacity sampling method is used. Models are available for
micro-turbulence velocities of v$_{\rm turb}$ = 0, 1, 2 and 5
km$\slash$s (for comparison purposes with ATLAS, we have only
considered a value of 2 km$\slash$s). The mixing length parameter value is $1/  H_{\rm p}$=1.5.
The models are also divided in several
metal abundance groups, out of which we consider the one with
abundances from \cite{grevesse07}. Molecular opacity sources
include HCN, H$_{\rm 2}$O, C$_{\rm 2}$, C$_{\rm 3}$, C$_{\rm
  2}$H$_{\rm 2}$, CH, CN, CaH, FeH, MgH, NH, OH, SiH, SiO, TiO, VO and
ZrO. Continuous absorption sources are H I, H$^-$, H$_{\rm 2}^-$,
H$_{\rm 2}^+$, He I, He$^-$, C I, C II, C$^-$, N I, N II, N$^-$, O I,
O II, O$^-$, Mg I, Mg II, Al I, Al II, Si I, Si II, Ca I, Ca II, Fe I,
Fe II, CH, OH, CO$^-$ and H$_{\rm 2}$O$^-$. The codes also include
collision-induces absorption from H I + H I, H I + He I, H$_{\rm 2}$ +
H I, H$_{\rm 2}$ + H$_{\rm 2}$, H$_{\rm 2}$ + He I; continuous
electron scattering and Rayleigh scattering for H I, H$_{\rm 2}$ and
He I. The authors suggest that the only significant difference using 
different molecular opacity sources comes from CO, H$_{\rm 2}$O and TiO.
CO sources adopted by \citep{gustafsson} (Table 2) are from \cite{goorvitch94}
and \cite{kurucz95}, H$_{\rm 2}$O from \cite{barber06} and TiO from \cite{plez98}. 
    
The {\sc Phoenix} models \citep{husser} are based on the \cite{hausch}
stellar atmosphere code. The gas-phase chemistry is treated with
  the Astrophysical Chemical Equilibrium Solver (ACES,
  \citealt{witte2011}).  Husser et al. note that while condensation is
  included as element sink in the equation of state, it is omitted
  from opacity calculations and additionally no dust settlement is
  included in any of the models. The gas opacity species (line and
  continuum) are the same like in {\sc Drfit-Phoenix} (see below).
  The code uses mixing length theory, with $1/ H_{\rm p} \sim 1.8
  \ldots 3.5$ for the M-dwarf parameter space.  The micro-turbulent
  velocity is linked to the convective velocity that results from
  mixing-lenth theory (MLT). However, the micro-turbulent velocity it
  is only considered in the calculations of the high-resolution
  spectra, but not used for the computation of the atmospheric
  structure. Based on this assumption, v$_{\rm turb}<$ 1 km$\slash$s
  for {\sc Phoenix} model atmospheres in the M-dwarf parameter range.
    
The {\sc Drift-Phoenix} models are aimed at brown dwarf and planet
atmospheres. They are a combination of the {\sc Phoenix} atmosphere
code \citep{hausch}, version 16.00.02A, and the {\sc Drift} module
(\citealt{witte}, \citealt{hellingc}) that models cloud
formation. {\sc Drift} solves a system of element conservation and
dust moment equations in phase non-equilibrium including the processes
of dust nucleation, growth and/or evaporation.  The influence of
gravitational settling and element replenishment by convective
overshooting is considered in relation to the formation processes. Six
main elements are considered in these processes - Ti, O, Al, Fe, Si
and Mg, together with the seven most important solids consisting of
these elements - TiO$_{\rm 2}$[s], Al$_{\rm 2}$O$_{\rm 3}$[s], Fe[s],
SiO$_{\rm 2}$[s], MgO[s], MgSiO$_{\rm 3}$[s] and Mg$_{\rm 2}$SiO$_{\rm
  4}$[s]. The line opacity sources considered include H$_{\rm
    2}$, CH, NH, OH, MgH, SiH, CN, SiO, CO$_{\rm 2}$, O$_{\rm 3}$,
  N$_{\rm 2}^{\rm -}$O, CH$_{\rm 4}$, SO$_{\rm 2}$, NH$_{\rm 3}$, HCl,
  N$_{\rm 2}$, VO, CaH, CrH and FeH. Collision induced absorption sources include H$_{\rm
    2}$ - H$_{\rm 2}$, H$_{\rm 2}$ - He, H$_{\rm 2}$ - CH$_{\rm 4}$,
  H$_{\rm 2}$ - N$_{\rm 2}$, N$_{\rm 2}$ - CH$_{\rm 4}$, N$_{\rm 2}$ -
  N$_{\rm 2}$, CH$_{\rm 4}$ - CH$_{\rm 4}$, CO$_{\rm 2}$ - CO$_{\rm
    2}$, Ar - H$_{\rm 2}$ and Ar - CH$_{\rm 4}$. CO lines are adopted
  from \cite{goorvitch94}, H$_{\rm 2}$O from \cite{barber08} and TiO
  from \cite{schwenke98}. Mie and effective medium theory are applied
  to calculate the cloud opacity of mixed grains including the above mentioned solid materials.

   Only the {\sc Phoenix} and {\sc Drift-Phoenix} model families use the same line lists for gas opacity sources, and only {\sc MARCS} and {\sc Drift-Phoenix} use
  the same values for the element abundances \citep{grevesse07}. It is therefore not surprising that
  the model fluxes differ particular for low T$_{\rm eff}$, where the
  influence of molecular and dust opacity is most prominent. In
  addition, there are differences in input parameter values such as
  the mixing length parameter.  \citealt{husser} (their
    Sect. 2.3.3) suggest that the micro-turbulent velocities does have
    no noticeable effect on the atmospheric structure computation
    results.

  We further note that \citep{gustafsson} presented a comparison of the
{\sc Marcs} model atmospheres to {\sc ATLAS} and {\sc Phoenix}
(NextGen) model atmospheres as available at the time. Their comparison
of the ($T_{\rm gas}$, $p_{\rm gas}$)-structures of {\sc Marcs}
and{\sc ATLAS} model atmospheres for giants and supergiants for
log(g)=4.5 and T$_{\rm eff}=4000, 5000, 6000, 7000$K did show a
respectiable agreement between the models. The same holds for their
test of varying metalicities, and for a comparison to NextGen {\sc
  Phoenix} models with (log(g), T$_{\rm eff}$)= (0.0, 3000K), (3.0,
5000K). No radiation fluxes were compared. \cite{plez11} presented
comparison of synthetic Johnson-Cousins UBVRIJHK photometry and
colours of {\sc Marcs}, {\sc ATLAS} and {\sc Phoenix} (NextGen) model
atmospheres for T$_{\rm eff}=3500\,\ldots\,8000$K.  \cite{plez11}
demonstrate that differences do increase with decreasing T$_{\rm eff}$
particulare for T$_{\rm eff}<4000$K. Our study presented in this paper
does support these findings and extend these early model comparisons
into the M-dwarf regime.

 \subsection{Comparing synthetic photometry and observations}
  
 We compare the synthetic photometry for all three model atmosphere
families with two observations:
\begin{itemize}
\item[i)] We compare the synthetic
  photometry results with observations for the M stars
  published in \cite{koen2010}
\item[ ii)] The B-V,  J-K and H-K colours of
two observed M-dwarf planet host stars (Kepler 42, \citealt{Kepler42};
GJ1214, \citealt{GJ1214}) are included in Fig.~\ref{colours} for comparison. 

\end{itemize}

\begin{figure*}

	\begin{tabular}{cc}
	    \vspace{5mm}
	    \includegraphics[width=0.30\linewidth, keepaspectratio, angle=90]{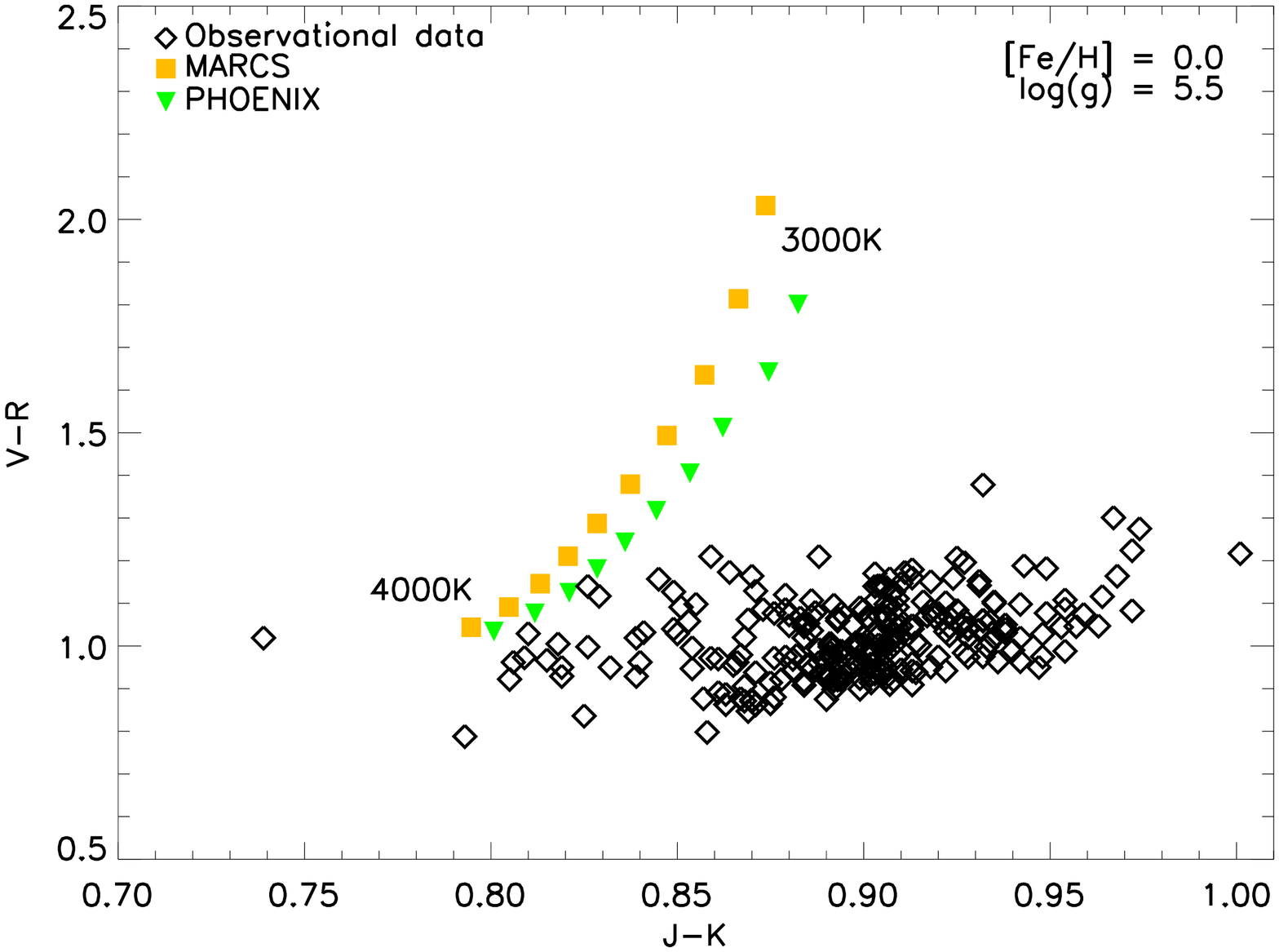} \hspace{5mm}
		\includegraphics[width=0.30\linewidth, keepaspectratio, angle=90]{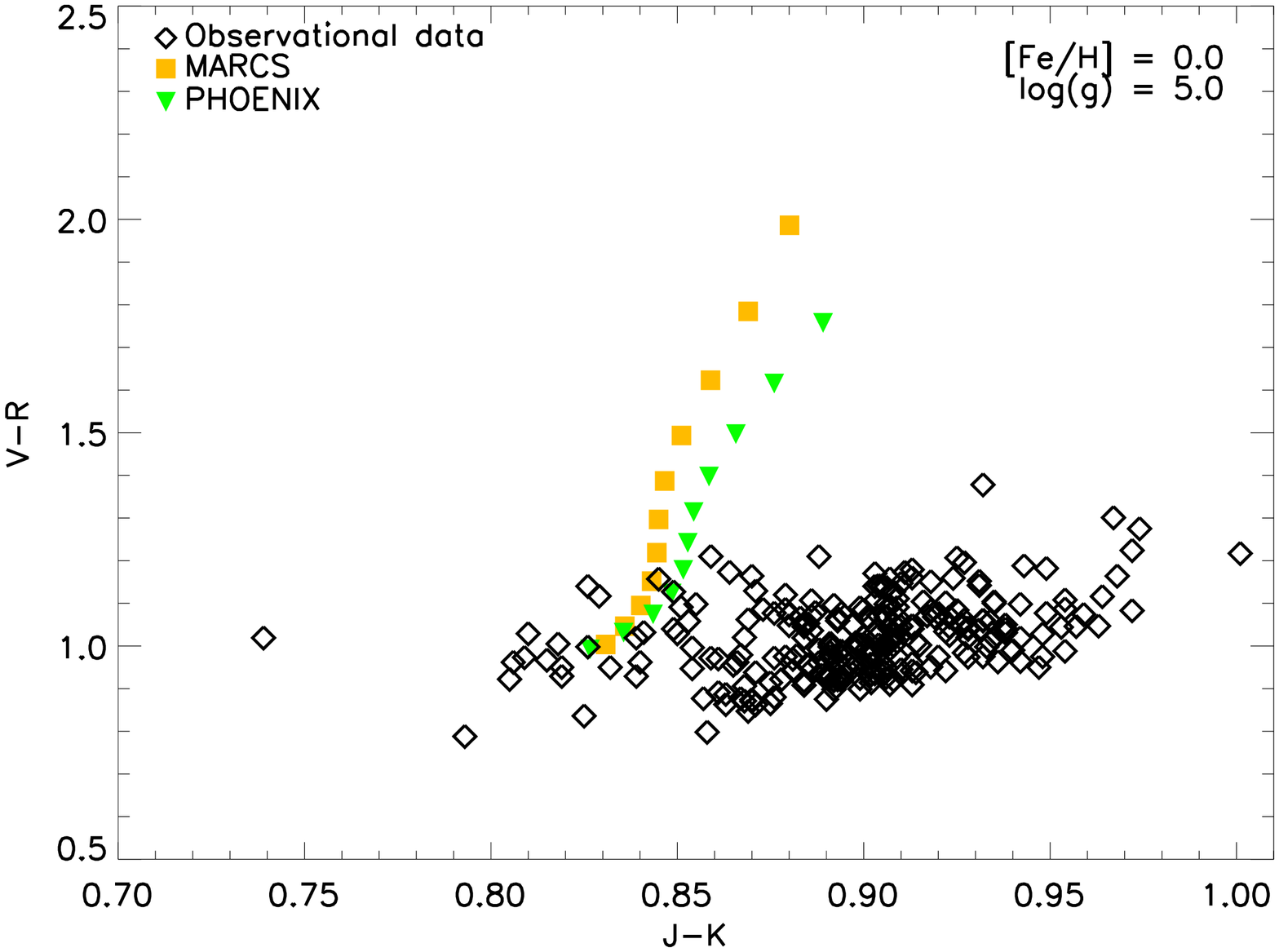} \\
		\includegraphics[width=0.30\linewidth, keepaspectratio, angle=90]{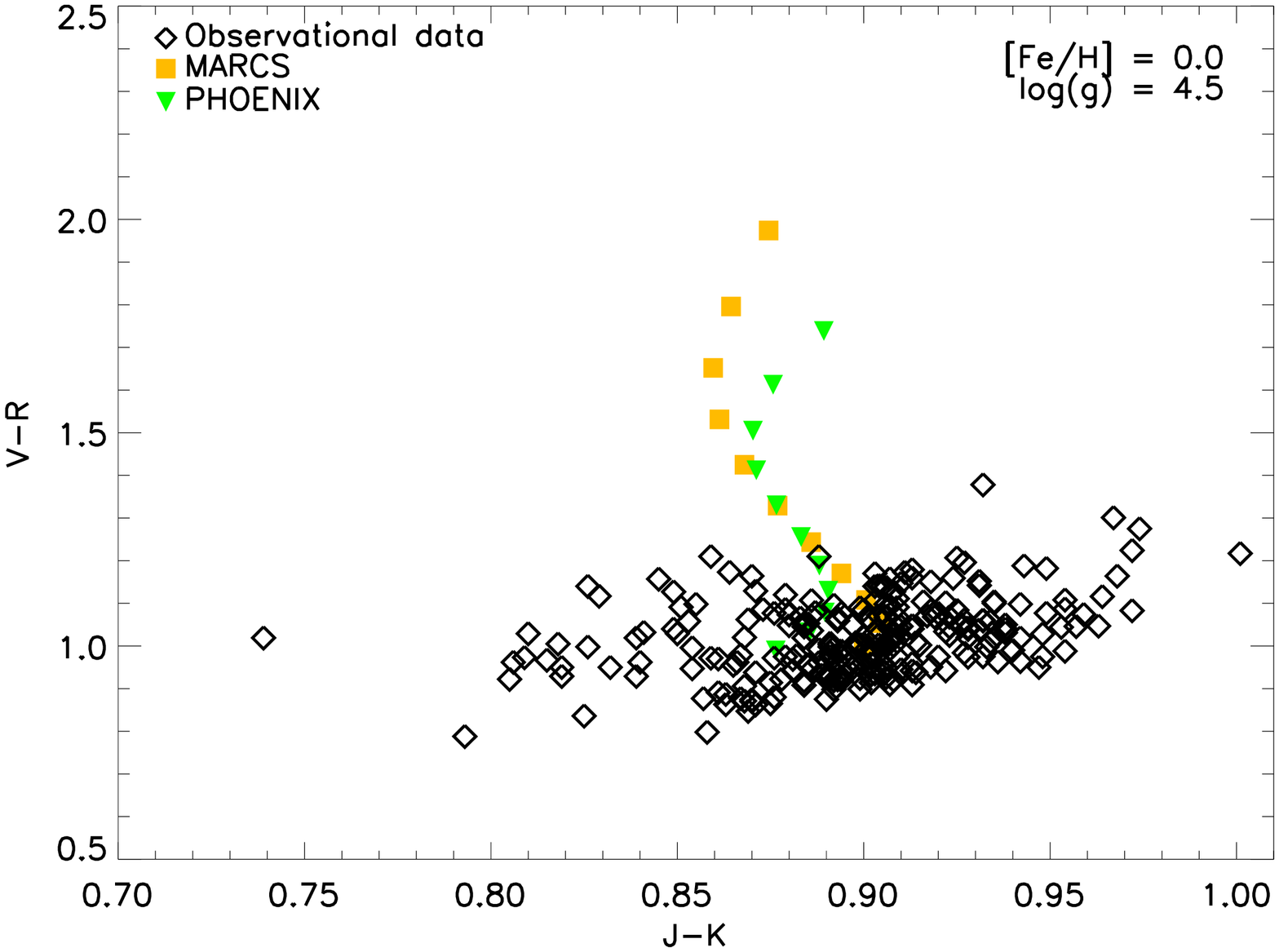}	
	\end{tabular}

	\caption{ Colour-colour plot for a set of observed M stars
          (black diamonds, \ref{TC5}) and synthetic photometry of {\sc
            MARCS} (yellow sqaures) and {\sc Phoenix} (green
          triangles) models for T$_{\rm eff}$ = 3000 - 4000K, [M/H] =
          0.0 and log(g) = 5.5 - 4.5. Each panel contains sets of
          models for a particular value of log(g). The model T$_{\rm
            eff}$ changes from 3000K for higher y-axis values to 4000K
          for lower y-axis values in each panel. Observed sample is
          adopted from the UBVRI, JHK photometry of
          \citet{koen2010}. It contains stars with various
          temperatures, surface gravity and metalicity values. The
          entire sample is plotted in all three panels. Typical
          observational uncertainty for the sample is $\sim$0.01mag
          for both, optical and ifrared. The
          median of the observed colours is well reproduced by the
          log(g) = 4.5 atmosphere models. Scatter in the observations
          is larger than the differences in the models would suggest.
        }
	\label{observed}
    \end{figure*}

 All objects of spectral type M, and for which optical and infrared
 photometry was available, were selected from Tables 2 \& 4
 \cite{koen2010} for our comparison. The majority of this sample of
 objects are early M stars with T$_{\rm eff}\approx$4000K and
 log(g)$\approx$4.5.  Table \ref{TC5} lists the names, associated
 photometric magnitudes and spectral types of all stars used for this
 comparison.  Only the {\sc MARCS} and the {\sc Phoenix} model
 families cover the respective parameter range. Figure \ref{observed}
 presents a colour plot of the photometry that compares the {\sc MARCS}
 and {\sc Phoenix} model results with the sample of observed M stars.

Early M dwarfs are represented by model atmospheres with T$_{\rm
  eff}\approx 4000$K, and the sample of observed M dwarfs does not
contain examples with T$_{\rm eff}<3500$K. Therefore, the
upper half of the plot is empty, hence, it does not imply the models
are giving incorrect predictions for these effective temperatures.
The spectral type of the observed targets explains the lack of objects
in the upper half of the plots and is not a mismatch between models
and observations. For log(g)$=4.5$ the median of the observed colours
is well reproduced by the atmosphere models. For higher log(g), the
observed colours are redder than predicted by atmosphere models.
However, the scatter in the observations is larger than the
differences in the models would suggest. The measurement uncertainties
$\sim$ 0.01 mag (\cite{koen2010}) are not big enough to account for the
scatter in the observed data. 
    
    The reason for the differences between models and observations is
    not obvious. One reason could be a mismatch between the
    metallicities of the stars and the (solar) metallicity in the
    models. Note that not all objects in Fig.~\ref{observed} have reliably
    measured metallicities, hence, the scatter of the observed data
    could be partly due to varying stellar metallicities. The comparison between the exoplanet host stars strengthens this hypothesis. While GJ1214 has approximately solar metallicity, Kepler-42 is reported to have sub-solar metallicity  ( [Fe/H]$= - 0.48\pm 0.17$ and [M/H]$= - 0.33\pm 0.12$, \citealt{Kepler42}). GJ1214 is significantly redder in near-infrared colours than Kepler-42, but still only marginally consistent with the H-K colour predicted from atmosphere models. Both objects are redder in B-V than all predictions from the models. 
    
    Alternatively, the mismatch could be caused by physical processes
    not included in the models considered here, for example, effects
    related to the presence of strong magnetic fields
    (e.g. \citealt{vidotto}). It has been shown that strong magnetic fields can alter the
fundamental properties of cool stars, in particular, suppress the
temperature and inflate the radius. A temperature suppression of up to
200-400\,K is realistic for early M-type stars, see
\citealt{Stassun}). This could possibly explain an increase of up to
0.1\,mag in the J-K colour (see Figs.~\ref{colours}). In summary, the best
explanation for the scatter in the observed datapoints in Fig.~\ref{observed} is probably 
a combination of a range of metallicities and the
presence of magnetic fields, whereas the contribution from measurement
uncertainties is only minor. 

  \subsection{Implications of host-star's uncertainties for exoplanets}

  Estimating exoplanetary mass and radius directly depends on
  knowledge of the host star's mass and radius. Most often, they are
  derived by comparison to evolutionary models which, however, already
  carry the uncertainties in model atmospheres discussed in the
  previous sections.  Stellar atmosphere models can provide values for
  surface gravity, log(g), but there is still a degeneracy in possible
  values for stellar mass and radius.
  
 An important property for a star-planet(s) system is the
  habitable zone (HZ). The habitable zone refers to the distance away
  from the star where liquid water could exist on the surface of a
  planet, provided sufficient atmospheric pressure. Detailed
  calculations for the extent of the HZ have been conducted by
  \cite{kasting}, \cite{jones} and \cite{kopparapu}. \cite{kane} uses
  their methods to estimate the uncertainty of the habitable zones
  location (resulting from stellar parameter (effective temperature,
  radius, surface gravity, mass) uncertainties) for confirmed
  exoplanetary host stars and Kepler candidate hosts. The author
  demonstrates that $\sim$ 5\% uncertainties in T$_{\rm eff}$ result
  in $\sim$ 10\% uncertainty in the HZ location. Furthermore, the HZ
  distance is shown to have a linear dependence on the stellar radius
  R$_*$ and hence proportional to $\sqrt{1 \slash \rm{g}}$ and
  $\sqrt{\rm {M_*}}$, where g and M$_*$ are the stellar surface
  gravity and mass. The system Kepler-27 is used as an example where
  the host star's parameters have large uncertainties. The associated
  error in the HZ region is demonstrated to be large enough, so that a
  planet in habitable zone may very well lie outside of it on a
  1-$\sigma$ level. The author further states that this is the case
  for the majority of Kepler candidates.

\cite{plavchan} compare transit durations of Kepler targets to a synthetic distribution crated based on eccentricities of exoplanets discovered by the radial velocity method. The authors find an over-abundance of Kepler targets with transit durations longer than expected and a median transit duration of $\sim$25\% longer than predicted. These effects are both attributed to under-estimates of the stellar radii. In addition, a statistically significant trend is found in the average transit duration as a function of stellar mass and radius which is explained by errors in determination of stellar radii as a function of spectral type.

A particularly underestmated factor for M-dwarfs is their strong
magnetic field activity. The magnetic activity of the host star can
have strong implications for the habitability of a
planet. \cite{vidotto}, \cite{vidotto2014} address planetary
magnetoshpere size in relation to the stellar magnetic fields and show
that for non-axisymmetric stellar magnetic field topologies, the size
of the planetary magnetosphere can expand/shrink by up to 20\% along
its orbit.  In addition the authors argue that planets in systems
around host stars with such magnetic field topologies will be better
shielded against galactic cosmic rays even in the absence of a thick
planetary atmosphere or a large planetary magnetosphere.

\section{Summary}

   We compared {\sc ATLAS9}, {\sc MARCS}, {\sc Phoenix} and {\sc
     Drift-Phoenix} atmosphere models in the M-dwarf parameter range
   that includes young M-dwarfs and also brown dwarfs. Our study
     has been inspired by the first model atmosphere comparsion in
     \cite{gustafsson} and in \cite{plez11} which focused on T$_{\rm
       eff}>3500$K, and by extensive studies for space missions as in
     \cite{sarro}. Our comparison of (T$_{\rm gas}$, p$_{\rm gas}$)
   structures for T$_{\rm eff}<3500$K reveals difference in local
   temperatures between the {\sc MARCS}, {\sc Phoenix}, and {\sc
     Drift-Phoenix} model atmosphere families of, on average, less
   than 300K. Such a variation becomes significant for low T$_{\rm
     eff}$ models, where dust condensation plays a major role for the
   shape of the SED.
  
  We compiled UKIDSS ZYJHK, 2MASS JHKs and Johnson UBVRI synthetic
  photometric data for the {\sc ATLAS}, {\sc MARCS}, {\sc
      Phoenix}, and {\sc Drift-Phoenix} model families. Colour
  indices differ between models by no more than 0.15 dex in the IR
  range. Both, atmospheric structure and synthetic photometry data,
  suggests that model atmospheres with higher surface gravity agree
  better between different models regardless of their T$_{\rm
    eff}$. Comparing to observational data, the difference in the
  models is smaller than the typical observational errors of 0.01
  mag. However a spread in the data is present which is not account
  for by the models, which may suggest a mismatch between model and
  stellar metallicities.

  The present paper demonstrated differences and similarities
   between various model atmosphere families which allows a better
   estimate of systematic uncertainty values that may result from our
   limitted capacity of modelling every aspect of atmosphere physics
   and chemistry in the best possible way, and from the tentativeness
   of the 'best possible way'. Optimally, more than one model family
   should be used when working with observational data. The need for
   model atmosphere diversity has been demonstrated, for example, with
   respect to disk detection \citep{sinclair} or determining planetary
   parameter \citep{southworth}.  Such studies suggest that a similar
   multi-model approach could be beneficial for studies as for example
   performed in \cite{sarro} who present a module that will be used to
   detect and characterise ultra-cool dwarfs in the Gaia database.

\section{Acknowledgments}

We thank the authors of the {\sc ATLAS9}, {\sc MARCS}, {\sc Phoenix}
and {\sc Drift-Phoenix} model atmosphere grids to allow us to use
their model results for this comparison study, and for their helpful
suggestions.  We thank Dr Tim-Oliver Husser (Institut f\"ur Astrophysik,
Georg-August-Universitat G\"ottingen) for his support with the {\sc
  Phoenix} models. We also thank S\"oren Witte and our referee for constructive feedback.
 ChH highlights financial support of the European
Community under the FP7 by an ERC starting grant.  IB thanks the
Physics Trust of the University of St Andrews for supporting her
summer placement.  Most literature search was performed using the
ADS. Our local computer support is highly acknowledged.

\clearpage 
\appendix
\section{Parameter values of models used}\label{appA}

	Tables A1 - A4 indicate availability of models of different families for various parameter value combinations. 

 \begin{table*}

     The tables below describe the parameter values for all the models used in this work.
     Empty cells indicate that a given set of parameter values was not used as the corresponding model was missing in some model family.
 
    \caption{Common models between {\sc ATLAS} and {\sc MARCS} for T$_{\rm eff}$ = 3500K}
    \label{TA1}
    
    \begin{tabular}{cccccccc}
	
	\hline
	
	\backslashbox{log(g)}{[M/H]}&
	-2.5 & -2.0 & -1.5 & -1.0 & -0.5 & 0.0& +0.5 \\
	
	\hline  3.0 & X & X &  & X& X & X& X\\
      
	\hline  3.5 & X & X & X & X& X & X& X\\
      
	\hline	4.0 & X & X & X & X& X & X& X\\
      
	\hline	4.5 & X & X & X & X& X & X& X\\
      
	\hline	5.0 & X & X & X & X& X & X& X\\

	\hline
	
    \end{tabular}

  \end{table*}

  \begin{table*}

      \caption{Common models between {\sc ATLAS} and {\sc MARCS} for T$_{\rm eff}$ = 4000K}
      \label{TA2}
      
      \begin{tabular}{|c|ccccccc|}
	  
	  \hline
	  
	  \backslashbox{log(g)}{[M/H]}&
	  -2.5 & -2.0 & -1.5 & -1.0 & -0.5 & 0.0& +0.5 \\
	  
	  \hline  3.0 & X & X & X & X& X & X& X\\
	
	  \hline  3.5 & X & X & X & X& X & X& X\\
	
	  \hline  4.0 & X & X & X & X& X & X& X\\
	
	  \hline  4.5 & X & X & X & X& X & X& X\\
	
	  \hline  5.0 & X & X & X & X& X & X& X\\

	  \hline
	  
      \end{tabular}

  \end{table*}

   \begin{table*}

      \caption{Common models between {\sc MARCS}, {\sc Phoenix} and {\sc Drift-Phoenix} for [M/H] = 0.0}
      \label{TA3}
      
      \begin{tabular}{|c|ccccccc|}
	  
	  \hline
	  
	  \backslashbox{log(g)}{T$_{\rm eff}$}&
	  2500 & 2600 & 2700 & 2700 & 2800 & 2900 & 3000 \\
	  
	  \hline  3.0 & X & X & X & X& X & X& X\\
	
	  \hline  3.5 & X & X & X & X& X & X& X\\
	
	  \hline  4.0 & X & X & X & X& X & X& X\\
	
	  \hline  4.5 & X & X & X & X& X & X& X\\
	
	  \hline  5.0 & X & X & X & X& X & X& X\\

	  \hline  5.5 & X & X & X & X& X & X& X\\
	  
	  \hline
	  
      \end{tabular}

  \end{table*} 

  \begin{table*}

      \caption{Additional common models between {\sc MARCS}, {\sc Phoenix} for [M/H] = 0.0}
      \label{TA4}
      
      \begin{tabular}{|c|cccccccccc|}
	  
	  \hline
	  
	  \backslashbox{log(g)}{T$_{\rm eff}$}&
	  3100 & 3200 & 3300 & 3400 & 3500 & 3600 & 3700 & 3800 & 3900 & 4000 \\
	  
	  \hline  3.0 & X & X & X & X& X & X & X & X & X & X\\
	
	  \hline  3.5 & X & X & X & X& X & X& X & X & X & X\\
	
	  \hline  4.0 & X & X & X & X& X & X& X & X & X & X\\
	
	  \hline  4.5 & X & X & X & X& X & X& X & X & X & X\\
	
	  \hline  5.0 & X & X & X & X& X & X& X & X & X & X\\

	  \hline  5.5 & X & X & X & X& X & X& X & X & X & X\\
	  
	  \hline
	  
      \end{tabular}
    
  \end{table*}
  
 \clearpage
\section{Complementary T-p structure and flux ration plots}\label{s:appB}
 Figure B1 presents a sample of plots illustrating the difference in temperature-pressure structures between {\sc ATLAS} and {\sc MARCS} models in the metallicity parameter space. Figure B2 gives synthetic flux ratios in the optical bands fro the {\sc ATLAS} and {\sc MARCS} models.

\begin{figure*}

    \begin{tabular}{cc}
	\vspace{6mm}
	\includegraphics[width=0.3\linewidth, keepaspectratio, angle=90]{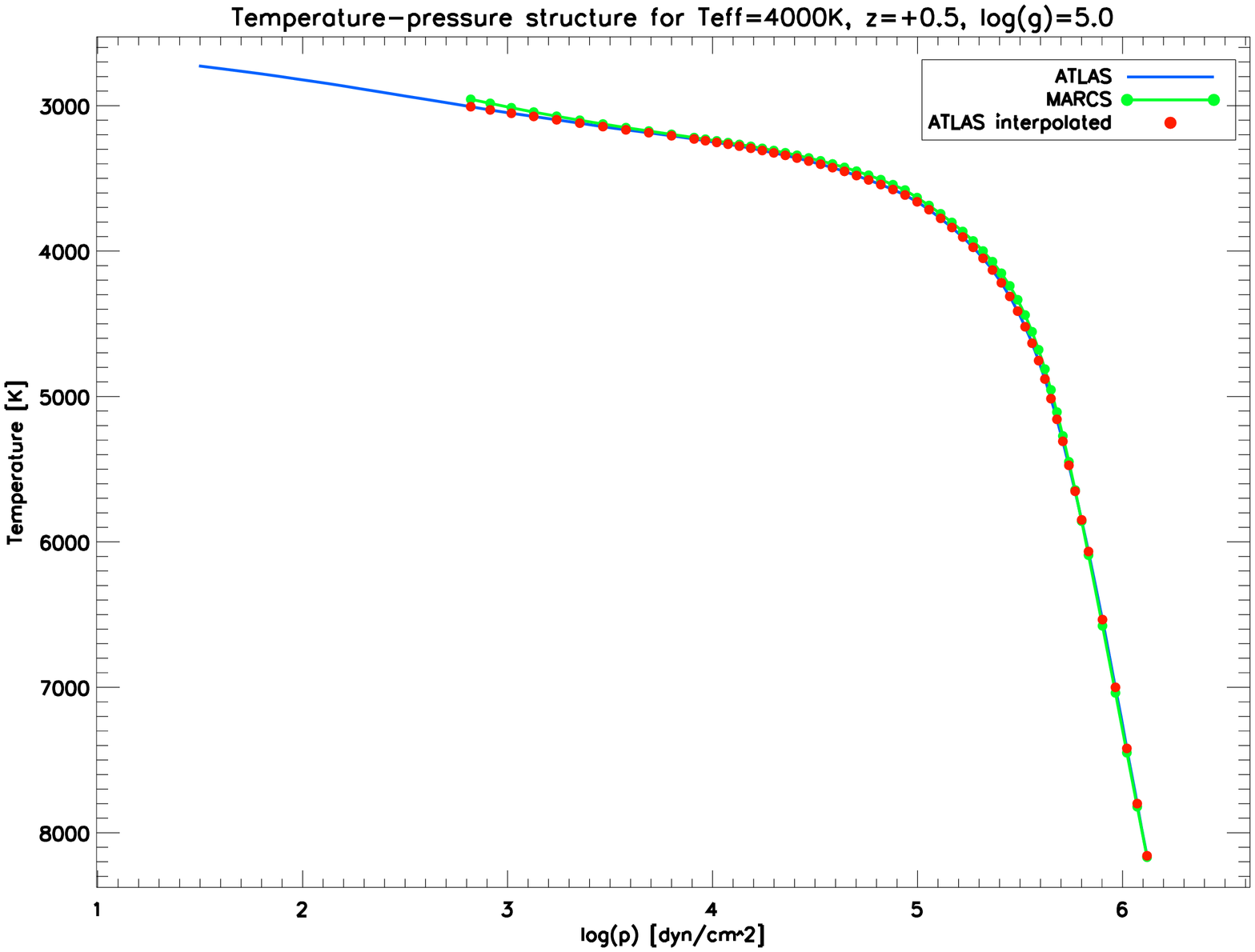}\hspace{7mm}
	\includegraphics[width=0.3\linewidth, keepaspectratio, angle=90]{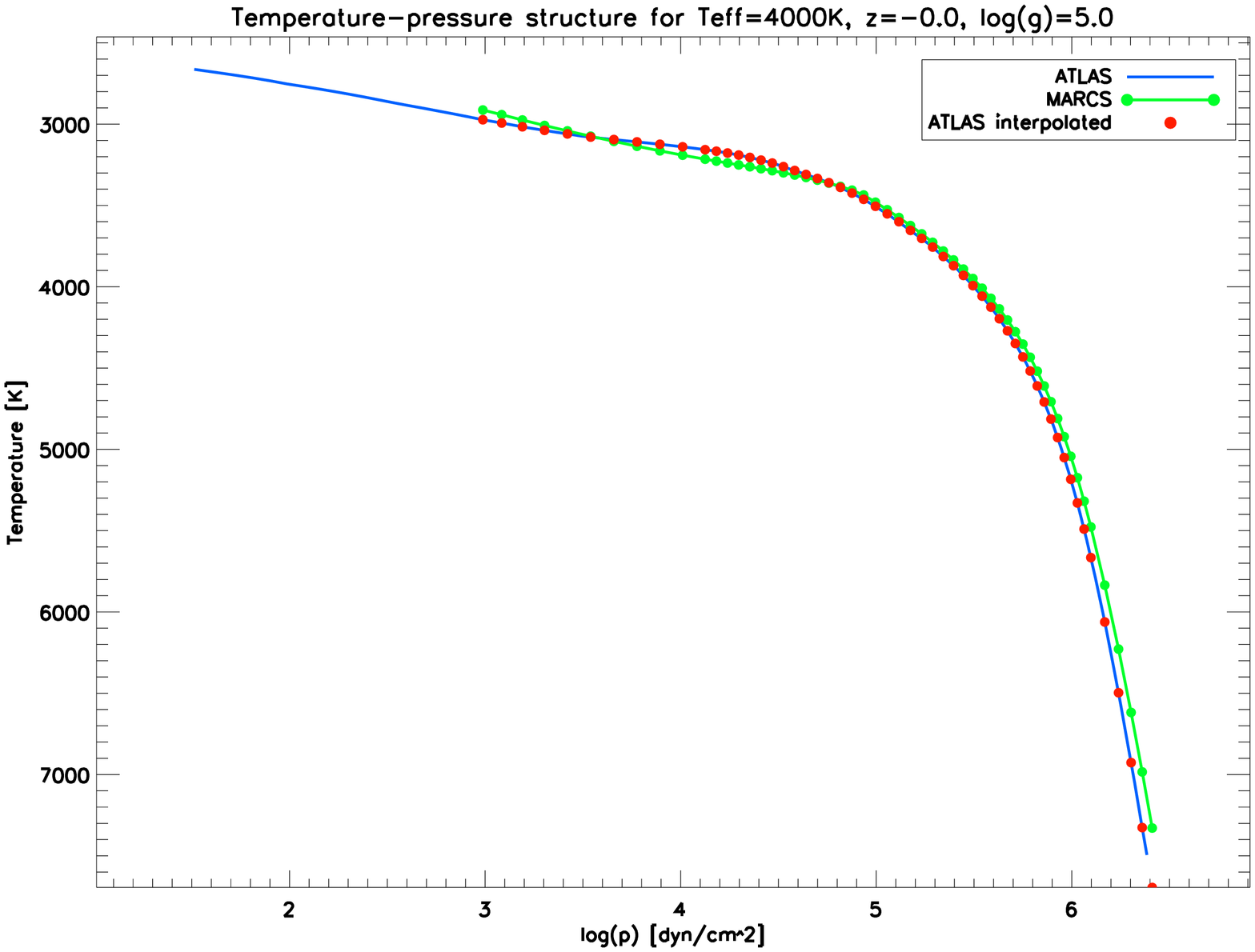}\\
	\vspace{6mm}	
	\includegraphics[width=0.3\linewidth, keepaspectratio, angle=90]{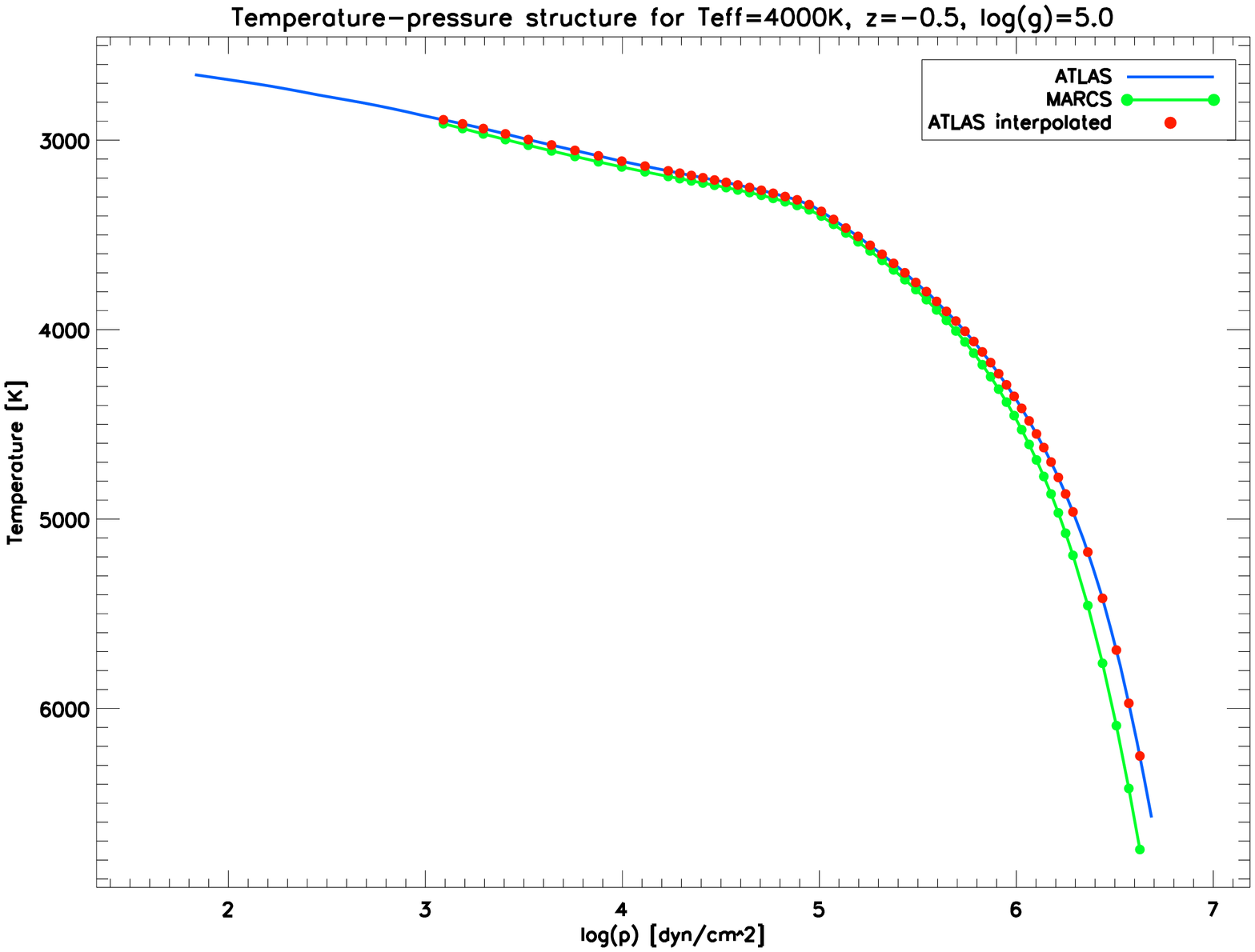}\hspace{7mm}
	\includegraphics[width=0.3\linewidth, keepaspectratio, angle=90]{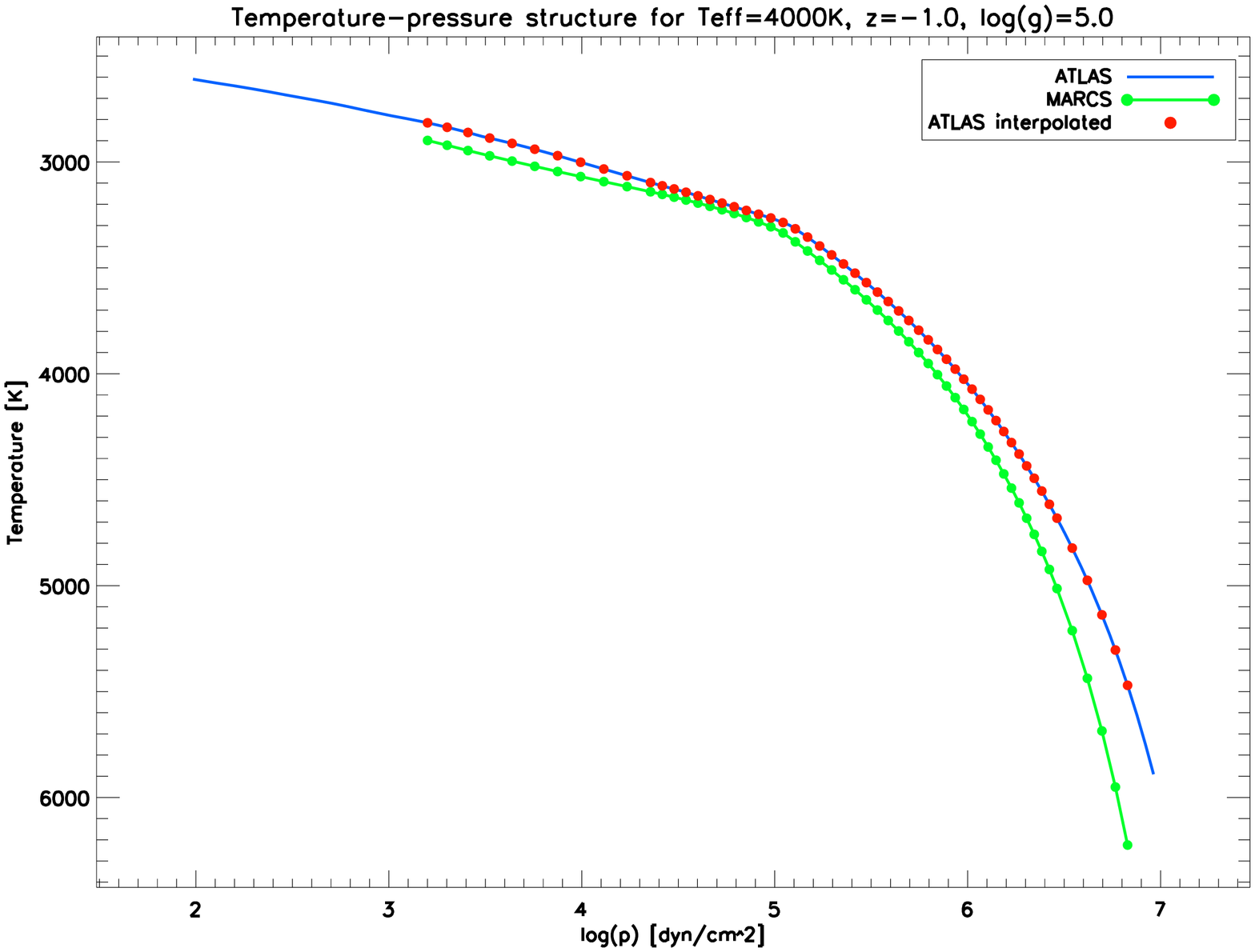}\\
	\vspace{6mm}	
	\includegraphics[width=0.3\linewidth, keepaspectratio, angle=90]{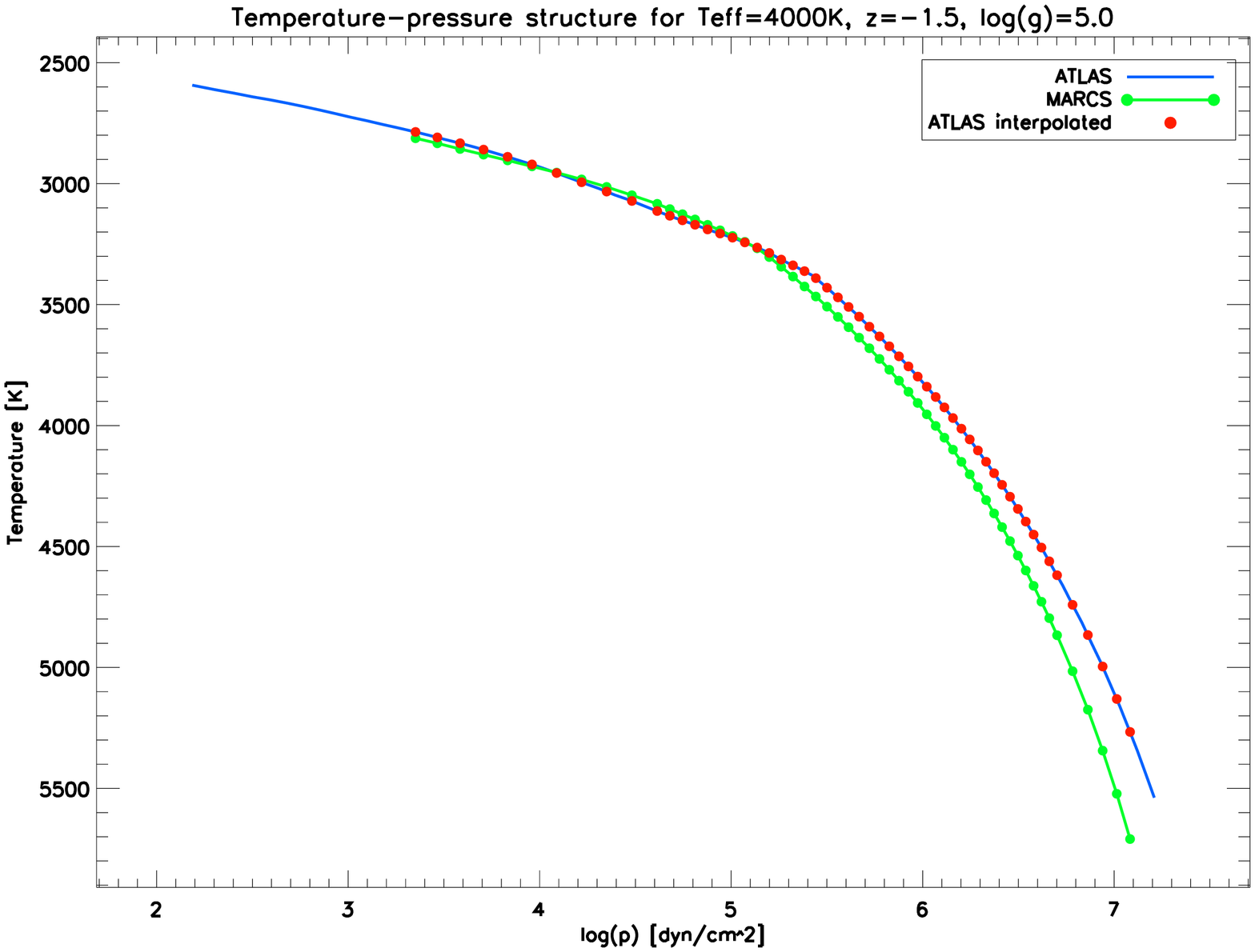}\hspace{7mm}
	\includegraphics[width=0.3\linewidth, keepaspectratio, angle=90]{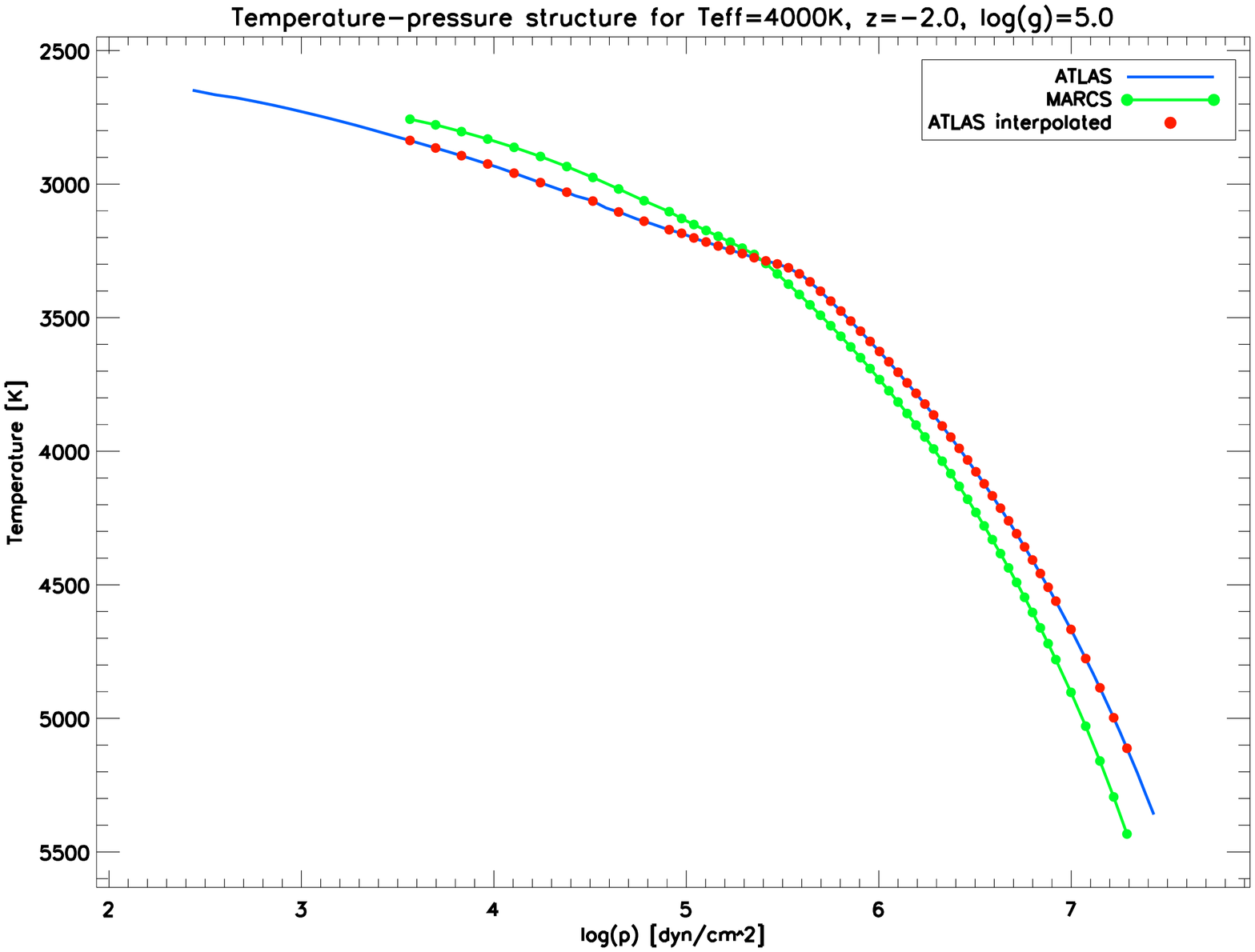}\\
	\includegraphics[width=0.3\linewidth, keepaspectratio, angle=90]{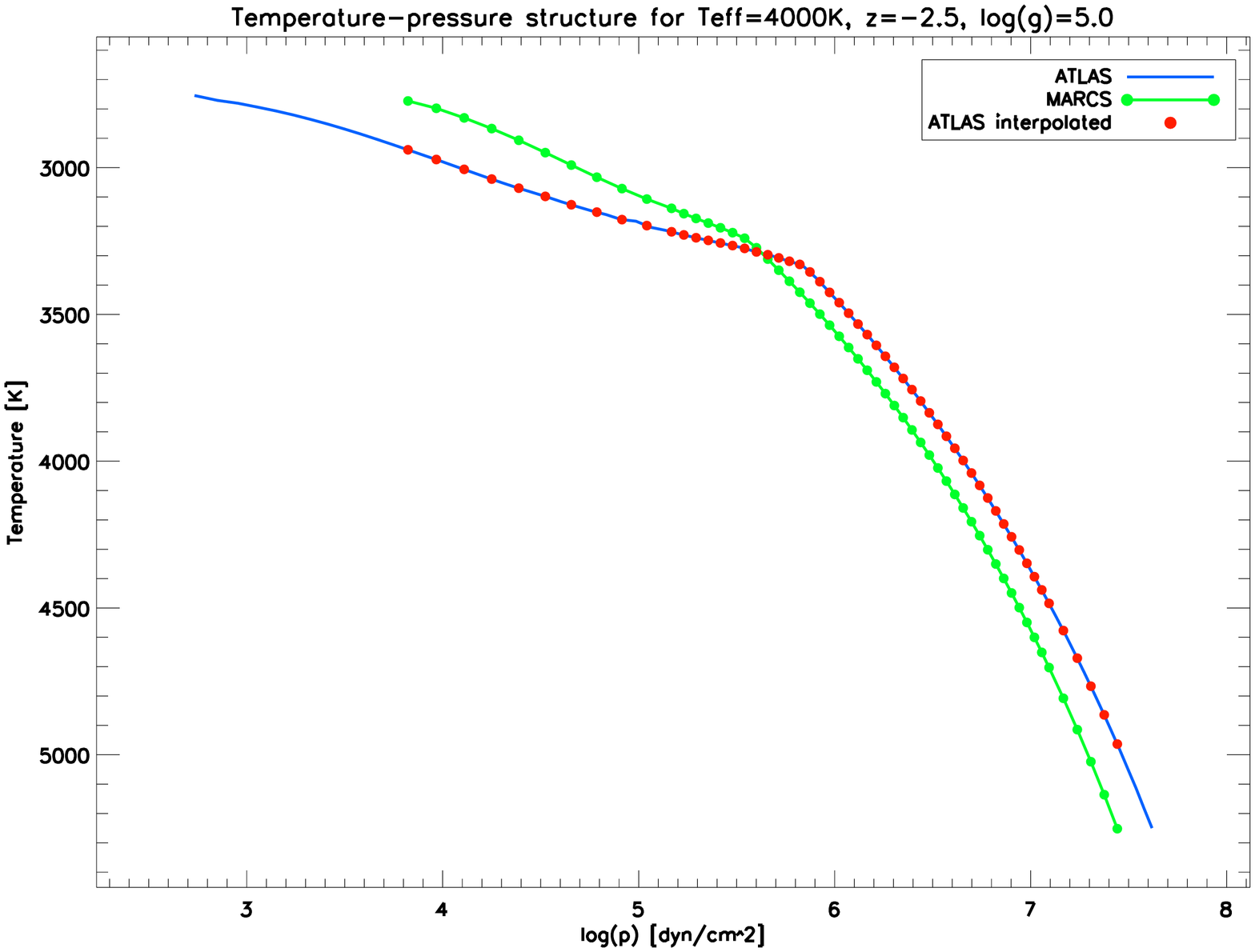}
    
    \end{tabular}

     \caption{
	      Local temperature-pressure structures of the {\sc ATLAS} and {\sc MARCS} models for T$_{\rm eff}$ = 4000K,
	      log(g) = 5.0 and various metallicity values.
	      }
    \label{B1}
\end{figure*}

\begin{figure*}

    \begin{tabular}{cc}
	\vspace{7mm}
	\includegraphics[width=0.30\linewidth, keepaspectratio, angle=90]{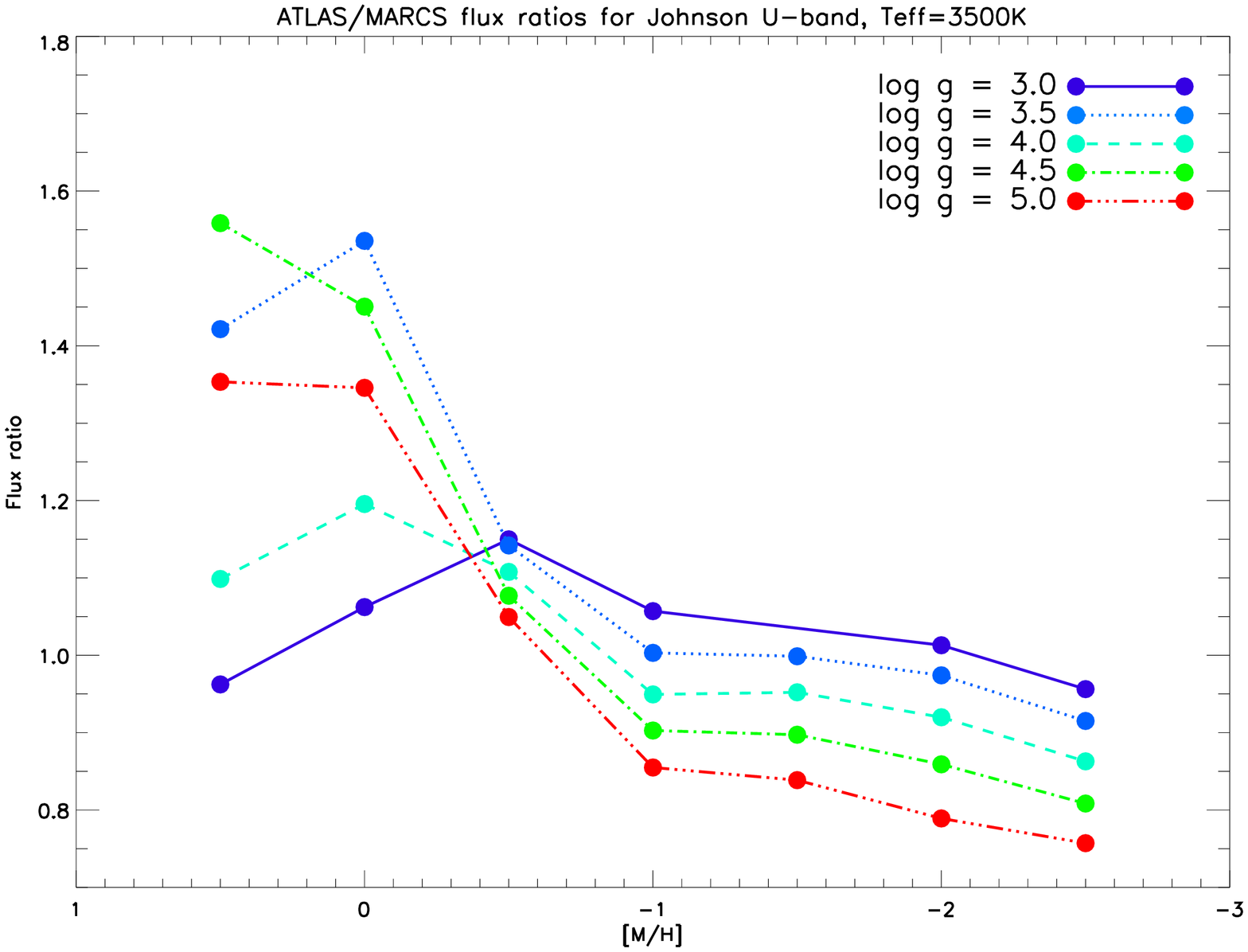}\hspace{5mm}
	\includegraphics[width=0.30\linewidth, keepaspectratio, angle=90]{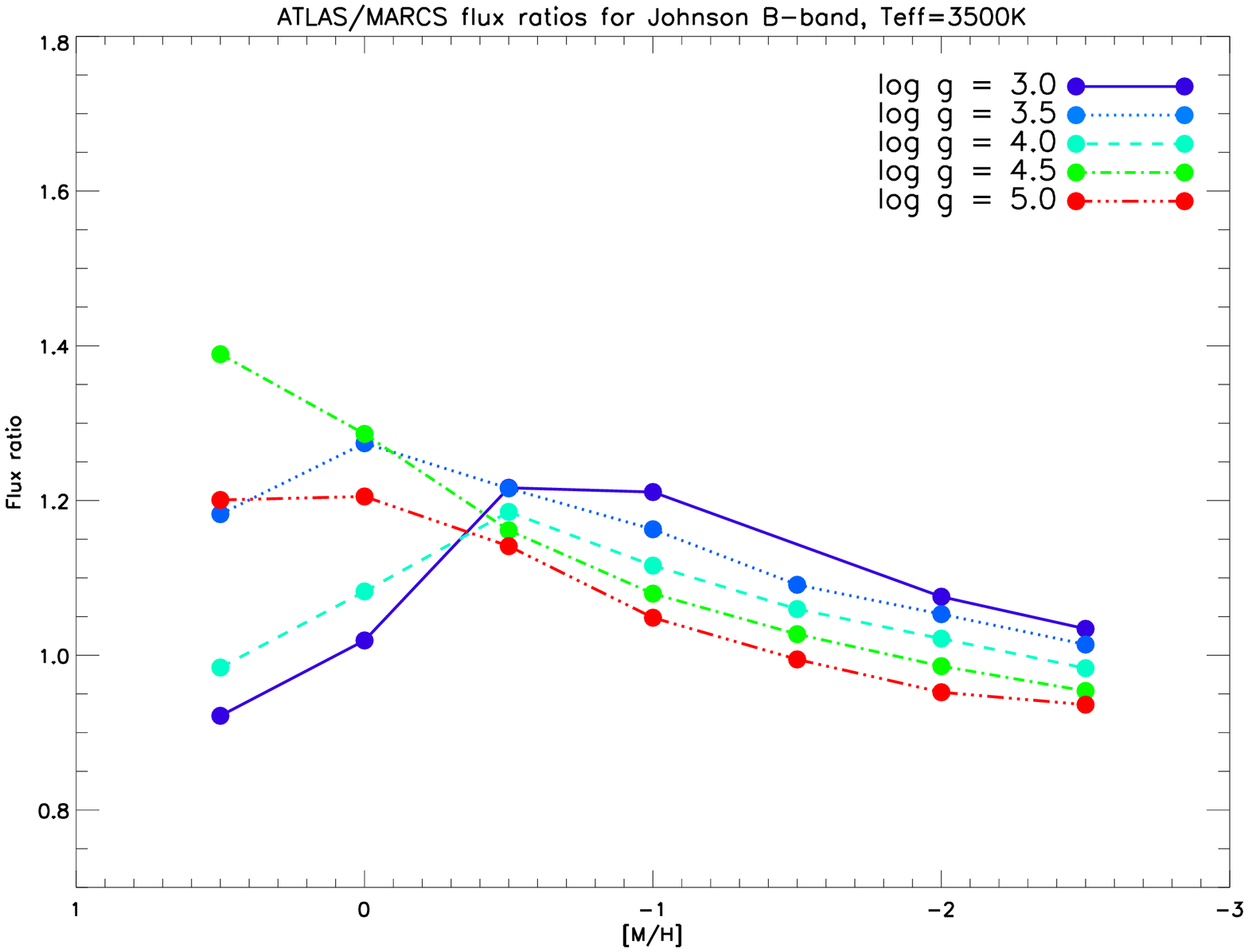}\\
	\includegraphics[width=0.30\linewidth, keepaspectratio, angle=90]{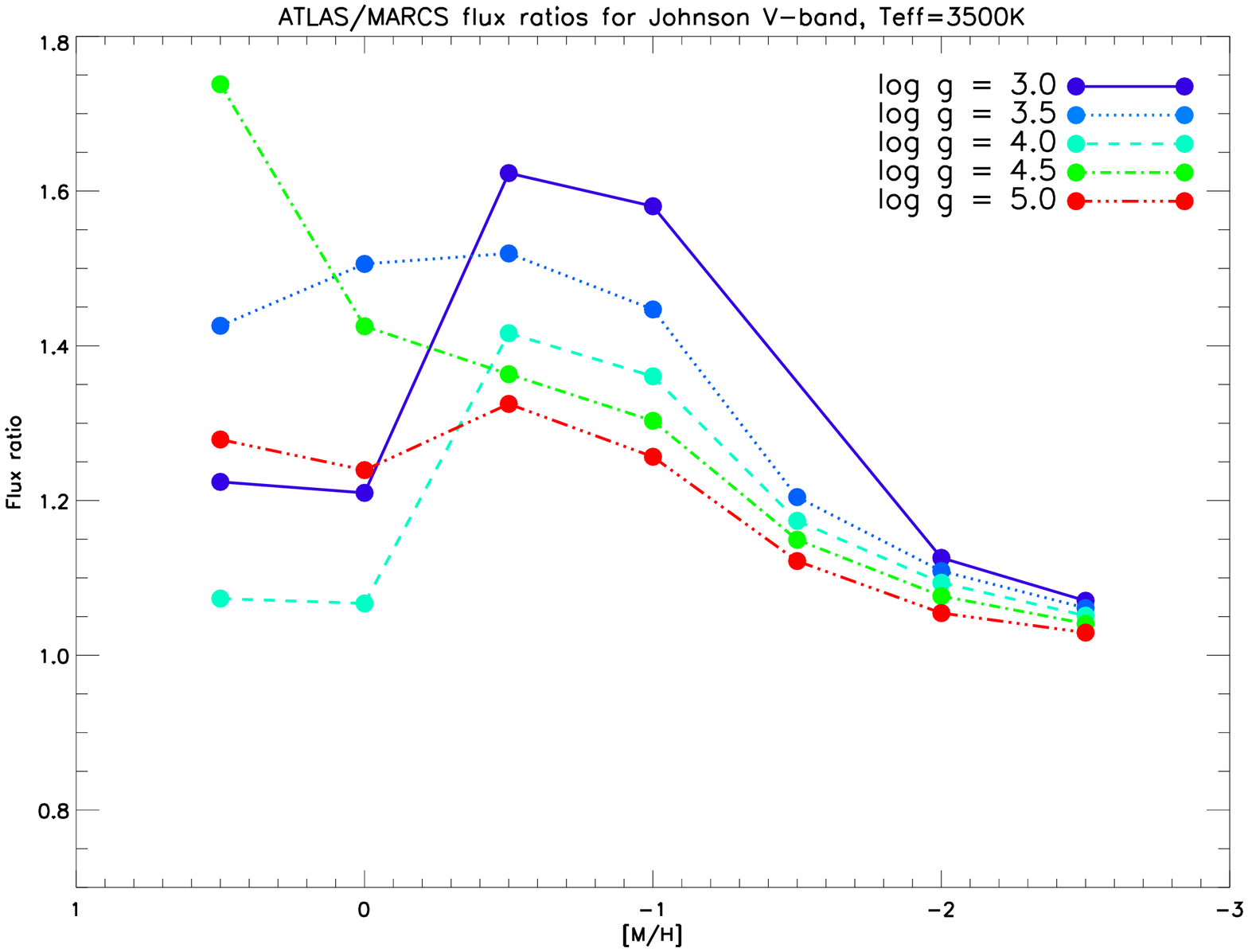}\hspace{5mm}
	\includegraphics[width=0.30\linewidth, keepaspectratio, angle=90]{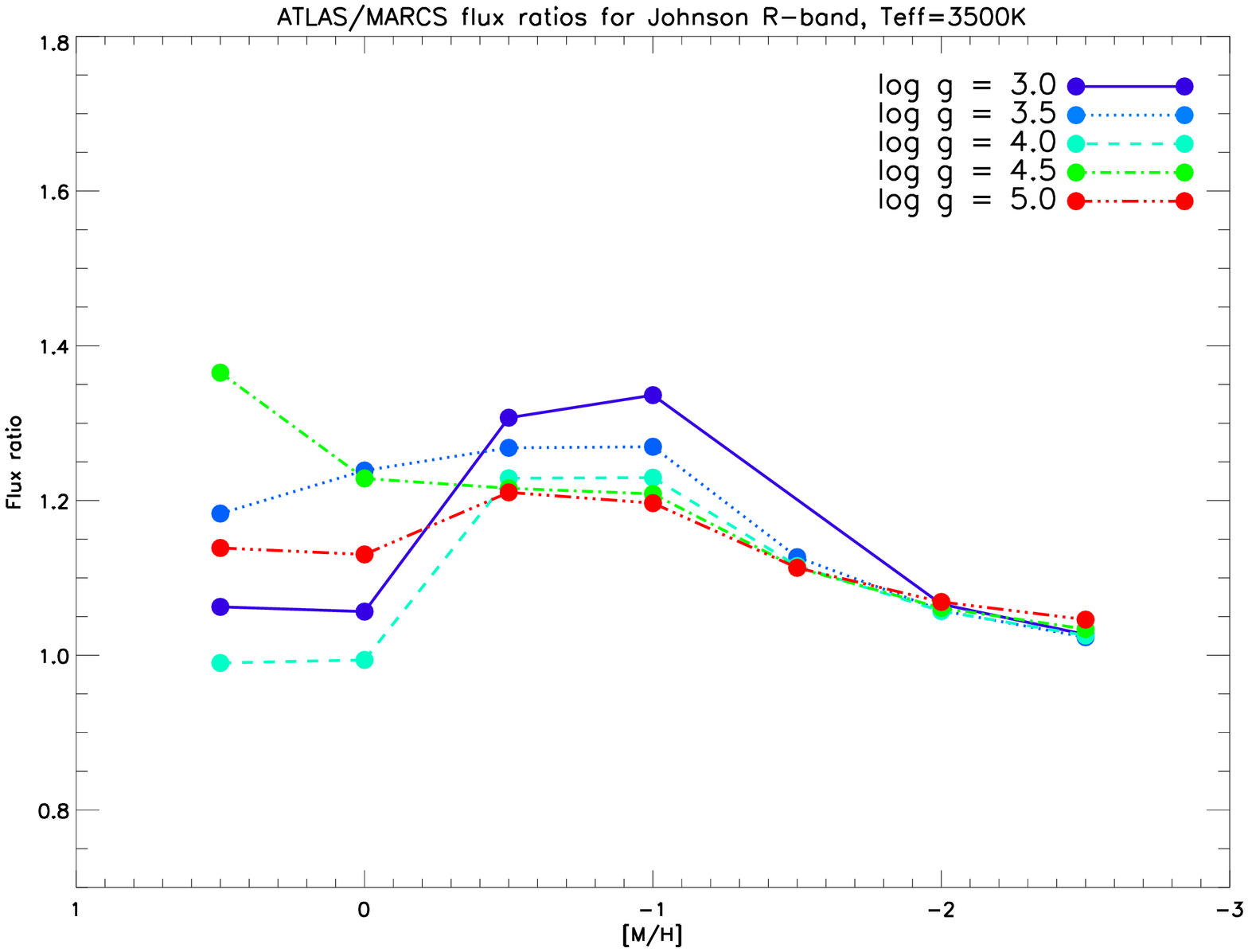}\\

    \end{tabular}

     \caption{
	      Synthetic flux ratios between for {\sc ATLAS/MARCS} models in the optical Johnson UBVR bandpasses for
	      T$_{\rm eff}$ = 3500 K  
	      }
    \label{B2}
\end{figure*}

\begin{figure*}

    \begin{tabular}{cc}
	\vspace{6mm}
	\includegraphics[width=0.30\linewidth, keepaspectratio, angle=90]{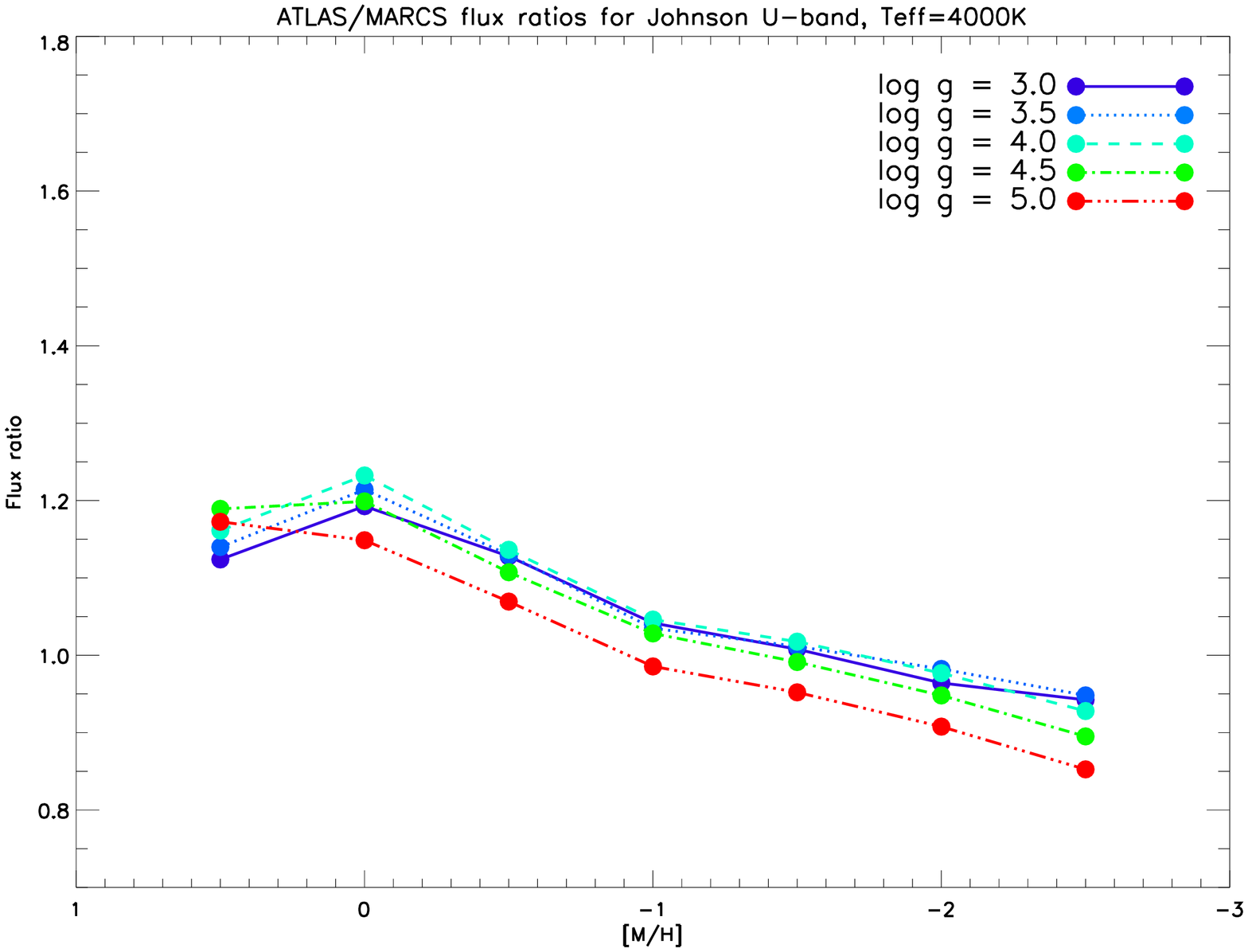}\hspace{5mm}
	\includegraphics[width=0.30\linewidth, keepaspectratio, angle=90]{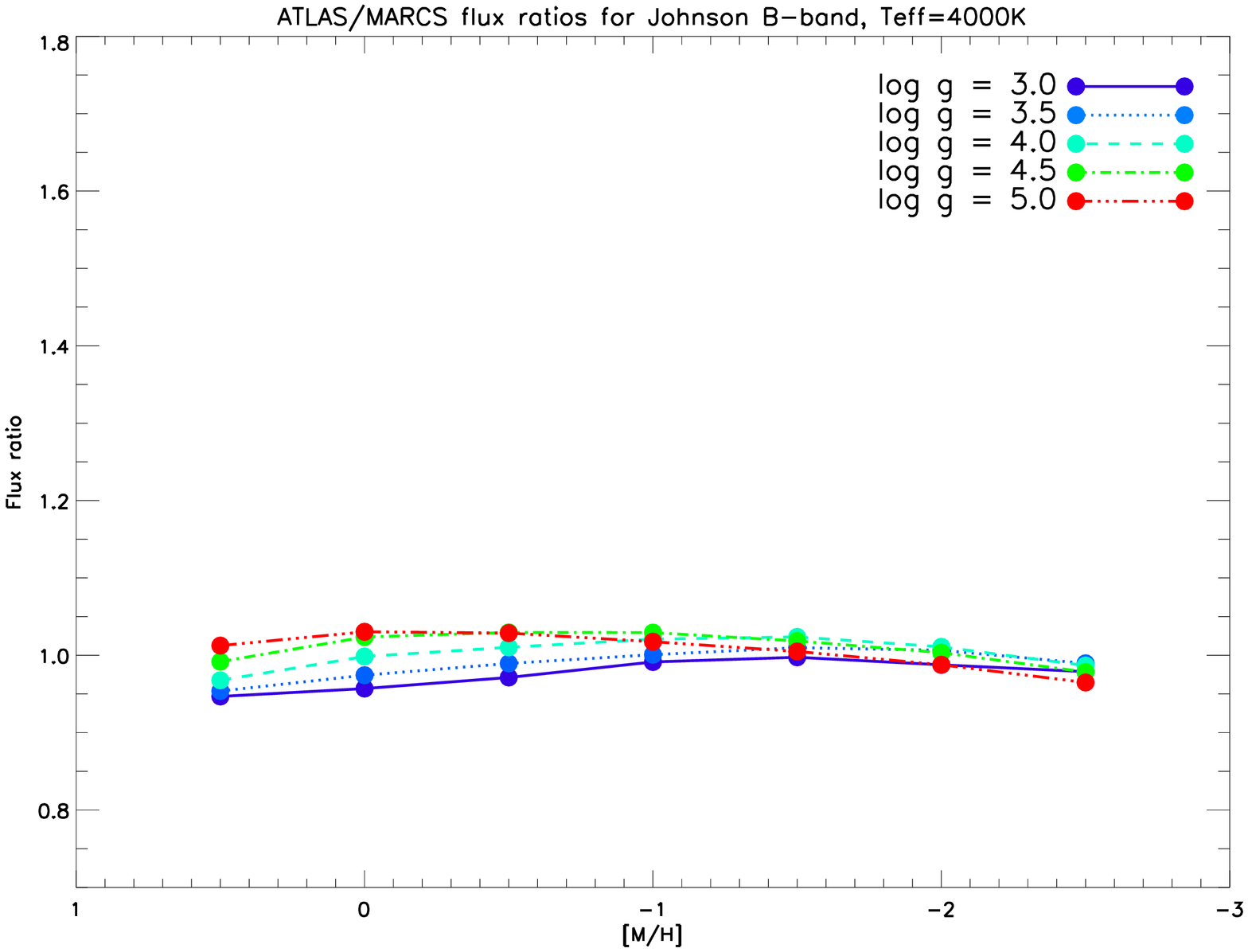}\\
	\includegraphics[width=0.30\linewidth, keepaspectratio, angle=90]{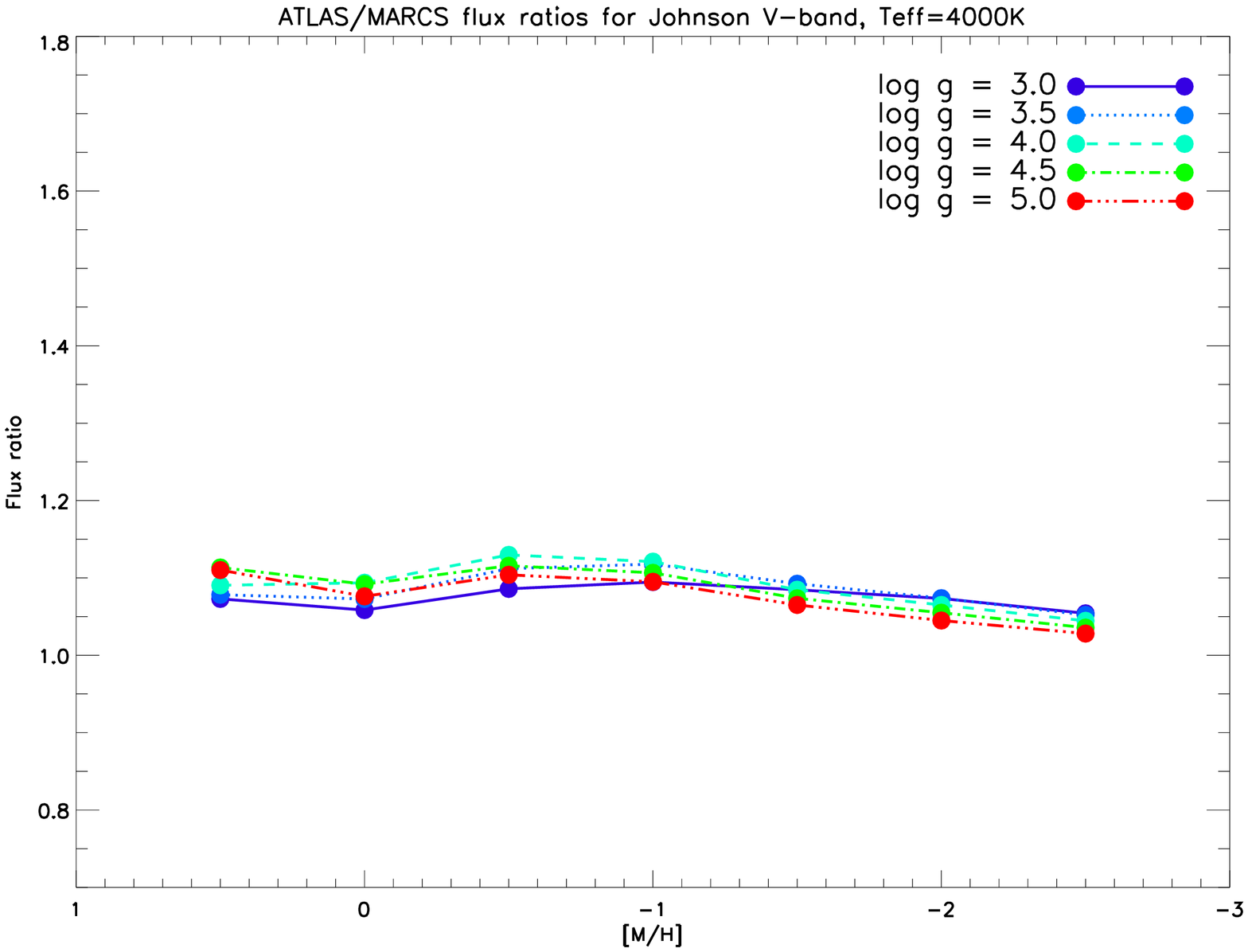}\hspace{5mm}
	\includegraphics[width=0.30\linewidth, keepaspectratio, angle=90]{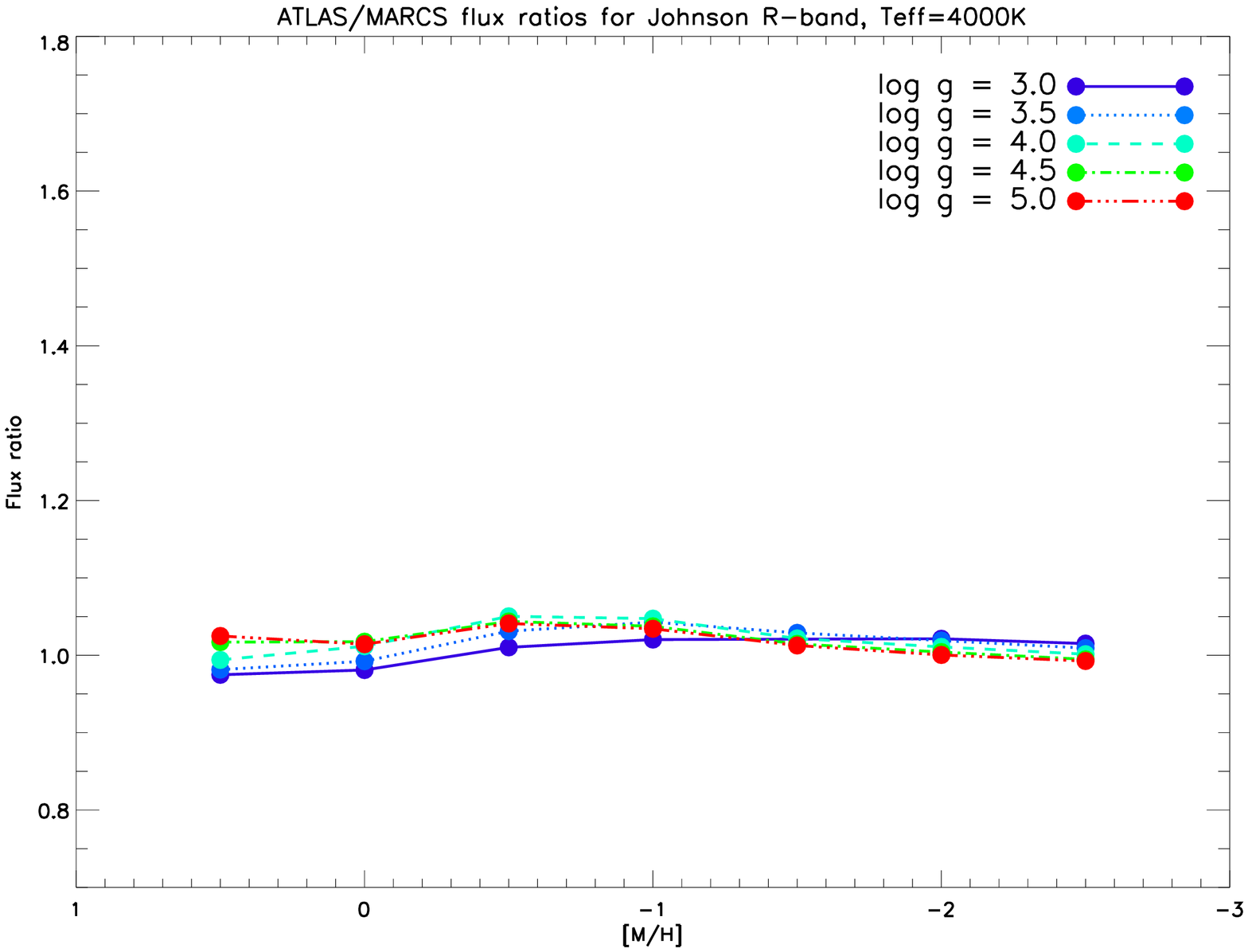}\\

    \end{tabular}

     \caption{
	      Synthetic flux ratios between for {\sc ATLAS/MARCS} models in the optical Johnson UBVR bandpasses for
	      T$_{\rm eff}$ = 4000 K  
	      }
    \label{B3}
\end{figure*}
 
 \clearpage
\section{Synthetic flux and colour data}\label{s:appC}

Table \ref{filters} summarises the filter wavelength ranges where throughput values are above 1\% for the systems used in this study. \ref{TC1}-\ref{TC4} provide a complete set of synthetic fluxes for all models discussed in this work.
Table \ref{TC5} contains  photometric data for the sample M-dwarfs used for the comparison with synthetic colours in Section 5.2. 

\begin{table*}
    \centering
      \caption{ Wavelength range for filters used for synthetic photometry. {\bf The range is given for throughput values above 1\%} }
      \label{filters}
      
      \begin{tabular}{|c|c|}
      
	    \hline    
	    Filter & Wavelength range \\  \hline
	    UKIDSS Z & 0.82$-$0.94 $\mu$m \\
	    UKIDSS Y & 0.96$-$1.10 $\mu$m \\
	    UKIDSS J & 1.15$-$1.35 $\mu$m \\
	    UKIDSS H & 1.45$-$1.82 $\mu$m \\
	    UKIDSS K & 1.96$-$2.44 $\mu$m \\
	    2MASS J & 1.08$-$1.41 $\mu$m \\
	    2MASS H & 1.48$-$1.82 $\mu$m \\
	    2MASS Ks & 1.95$-$2.36 $\mu$m \\
	    Johnson U & 3050$-$4100 $\AA$ \\
	    Johnson B & 3700$-$5500 $\AA$ \\ 
	    Johnson V & 4700$-$7300 $\AA$ \\
	    Johnson R & 5250$-$9450 $\AA$ \\
	    Johnson I & 6900$-$11800 $\AA$ \\ \hline

      \end{tabular}

  \end{table*}

\begin{table*}
	\caption{Complete set of {\sc Phoenix} synthetic fluxes for models of solar metalicity. The values in the filter columns correspond to $-2.5\rm{ log( {F_{\rm R1}} \slash {F_{\rm R1,Vega}}} )$, where R1 is the corresponding filter.  Column 3-7 refer to the Johnson and columns 11-15 to the UKIDSS filters.}
	\label{TC1}
	
	\begin{tabular}{ccccccccccccccc}
		
	\hline
	
\bf{ T$_{\rm eff}$}  & \bf{log(g)} & \bf{U} & \bf{B} & \bf{V} & \bf{R} & \bf{I} & \bf{2MASS H} &\bf{2MASS J} & \bf{2MASS Ks} & \bf{H} & \bf{J} & \bf{K} & \bf{Y} & \bf{Z} \\
              	
\hline	
	
2500	&	3.0	&	11.50	&	10.99	&	9.83	&	7.32	&	5.05	&	2.51	&	3.10	&	2.12	&	2.57	&	2.97	&	2.07	&	2.97	&	2.07	\\
2500	&	3.5	&	11.74	&	11.03	&	9.88	&	7.32	&	5.02	&	2.47	&	3.08	&	2.10	&	2.53	&	2.95	&	2.05	&	2.95	&	2.05	\\
2500	&	4.0	&	12.14	&	11.14	&	9.95	&	7.32	&	5.01	&	2.43	&	3.05	&	2.08	&	2.49	&	2.93	&	2.03	&	2.93	&	2.03	\\
2500	&	4.5	&	12.72	&	11.33	&	9.99	&	7.30	&	5.00	&	2.39	&	3.03	&	2.07	&	2.45	&	2.90	&	2.02	&	2.90	&	2.02	\\
2500	&	5.0	&	13.46	&	11.52	&	9.99	&	7.24	&	4.96	&	2.34	&	3.00	&	2.06	&	2.40	&	2.87	&	2.03	&	2.87	&	2.03	\\
2500	&	5.5	&	14.43	&	11.63	&	9.79	&	7.03	&	4.83	&	2.33	&	2.98	&	2.11	&	2.38	&	2.85	&	2.08	&	2.85	&	2.08	\\
2600	&	3.0	&	10.97	&	10.44	&	9.23	&	6.89	&	4.77	&	2.37	&	2.96	&	2.00	&	2.43	&	2.83	&	1.96	&	2.83	&	1.96	\\
2600	&	3.5	&	11.15	&	10.44	&	9.23	&	6.87	&	4.74	&	2.34	&	2.94	&	1.98	&	2.40	&	2.82	&	1.94	&	2.82	&	1.94	\\
2600	&	4.0	&	11.46	&	10.52	&	9.29	&	6.87	&	4.72	&	2.30	&	2.92	&	1.96	&	2.36	&	2.80	&	1.92	&	2.80	&	1.92	\\
2600	&	4.5	&	11.94	&	10.68	&	9.40	&	6.89	&	4.71	&	2.26	&	2.89	&	1.95	&	2.32	&	2.77	&	1.91	&	2.77	&	1.91	\\
2600	&	5.0	&	12.59	&	10.91	&	9.48	&	6.89	&	4.70	&	2.22	&	2.87	&	1.94	&	2.27	&	2.75	&	1.90	&	2.75	&	1.90	\\
2600	&	5.5	&	13.43	&	11.14	&	9.48	&	6.84	&	4.66	&	2.19	&	2.85	&	1.96	&	2.24	&	2.72	&	1.93	&	2.72	&	1.93	\\
2700	&	3.0	&	10.50	&	9.96	&	8.73	&	6.53	&	4.53	&	2.24	&	2.82	&	1.89	&	2.29	&	2.70	&	1.85	&	2.70	&	1.85	\\
2700	&	3.5	&	10.62	&	9.89	&	8.65	&	6.46	&	4.48	&	2.21	&	2.81	&	1.87	&	2.26	&	2.69	&	1.83	&	2.69	&	1.83	\\
2700	&	4.0	&	10.86	&	9.94	&	8.67	&	6.45	&	4.45	&	2.18	&	2.79	&	1.86	&	2.23	&	2.67	&	1.81	&	2.67	&	1.81	\\
2700	&	4.5	&	11.25	&	10.07	&	8.75	&	6.47	&	4.44	&	2.14	&	2.77	&	1.84	&	2.19	&	2.65	&	1.80	&	2.65	&	1.80	\\
2700	&	5.0	&	11.79	&	10.26	&	8.87	&	6.50	&	4.43	&	2.10	&	2.74	&	1.83	&	2.15	&	2.62	&	1.79	&	2.62	&	1.79	\\
2700	&	5.5	&	12.53	&	10.52	&	8.97	&	6.51	&	4.41	&	2.07	&	2.72	&	1.83	&	2.12	&	2.60	&	1.80	&	2.60	&	1.80	\\
2800	&	3.0	&	10.06	&	9.51	&	8.30	&	6.22	&	4.30	&	2.08	&	2.68	&	1.77	&	2.13	&	2.56	&	1.73	&	2.56	&	1.73	\\
2800	&	3.5	&	10.12	&	9.39	&	8.13	&	6.10	&	4.24	&	2.08	&	2.68	&	1.76	&	2.13	&	2.56	&	1.72	&	2.56	&	1.72	\\
2800	&	4.0	&	10.31	&	9.39	&	8.10	&	6.06	&	4.20	&	2.06	&	2.67	&	1.75	&	2.11	&	2.55	&	1.71	&	2.55	&	1.71	\\
2800	&	4.5	&	10.62	&	9.48	&	8.14	&	6.07	&	4.19	&	2.03	&	2.65	&	1.73	&	2.08	&	2.54	&	1.70	&	2.54	&	1.70	\\
2800	&	5.0	&	11.08	&	9.65	&	8.23	&	6.09	&	4.18	&	1.99	&	2.63	&	1.72	&	2.04	&	2.52	&	1.69	&	2.52	&	1.69	\\
2800	&	5.5	&	11.71	&	9.87	&	8.34	&	6.12	&	4.16	&	1.95	&	2.61	&	1.72	&	2.00	&	2.49	&	1.69	&	2.49	&	1.69	\\
2900	&	3.0	&	9.68	&	9.10	&	7.94	&	5.94	&	4.09	&	1.92	&	2.54	&	1.64	&	1.96	&	2.42	&	1.60	&	2.42	&	1.60	\\
2900	&	3.5	&	9.68	&	8.94	&	7.69	&	5.78	&	4.01	&	1.94	&	2.54	&	1.65	&	1.99	&	2.43	&	1.61	&	2.43	&	1.61	\\
2900	&	4.0	&	9.81	&	8.90	&	7.60	&	5.72	&	3.98	&	1.94	&	2.54	&	1.64	&	1.98	&	2.43	&	1.60	&	2.43	&	1.60	\\
2900	&	4.5	&	10.06	&	8.95	&	7.59	&	5.70	&	3.96	&	1.92	&	2.54	&	1.63	&	1.96	&	2.43	&	1.60	&	2.43	&	1.60	\\
2900	&	5.0	&	10.44	&	9.07	&	7.64	&	5.71	&	3.94	&	1.88	&	2.52	&	1.62	&	1.93	&	2.41	&	1.59	&	2.41	&	1.59	\\
2900	&	5.5	&	10.96	&	9.27	&	7.73	&	5.74	&	3.93	&	1.85	&	2.50	&	1.61	&	1.89	&	2.39	&	1.58	&	2.39	&	1.58	\\
3000	&	3.0	&	9.33	&	8.72	&	7.61	&	5.69	&	3.90	&	1.73	&	2.41	&	1.49	&	1.78	&	2.30	&	1.46	&	2.30	&	1.46	\\
3000	&	3.5	&	9.27	&	8.52	&	7.30	&	5.50	&	3.81	&	1.79	&	2.41	&	1.52	&	1.83	&	2.30	&	1.49	&	2.30	&	1.49	\\
3000	&	4.0	&	9.35	&	8.45	&	7.15	&	5.40	&	3.76	&	1.81	&	2.42	&	1.53	&	1.85	&	2.31	&	1.50	&	2.31	&	1.50	\\
3000	&	4.5	&	9.54	&	8.46	&	7.10	&	5.37	&	3.74	&	1.80	&	2.42	&	1.53	&	1.84	&	2.31	&	1.50	&	2.31	&	1.50	\\
3000	&	5.0	&	9.85	&	8.55	&	7.12	&	5.36	&	3.73	&	1.78	&	2.41	&	1.52	&	1.82	&	2.30	&	1.49	&	2.30	&	1.49	\\
3000	&	5.5	&	10.30	&	8.71	&	7.18	&	5.38	&	3.72	&	1.75	&	2.40	&	1.51	&	1.79	&	2.29	&	1.48	&	2.29	&	1.48	\\
3100	&	3.0	&	9.02	&	8.35	&	7.30	&	5.46	&	3.71	&	1.54	&	2.29	&	1.33	&	1.59	&	2.18	&	1.30	&	2.18	&	1.30	\\
3100	&	3.5	&	8.92	&	8.13	&	6.95	&	5.24	&	3.61	&	1.63	&	2.29	&	1.39	&	1.67	&	2.19	&	1.36	&	2.19	&	1.36	\\
3100	&	4.0	&	8.94	&	8.04	&	6.76	&	5.12	&	3.57	&	1.67	&	2.30	&	1.42	&	1.71	&	2.20	&	1.39	&	2.20	&	1.39	\\
3100	&	4.5	&	9.08	&	8.03	&	6.68	&	5.06	&	3.55	&	1.69	&	2.31	&	1.43	&	1.73	&	2.20	&	1.40	&	2.20	&	1.40	\\
3100	&	5.0	&	9.33	&	8.09	&	6.66	&	5.04	&	3.53	&	1.68	&	2.30	&	1.43	&	1.72	&	2.20	&	1.40	&	2.20	&	1.40	\\
3100	&	5.5	&	9.69	&	8.20	&	6.69	&	5.05	&	3.52	&	1.65	&	2.30	&	1.42	&	1.69	&	2.19	&	1.39	&	2.19	&	1.39	\\
3200	&	3.0	&	8.73	&	8.00	&	7.00	&	5.23	&	3.54	&	1.36	&	2.17	&	1.18	&	1.41	&	2.06	&	1.15	&	2.06	&	1.15	\\
3200	&	3.5	&	8.60	&	7.77	&	6.62	&	5.00	&	3.44	&	1.47	&	2.18	&	1.25	&	1.52	&	2.08	&	1.23	&	2.08	&	1.23	\\
3200	&	4.0	&	8.58	&	7.66	&	6.40	&	4.86	&	3.39	&	1.54	&	2.19	&	1.30	&	1.58	&	2.09	&	1.27	&	2.09	&	1.27	\\
3200	&	4.5	&	8.67	&	7.63	&	6.29	&	4.79	&	3.36	&	1.57	&	2.20	&	1.33	&	1.61	&	2.09	&	1.30	&	2.09	&	1.30	\\
3200	&	5.0	&	8.86	&	7.66	&	6.25	&	4.75	&	3.35	&	1.57	&	2.20	&	1.33	&	1.61	&	2.10	&	1.30	&	2.10	&	1.30	\\
3200	&	5.5	&	9.15	&	7.75	&	6.25	&	4.74	&	3.34	&	1.56	&	2.19	&	1.33	&	1.60	&	2.09	&	1.30	&	2.09	&	1.30	\\
3300	&	3.0	&	8.51	&	7.68	&	6.66	&	5.00	&	3.39	&	1.22	&	2.08	&	1.05	&	1.27	&	1.97	&	1.03	&	1.97	&	1.03	\\
3300	&	3.5	&	8.34	&	7.45	&	6.30	&	4.77	&	3.28	&	1.33	&	2.08	&	1.13	&	1.37	&	1.98	&	1.10	&	1.98	&	1.10	\\
3300	&	4.0	&	8.27	&	7.32	&	6.07	&	4.62	&	3.22	&	1.41	&	2.09	&	1.19	&	1.45	&	1.99	&	1.16	&	1.99	&	1.16	\\
3300	&	4.5	&	8.30	&	7.27	&	5.94	&	4.53	&	3.19	&	1.45	&	2.10	&	1.22	&	1.49	&	1.99	&	1.20	&	1.99	&	1.20	\\
3300	&	5.0	&	8.44	&	7.28	&	5.88	&	4.48	&	3.18	&	1.47	&	2.10	&	1.24	&	1.51	&	2.00	&	1.21	&	2.00	&	1.21	\\
3300	&	5.5	&	8.67	&	7.34	&	5.87	&	4.46	&	3.17	&	1.47	&	2.10	&	1.24	&	1.50	&	2.00	&	1.22	&	2.00	&	1.22	\\
3400	&	3.0	&	8.34	&	7.34	&	6.22	&	4.70	&	3.23	&	1.11	&	2.01	&	0.94	&	1.16	&	1.91	&	0.93	&	1.91	&	0.93	\\
3400	&	3.5	&	8.12	&	7.15	&	5.97	&	4.53	&	3.14	&	1.21	&	1.99	&	1.01	&	1.25	&	1.89	&	0.99	&	1.89	&	0.99	\\
3400	&	4.0	&	8.00	&	7.01	&	5.75	&	4.38	&	3.07	&	1.29	&	1.99	&	1.08	&	1.33	&	1.89	&	1.05	&	1.89	&	1.05	\\
3400	&	4.5	&	7.99	&	6.94	&	5.62	&	4.29	&	3.04	&	1.34	&	2.00	&	1.12	&	1.38	&	1.90	&	1.10	&	1.90	&	1.10	\\
3400	&	5.0	&	8.07	&	6.93	&	5.55	&	4.24	&	3.02	&	1.37	&	2.00	&	1.15	&	1.40	&	1.90	&	1.12	&	1.90	&	1.12	\\
3400	&	5.5	&	8.24	&	6.97	&	5.53	&	4.21	&	3.01	&	1.38	&	2.00	&	1.16	&	1.41	&	1.91	&	1.13	&	1.91	&	1.13	\\

	\hline
	\end{tabular}
\end{table*}

\begin{table*}

	\contcaption{}
	
	\begin{tabular}{ccccccccccccccc}
		
	\hline
	
\bf{ T$_{\rm eff}$}  & \bf{log(g)} & \bf{U} & \bf{B} & \bf{V} & \bf{R} & \bf{I} & \bf{2MASS H} &\bf{2MASS J} & \bf{2MASS Ks} & \bf{H} & \bf{J} & \bf{K} & \bf{Y} & \bf{Z} \\
              	
\hline	

3500	&	3.0	&	8.14	&	6.99	&	5.81	&	4.42	&	3.07	&	1.01	&	1.93	&	0.84	&	1.05	&	1.83	&	0.83	&	1.83	&	0.83	\\
3500	&	3.5	&	7.92	&	6.84	&	5.61	&	4.28	&	3.00	&	1.09	&	1.91	&	0.91	&	1.13	&	1.81	&	0.90	&	1.81	&	0.90	\\
3500	&	4.0	&	7.76	&	6.71	&	5.43	&	4.14	&	2.93	&	1.18	&	1.91	&	0.98	&	1.22	&	1.81	&	0.96	&	1.81	&	0.96	\\
3500	&	4.5	&	7.71	&	6.64	&	5.31	&	4.06	&	2.90	&	1.24	&	1.91	&	1.03	&	1.27	&	1.81	&	1.00	&	1.81	&	1.00	\\
3500	&	5.0	&	7.73	&	6.62	&	5.25	&	4.01	&	2.88	&	1.27	&	1.91	&	1.06	&	1.31	&	1.81	&	1.03	&	1.81	&	1.03	\\
3500	&	5.5	&	7.85	&	6.64	&	5.22	&	3.98	&	2.87	&	1.29	&	1.91	&	1.08	&	1.32	&	1.82	&	1.05	&	1.82	&	1.05	\\
3600	&	3.0	&	7.93	&	6.63	&	5.39	&	4.12	&	2.93	&	0.92	&	1.85	&	0.76	&	0.96	&	1.75	&	0.75	&	1.75	&	0.75	\\
3600	&	3.5	&	7.72	&	6.53	&	5.27	&	4.03	&	2.86	&	0.99	&	1.84	&	0.82	&	1.03	&	1.74	&	0.81	&	1.74	&	0.81	\\
3600	&	4.0	&	7.54	&	6.42	&	5.13	&	3.92	&	2.80	&	1.07	&	1.83	&	0.88	&	1.11	&	1.73	&	0.87	&	1.73	&	0.87	\\
3600	&	4.5	&	7.45	&	6.35	&	5.02	&	3.84	&	2.77	&	1.14	&	1.82	&	0.94	&	1.17	&	1.73	&	0.92	&	1.73	&	0.92	\\
3600	&	5.0	&	7.44	&	6.32	&	4.97	&	3.79	&	2.74	&	1.18	&	1.82	&	0.97	&	1.21	&	1.73	&	0.95	&	1.73	&	0.95	\\
3600	&	5.5	&	7.51	&	6.34	&	4.95	&	3.77	&	2.73	&	1.20	&	1.82	&	1.00	&	1.23	&	1.73	&	0.97	&	1.73	&	0.97	\\
3700	&	3.0	&	7.69	&	6.28	&	5.00	&	3.84	&	2.78	&	0.85	&	1.77	&	0.69	&	0.89	&	1.67	&	0.69	&	1.67	&	0.69	\\
3700	&	3.5	&	7.50	&	6.23	&	4.94	&	3.79	&	2.73	&	0.89	&	1.75	&	0.74	&	0.93	&	1.66	&	0.73	&	1.66	&	0.73	\\
3700	&	4.0	&	7.33	&	6.15	&	4.84	&	3.70	&	2.68	&	0.97	&	1.75	&	0.80	&	1.01	&	1.65	&	0.78	&	1.65	&	0.78	\\
3700	&	4.5	&	7.21	&	6.08	&	4.76	&	3.63	&	2.64	&	1.04	&	1.74	&	0.85	&	1.08	&	1.65	&	0.84	&	1.65	&	0.84	\\
3700	&	5.0	&	7.16	&	6.05	&	4.71	&	3.58	&	2.62	&	1.09	&	1.74	&	0.89	&	1.12	&	1.64	&	0.87	&	1.64	&	0.87	\\
3700	&	5.5	&	7.19	&	6.05	&	4.69	&	3.56	&	2.60	&	1.12	&	1.74	&	0.92	&	1.15	&	1.64	&	0.90	&	1.64	&	0.90	\\
3800	&	3.0	&	7.42	&	5.96	&	4.66	&	3.58	&	2.65	&	0.79	&	1.69	&	0.64	&	0.83	&	1.60	&	0.64	&	1.60	&	0.64	\\
3800	&	3.5	&	7.26	&	5.93	&	4.64	&	3.56	&	2.61	&	0.81	&	1.67	&	0.66	&	0.85	&	1.58	&	0.66	&	1.58	&	0.66	\\
3800	&	4.0	&	7.11	&	5.88	&	4.58	&	3.50	&	2.57	&	0.87	&	1.66	&	0.72	&	0.91	&	1.57	&	0.71	&	1.57	&	0.71	\\
3800	&	4.5	&	6.98	&	5.82	&	4.51	&	3.43	&	2.53	&	0.95	&	1.66	&	0.77	&	0.98	&	1.57	&	0.76	&	1.57	&	0.76	\\
3800	&	5.0	&	6.90	&	5.79	&	4.46	&	3.39	&	2.50	&	1.01	&	1.66	&	0.81	&	1.04	&	1.56	&	0.80	&	1.56	&	0.80	\\
3800	&	5.5	&	6.90	&	5.79	&	4.45	&	3.37	&	2.48	&	1.04	&	1.66	&	0.84	&	1.07	&	1.56	&	0.82	&	1.56	&	0.82	\\
3900	&	3.0	&	7.12	&	5.64	&	4.35	&	3.34	&	2.52	&	0.74	&	1.61	&	0.60	&	0.77	&	1.52	&	0.60	&	1.52	&	0.60	\\
3900	&	3.5	&	7.01	&	5.64	&	4.36	&	3.33	&	2.49	&	0.74	&	1.60	&	0.60	&	0.78	&	1.50	&	0.60	&	1.50	&	0.60	\\
3900	&	4.0	&	6.88	&	5.62	&	4.33	&	3.30	&	2.45	&	0.79	&	1.59	&	0.64	&	0.82	&	1.49	&	0.64	&	1.49	&	0.64	\\
3900	&	4.5	&	6.75	&	5.58	&	4.28	&	3.25	&	2.42	&	0.86	&	1.58	&	0.70	&	0.89	&	1.49	&	0.68	&	1.49	&	0.68	\\
3900	&	5.0	&	6.65	&	5.55	&	4.24	&	3.21	&	2.39	&	0.92	&	1.58	&	0.74	&	0.95	&	1.48	&	0.73	&	1.48	&	0.73	\\
3900	&	5.5	&	6.62	&	5.54	&	4.22	&	3.19	&	2.37	&	0.96	&	1.57	&	0.77	&	0.99	&	1.48	&	0.76	&	1.48	&	0.76	\\
4000	&	3.0	&	6.82	&	5.37	&	4.09	&	3.13	&	2.39	&	0.69	&	1.53	&	0.56	&	0.73	&	1.44	&	0.56	&	1.44	&	0.56	\\
4000	&	3.5	&	6.72	&	5.36	&	4.10	&	3.13	&	2.37	&	0.68	&	1.52	&	0.56	&	0.72	&	1.42	&	0.55	&	1.42	&	0.55	\\
4000	&	4.0	&	6.62	&	5.35	&	4.10	&	3.11	&	2.34	&	0.71	&	1.51	&	0.58	&	0.74	&	1.41	&	0.57	&	1.41	&	0.57	\\
4000	&	4.5	&	6.51	&	5.34	&	4.07	&	3.08	&	2.31	&	0.77	&	1.50	&	0.62	&	0.80	&	1.41	&	0.62	&	1.41	&	0.62	\\
4000	&	5.0	&	6.41	&	5.31	&	4.03	&	3.04	&	2.28	&	0.83	&	1.50	&	0.67	&	0.86	&	1.40	&	0.66	&	1.40	&	0.66	\\
	
	\hline
	\end{tabular}
\end{table*}

\begin{table*}

	\caption{Complete set of {\sc MARCS} synthetic fluxes for models of solar metalicity. The values in the filter columns correspond to $-2.5\rm{ log( {F_{\rm R1}} \slash {F_{\rm R1,Vega}}} )$, where R1 is the corresponding filter. Column 3-7 refer to the Johnson and columns 11-15 to the UKIDSS filters.}
	\label{TC2}
	
	\begin{tabular}{ccccccccccccccc}
		
	\hline
	
\bf{ T$_{\rm eff}$}  & \bf{log(g)} & \bf{U} & \bf{B} & \bf{V} & \bf{R} & \bf{I} & \bf{2MASS H} &\bf{2MASS J} & \bf{2MASS Ks} & \bf{H} & \bf{J} & \bf{K} & \bf{Y} & \bf{Z} \\
              	
\hline	
2500	&	3.0	&	11.19	&	11.16	&	10.96	&	7.61	&	5.03	&	2.43	&	3.00	&	2.09	&	2.49	&	2.87	&	2.04	&	2.87	&	2.04	\\
2500	&	3.5	&	11.64	&	11.32	&	10.90	&	7.57	&	5.02	&	2.42	&	3.03	&	2.08	&	2.49	&	2.89	&	2.04	&	2.89	&	2.04	\\
2500	&	4.0	&	12.26	&	11.60	&	10.97	&	7.59	&	5.02	&	2.41	&	3.03	&	2.08	&	2.47	&	2.90	&	2.04	&	2.90	&	2.04	\\
2500	&	4.5	&	13.14	&	12.00	&	11.08	&	7.64	&	5.03	&	2.39	&	3.04	&	2.09	&	2.45	&	2.91	&	2.05	&	2.91	&	2.05	\\
2500	&	5.0	&	14.30	&	12.43	&	11.16	&	7.70	&	5.02	&	2.38	&	3.04	&	2.13	&	2.44	&	2.91	&	2.09	&	2.91	&	2.09	\\
2500	&	5.5	&	15.53	&	12.78	&	11.20	&	7.77	&	4.99	&	2.38	&	3.04	&	2.20	&	2.43	&	2.91	&	2.17	&	2.91	&	2.17	\\
2600	&	3.0	&	10.66	&	10.55	&	10.23	&	7.16	&	4.74	&	2.28	&	2.85	&	1.97	&	2.34	&	2.72	&	1.92	&	2.72	&	1.92	\\
2600	&	3.5	&	11.02	&	10.65	&	10.11	&	7.10	&	4.72	&	2.28	&	2.88	&	1.96	&	2.34	&	2.75	&	1.92	&	2.75	&	1.92	\\
2600	&	4.0	&	11.51	&	10.85	&	10.14	&	7.10	&	4.71	&	2.27	&	2.89	&	1.96	&	2.33	&	2.76	&	1.91	&	2.76	&	1.91	\\
2600	&	4.5	&	12.21	&	11.16	&	10.25	&	7.14	&	4.72	&	2.25	&	2.89	&	1.96	&	2.31	&	2.77	&	1.92	&	2.77	&	1.92	\\
2600	&	5.0	&	13.18	&	11.54	&	10.35	&	7.20	&	4.71	&	2.24	&	2.90	&	1.98	&	2.29	&	2.77	&	1.94	&	2.77	&	1.94	\\
2600	&	5.5	&	14.32	&	11.90	&	10.41	&	7.26	&	4.70	&	2.23	&	2.90	&	2.04	&	2.29	&	2.77	&	2.00	&	2.77	&	2.00	\\
2700	&	3.0	&	10.20	&	10.02	&	9.60	&	6.77	&	4.47	&	2.12	&	2.70	&	1.84	&	2.18	&	2.57	&	1.80	&	2.57	&	1.80	\\
2700	&	3.5	&	10.47	&	10.04	&	9.39	&	6.67	&	4.44	&	2.15	&	2.74	&	1.85	&	2.20	&	2.61	&	1.80	&	2.61	&	1.80	\\
2700	&	4.0	&	10.85	&	10.17	&	9.37	&	6.65	&	4.43	&	2.14	&	2.75	&	1.84	&	2.19	&	2.63	&	1.80	&	2.63	&	1.80	\\
2700	&	4.5	&	11.40	&	10.40	&	9.45	&	6.68	&	4.43	&	2.12	&	2.76	&	1.84	&	2.18	&	2.64	&	1.80	&	2.64	&	1.80	\\
2700	&	5.0	&	12.20	&	10.72	&	9.56	&	6.73	&	4.43	&	2.11	&	2.76	&	1.85	&	2.16	&	2.64	&	1.81	&	2.64	&	1.81	\\
2700	&	5.5	&	13.20	&	11.06	&	9.65	&	6.79	&	4.42	&	2.10	&	2.76	&	1.89	&	2.15	&	2.64	&	1.85	&	2.64	&	1.85	\\
2800	&	3.0	&	9.81	&	9.56	&	9.07	&	6.43	&	4.24	&	1.95	&	2.56	&	1.71	&	2.00	&	2.44	&	1.67	&	2.44	&	1.67	\\
2800	&	3.5	&	9.99	&	9.49	&	8.77	&	6.29	&	4.19	&	2.00	&	2.60	&	1.73	&	2.06	&	2.48	&	1.69	&	2.48	&	1.69	\\
2800	&	4.0	&	10.28	&	9.55	&	8.68	&	6.24	&	4.17	&	2.01	&	2.62	&	1.73	&	2.06	&	2.51	&	1.69	&	2.51	&	1.69	\\
2800	&	4.5	&	10.71	&	9.72	&	8.72	&	6.25	&	4.16	&	2.00	&	2.63	&	1.73	&	2.05	&	2.52	&	1.69	&	2.52	&	1.69	\\
2800	&	5.0	&	11.34	&	9.97	&	8.81	&	6.29	&	4.16	&	1.98	&	2.63	&	1.73	&	2.03	&	2.52	&	1.70	&	2.52	&	1.70	\\
2800	&	5.5	&	12.19	&	10.28	&	8.91	&	6.34	&	4.16	&	1.97	&	2.64	&	1.76	&	2.02	&	2.52	&	1.72	&	2.52	&	1.72	\\
2900	&	3.0	&	9.49	&	9.16	&	8.62	&	6.14	&	4.04	&	1.77	&	2.44	&	1.57	&	1.82	&	2.32	&	1.53	&	2.32	&	1.53	\\
2900	&	3.5	&	9.57	&	9.01	&	8.24	&	5.96	&	3.96	&	1.86	&	2.47	&	1.61	&	1.91	&	2.36	&	1.57	&	2.36	&	1.57	\\
2900	&	4.0	&	9.77	&	9.01	&	8.08	&	5.88	&	3.93	&	1.88	&	2.50	&	1.62	&	1.93	&	2.38	&	1.58	&	2.38	&	1.58	\\
2900	&	4.5	&	10.10	&	9.11	&	8.06	&	5.86	&	3.92	&	1.88	&	2.51	&	1.62	&	1.93	&	2.40	&	1.58	&	2.40	&	1.58	\\
2900	&	5.0	&	10.60	&	9.30	&	8.12	&	5.88	&	3.92	&	1.87	&	2.51	&	1.62	&	1.91	&	2.40	&	1.59	&	2.40	&	1.59	\\
2900	&	5.5	&	11.30	&	9.56	&	8.21	&	5.92	&	3.92	&	1.86	&	2.51	&	1.64	&	1.90	&	2.40	&	1.60	&	2.40	&	1.60	\\
3000	&	3.0	&	9.21	&	8.81	&	8.22	&	5.88	&	3.86	&	1.58	&	2.33	&	1.42	&	1.64	&	2.22	&	1.39	&	2.22	&	1.39	\\
3000	&	3.5	&	9.20	&	8.59	&	7.79	&	5.66	&	3.76	&	1.70	&	2.35	&	1.49	&	1.75	&	2.24	&	1.45	&	2.24	&	1.45	\\
3000	&	4.0	&	9.33	&	8.53	&	7.56	&	5.55	&	3.72	&	1.75	&	2.38	&	1.51	&	1.80	&	2.27	&	1.48	&	2.27	&	1.48	\\
3000	&	4.5	&	9.57	&	8.57	&	7.48	&	5.51	&	3.71	&	1.76	&	2.39	&	1.52	&	1.81	&	2.28	&	1.48	&	2.28	&	1.48	\\
3000	&	5.0	&	9.96	&	8.70	&	7.50	&	5.51	&	3.70	&	1.76	&	2.40	&	1.52	&	1.80	&	2.29	&	1.48	&	2.29	&	1.48	\\
3000	&	5.5	&	10.52	&	8.91	&	7.57	&	5.54	&	3.70	&	1.75	&	2.40	&	1.53	&	1.79	&	2.29	&	1.49	&	2.29	&	1.49	\\
3100	&	3.0	&	8.98	&	8.47	&	7.83	&	5.63	&	3.70	&	1.41	&	2.23	&	1.27	&	1.47	&	2.12	&	1.24	&	2.12	&	1.24	\\
3100	&	3.5	&	8.89	&	8.22	&	7.40	&	5.40	&	3.58	&	1.55	&	2.24	&	1.36	&	1.60	&	2.13	&	1.32	&	2.13	&	1.32	\\
3100	&	4.0	&	8.94	&	8.10	&	7.11	&	5.25	&	3.53	&	1.62	&	2.26	&	1.40	&	1.66	&	2.16	&	1.37	&	2.16	&	1.37	\\
3100	&	4.5	&	9.11	&	8.10	&	6.98	&	5.18	&	3.51	&	1.65	&	2.28	&	1.42	&	1.69	&	2.17	&	1.38	&	2.17	&	1.38	\\
3100	&	5.0	&	9.40	&	8.18	&	6.95	&	5.16	&	3.50	&	1.65	&	2.29	&	1.42	&	1.69	&	2.18	&	1.39	&	2.18	&	1.39	\\
3100	&	5.5	&	9.84	&	8.34	&	6.99	&	5.18	&	3.50	&	1.64	&	2.29	&	1.42	&	1.68	&	2.19	&	1.39	&	2.19	&	1.39	\\
3200	&	3.0	&	8.76	&	8.15	&	7.43	&	5.38	&	3.54	&	1.26	&	2.14	&	1.13	&	1.32	&	2.03	&	1.11	&	2.03	&	1.11	\\
3200	&	3.5	&	8.62	&	7.88	&	7.03	&	5.15	&	3.42	&	1.40	&	2.14	&	1.22	&	1.45	&	2.04	&	1.19	&	2.04	&	1.19	\\
3200	&	4.0	&	8.61	&	7.72	&	6.72	&	4.98	&	3.36	&	1.49	&	2.16	&	1.28	&	1.53	&	2.05	&	1.25	&	2.05	&	1.25	\\
3200	&	4.5	&	8.70	&	7.68	&	6.54	&	4.89	&	3.33	&	1.53	&	2.17	&	1.31	&	1.57	&	2.07	&	1.28	&	2.07	&	1.28	\\
3200	&	5.0	&	8.92	&	7.72	&	6.48	&	4.85	&	3.32	&	1.54	&	2.18	&	1.32	&	1.58	&	2.08	&	1.29	&	2.08	&	1.29	\\
3200	&	5.5	&	9.26	&	7.84	&	6.48	&	4.85	&	3.31	&	1.54	&	2.19	&	1.33	&	1.58	&	2.09	&	1.30	&	2.09	&	1.30	\\
3300	&	3.0	&	8.56	&	7.81	&	6.99	&	5.11	&	3.38	&	1.14	&	2.06	&	1.00	&	1.19	&	1.95	&	0.99	&	1.95	&	0.99	\\
3300	&	3.5	&	8.39	&	7.57	&	6.66	&	4.91	&	3.27	&	1.25	&	2.05	&	1.09	&	1.30	&	1.95	&	1.07	&	1.95	&	1.07	\\
3300	&	4.0	&	8.31	&	7.38	&	6.35	&	4.73	&	3.20	&	1.36	&	2.06	&	1.17	&	1.40	&	1.96	&	1.14	&	1.96	&	1.14	\\
3300	&	4.5	&	8.35	&	7.31	&	6.16	&	4.63	&	3.17	&	1.42	&	2.07	&	1.21	&	1.46	&	1.97	&	1.18	&	1.97	&	1.18	\\
3300	&	5.0	&	8.49	&	7.32	&	6.06	&	4.57	&	3.15	&	1.44	&	2.08	&	1.23	&	1.48	&	1.98	&	1.20	&	1.98	&	1.20	\\
3300	&	5.5	&	8.75	&	7.40	&	6.04	&	4.55	&	3.15	&	1.45	&	2.09	&	1.24	&	1.48	&	1.99	&	1.21	&	1.99	&	1.21	\\
3400	&	3.0	&	8.36	&	7.44	&	6.51	&	4.81	&	3.22	&	1.04	&	1.98	&	0.89	&	1.09	&	1.88	&	0.88	&	1.88	&	0.88	\\
3400	&	3.5	&	8.19	&	7.26	&	6.28	&	4.67	&	3.14	&	1.12	&	1.97	&	0.96	&	1.16	&	1.87	&	0.95	&	1.87	&	0.95	\\
3400	&	4.0	&	8.06	&	7.06	&	5.98	&	4.48	&	3.05	&	1.24	&	1.97	&	1.05	&	1.28	&	1.87	&	1.03	&	1.87	&	1.03	\\
3400	&	4.5	&	8.04	&	6.97	&	5.80	&	4.37	&	3.02	&	1.31	&	1.98	&	1.11	&	1.35	&	1.88	&	1.08	&	1.88	&	1.08	\\
3400	&	5.0	&	8.12	&	6.96	&	5.70	&	4.31	&	3.00	&	1.34	&	1.99	&	1.14	&	1.38	&	1.89	&	1.11	&	1.89	&	1.11	\\
3400	&	5.5	&	8.30	&	7.01	&	5.66	&	4.28	&	2.99	&	1.35	&	1.99	&	1.15	&	1.39	&	1.89	&	1.12	&	1.89	&	1.12	\\

\hline
	\end{tabular}
\end{table*}

\begin{table*}

	\contcaption{}
	
	\begin{tabular}{ccccccccccccccc}
		
	\hline
	
\bf{ T$_{\rm eff}$}  & \bf{log(g)} & \bf{U} & \bf{B} & \bf{V} & \bf{R} & \bf{I} & \bf{2MASS H} &\bf{2MASS J} & \bf{2MASS Ks} & \bf{H} & \bf{J} & \bf{K} & \bf{Y} & \bf{Z} \\
              	
\hline	
3500	&	3.0	&	8.14	&	7.05	&	6.00	&	4.48	&	3.05	&	0.96	&	1.90	&	0.80	&	1.00	&	1.80	&	0.80	&	1.80	&	0.80	\\
3500	&	3.5	&	7.99	&	6.94	&	5.88	&	4.40	&	3.00	&	1.00	&	1.89	&	0.85	&	1.05	&	1.79	&	0.84	&	1.79	&	0.84	\\
3500	&	4.0	&	7.83	&	6.77	&	5.63	&	4.24	&	2.92	&	1.12	&	1.89	&	0.95	&	1.16	&	1.79	&	0.93	&	1.79	&	0.93	\\
3500	&	4.5	&	7.76	&	6.66	&	5.46	&	4.13	&	2.88	&	1.21	&	1.89	&	1.01	&	1.24	&	1.79	&	0.99	&	1.79	&	0.99	\\
3500	&	5.0	&	7.79	&	6.64	&	5.36	&	4.07	&	2.86	&	1.25	&	1.89	&	1.05	&	1.28	&	1.80	&	1.02	&	1.80	&	1.02	\\
3500	&	5.5	&	7.91	&	6.67	&	5.32	&	4.03	&	2.85	&	1.27	&	1.90	&	1.07	&	1.30	&	1.80	&	1.04	&	1.80	&	1.04	\\
3600	&	3.0	&	7.92	&	6.67	&	5.51	&	4.16	&	2.90	&	0.89	&	1.83	&	0.73	&	0.93	&	1.73	&	0.72	&	1.73	&	0.72	\\
3600	&	3.5	&	7.78	&	6.61	&	5.47	&	4.13	&	2.86	&	0.91	&	1.81	&	0.76	&	0.95	&	1.71	&	0.75	&	1.71	&	0.75	\\
3600	&	4.0	&	7.62	&	6.48	&	5.29	&	4.00	&	2.80	&	1.01	&	1.81	&	0.84	&	1.05	&	1.71	&	0.83	&	1.71	&	0.83	\\
3600	&	4.5	&	7.51	&	6.38	&	5.14	&	3.90	&	2.75	&	1.11	&	1.81	&	0.92	&	1.14	&	1.71	&	0.90	&	1.71	&	0.90	\\
3600	&	5.0	&	7.49	&	6.34	&	5.06	&	3.84	&	2.73	&	1.16	&	1.81	&	0.96	&	1.19	&	1.71	&	0.94	&	1.71	&	0.94	\\
3600	&	5.5	&	7.56	&	6.35	&	5.02	&	3.80	&	2.71	&	1.18	&	1.81	&	0.99	&	1.21	&	1.71	&	0.96	&	1.71	&	0.96	\\
3700	&	3.0	&	7.67	&	6.32	&	5.07	&	3.85	&	2.75	&	0.83	&	1.75	&	0.67	&	0.87	&	1.65	&	0.67	&	1.65	&	0.67	\\
3700	&	3.5	&	7.56	&	6.28	&	5.08	&	3.85	&	2.73	&	0.83	&	1.73	&	0.69	&	0.87	&	1.64	&	0.68	&	1.64	&	0.68	\\
3700	&	4.0	&	7.42	&	6.21	&	4.98	&	3.78	&	2.68	&	0.90	&	1.73	&	0.75	&	0.94	&	1.63	&	0.74	&	1.63	&	0.74	\\
3700	&	4.5	&	7.28	&	6.12	&	4.84	&	3.67	&	2.63	&	1.01	&	1.73	&	0.83	&	1.05	&	1.63	&	0.82	&	1.63	&	0.82	\\
3700	&	5.0	&	7.22	&	6.07	&	4.77	&	3.62	&	2.60	&	1.07	&	1.72	&	0.88	&	1.10	&	1.63	&	0.86	&	1.63	&	0.86	\\
3700	&	5.5	&	7.24	&	6.07	&	4.74	&	3.59	&	2.59	&	1.10	&	1.72	&	0.91	&	1.13	&	1.63	&	0.89	&	1.63	&	0.89	\\
3800	&	3.0	&	7.42	&	5.99	&	4.68	&	3.57	&	2.62	&	0.77	&	1.67	&	0.62	&	0.81	&	1.57	&	0.62	&	1.57	&	0.62	\\
3800	&	3.5	&	7.31	&	5.97	&	4.71	&	3.59	&	2.60	&	0.77	&	1.66	&	0.63	&	0.81	&	1.56	&	0.63	&	1.56	&	0.63	\\
3800	&	4.0	&	7.20	&	5.94	&	4.68	&	3.56	&	2.57	&	0.81	&	1.65	&	0.67	&	0.85	&	1.55	&	0.66	&	1.55	&	0.66	\\
3800	&	4.5	&	7.06	&	5.87	&	4.58	&	3.47	&	2.52	&	0.91	&	1.65	&	0.75	&	0.95	&	1.55	&	0.73	&	1.55	&	0.73	\\
3800	&	5.0	&	6.96	&	5.82	&	4.51	&	3.42	&	2.49	&	0.98	&	1.64	&	0.80	&	1.02	&	1.55	&	0.79	&	1.55	&	0.79	\\
3800	&	5.5	&	6.95	&	5.81	&	4.48	&	3.39	&	2.47	&	1.02	&	1.64	&	0.84	&	1.05	&	1.55	&	0.82	&	1.55	&	0.82	\\
3900	&	3.0	&	7.15	&	5.69	&	4.36	&	3.33	&	2.49	&	0.72	&	1.59	&	0.57	&	0.76	&	1.49	&	0.57	&	1.49	&	0.57	\\
3900	&	3.5	&	7.05	&	5.68	&	4.39	&	3.34	&	2.48	&	0.72	&	1.58	&	0.58	&	0.76	&	1.48	&	0.58	&	1.48	&	0.58	\\
3900	&	4.0	&	6.97	&	5.67	&	4.40	&	3.34	&	2.46	&	0.73	&	1.57	&	0.60	&	0.77	&	1.48	&	0.59	&	1.48	&	0.59	\\
3900	&	4.5	&	6.84	&	5.63	&	4.34	&	3.28	&	2.41	&	0.81	&	1.57	&	0.66	&	0.85	&	1.47	&	0.65	&	1.47	&	0.65	\\
3900	&	5.0	&	6.72	&	5.58	&	4.27	&	3.23	&	2.37	&	0.90	&	1.56	&	0.73	&	0.93	&	1.47	&	0.71	&	1.47	&	0.71	\\
3900	&	5.5	&	6.68	&	5.56	&	4.25	&	3.20	&	2.35	&	0.94	&	1.56	&	0.77	&	0.97	&	1.47	&	0.75	&	1.47	&	0.75	\\
4000	&	3.0	&	6.86	&	5.41	&	4.08	&	3.11	&	2.37	&	0.67	&	1.51	&	0.53	&	0.71	&	1.41	&	0.54	&	1.41	&	0.54	\\
4000	&	3.5	&	6.77	&	5.40	&	4.11	&	3.12	&	2.36	&	0.67	&	1.50	&	0.54	&	0.71	&	1.40	&	0.54	&	1.40	&	0.54	\\
4000	&	4.0	&	6.71	&	5.40	&	4.13	&	3.14	&	2.34	&	0.67	&	1.49	&	0.54	&	0.71	&	1.40	&	0.54	&	1.40	&	0.54	\\
4000	&	4.5	&	6.62	&	5.39	&	4.11	&	3.11	&	2.31	&	0.72	&	1.49	&	0.59	&	0.76	&	1.39	&	0.58	&	1.39	&	0.58	\\
4000	&	5.0	&	6.49	&	5.35	&	4.06	&	3.06	&	2.27	&	0.81	&	1.48	&	0.65	&	0.84	&	1.39	&	0.64	&	1.39	&	0.64	\\
\hline
	\end{tabular}
\end{table*}

\begin{table*}

	\caption{Complete set of {\sc Drift-Phoenix} synthetic fluxes for models of solar metalicity. The values in the filter columns correspond to $-2.5\rm{ log( {F_{\rm R1}} \slash {F_{\rm R1,Vega}}} )$, where R1 is the corresponding filter. Column 3-7 refer to the Johnson and columns 11-15 to the UKIDSS filters.}
	\label{TC3}
	
	\begin{tabular}{ccccccccccccccc}
		
	\hline
	
\bf{ T$_{\rm eff}$}  & \bf{log(g)} & \bf{U} & \bf{B} & \bf{V} & \bf{R} & \bf{I} & \bf{2MASS H} &\bf{2MASS J} & \bf{2MASS Ks} & \bf{H} & \bf{J} & \bf{K} & \bf{Y} & \bf{Z} \\
              	
\hline
2500	&	3.0	&	10.97	&	11.16	&	10.97	&	7.85	&	5.10	&	2.43	&	2.93	&	2.10	&	2.49	&	2.79	&	2.06	&	2.79	&	2.06	\\
2500	&	3.5	&	11.41	&	11.35	&	10.85	&	7.76	&	5.08	&	2.40	&	2.93	&	2.08	&	2.47	&	2.79	&	2.04	&	2.79	&	2.04	\\
2500	&	4.0	&	12.11	&	11.68	&	10.76	&	7.68	&	5.07	&	2.37	&	2.93	&	2.07	&	2.43	&	2.80	&	2.03	&	2.80	&	2.03	\\
2500	&	4.5	&	12.98	&	12.05	&	10.64	&	7.59	&	5.05	&	2.33	&	2.94	&	2.06	&	2.40	&	2.80	&	2.02	&	2.80	&	2.02	\\
2500	&	5.0	&	13.95	&	12.40	&	10.53	&	7.51	&	5.03	&	2.30	&	2.93	&	2.08	&	2.36	&	2.80	&	2.05	&	2.80	&	2.05	\\
2500	&	5.5	&	14.88	&	12.58	&	10.41	&	7.44	&	4.99	&	2.28	&	2.94	&	2.13	&	2.34	&	2.80	&	2.11	&	2.80	&	2.11	\\
2600	&	3.0	&	10.47	&	10.61	&	10.42	&	7.43	&	4.82	&	2.27	&	2.77	&	1.98	&	2.33	&	2.63	&	1.94	&	2.63	&	1.94	\\
2600	&	3.5	&	10.81	&	10.73	&	10.24	&	7.31	&	4.78	&	2.26	&	2.79	&	1.96	&	2.32	&	2.65	&	1.92	&	2.65	&	1.92	\\
2600	&	4.0	&	11.39	&	10.99	&	10.15	&	7.24	&	4.77	&	2.23	&	2.79	&	1.95	&	2.29	&	2.66	&	1.91	&	2.66	&	1.91	\\
2600	&	4.5	&	12.18	&	11.33	&	10.09	&	7.17	&	4.76	&	2.20	&	2.79	&	1.94	&	2.26	&	2.66	&	1.90	&	2.66	&	1.90	\\
2600	&	5.0	&	13.11	&	11.71	&	10.04	&	7.12	&	4.74	&	2.16	&	2.79	&	1.94	&	2.22	&	2.65	&	1.91	&	2.65	&	1.91	\\
2600	&	5.5	&	14.03	&	11.96	&	9.97	&	7.07	&	4.71	&	2.14	&	2.79	&	1.98	&	2.20	&	2.66	&	1.95	&	2.66	&	1.95	\\
2700	&	3.0	&	10.06	&	10.13	&	9.86	&	7.03	&	4.56	&	2.10	&	2.63	&	1.85	&	2.17	&	2.50	&	1.81	&	2.50	&	1.81	\\
2700	&	3.5	&	10.29	&	10.16	&	9.63	&	6.88	&	4.50	&	2.12	&	2.65	&	1.85	&	2.18	&	2.52	&	1.81	&	2.52	&	1.81	\\
2700	&	4.0	&	10.74	&	10.34	&	9.52	&	6.80	&	4.48	&	2.10	&	2.66	&	1.83	&	2.16	&	2.53	&	1.79	&	2.53	&	1.79	\\
2700	&	4.5	&	11.41	&	10.62	&	9.46	&	6.75	&	4.47	&	2.07	&	2.66	&	1.82	&	2.13	&	2.54	&	1.78	&	2.54	&	1.78	\\
2700	&	5.0	&	12.22	&	10.93	&	9.42	&	6.70	&	4.45	&	2.04	&	2.66	&	1.82	&	2.10	&	2.54	&	1.78	&	2.54	&	1.78	\\
2700	&	5.5	&	13.14	&	11.27	&	9.45	&	6.69	&	4.44	&	2.02	&	2.66	&	1.84	&	2.07	&	2.54	&	1.81	&	2.54	&	1.81	\\
2800	&	3.0	&	9.73	&	9.70	&	9.30	&	6.66	&	4.33	&	1.94	&	2.51	&	1.72	&	2.00	&	2.38	&	1.68	&	2.38	&	1.68	\\
2800	&	3.5	&	9.85	&	9.63	&	9.02	&	6.48	&	4.24	&	1.98	&	2.52	&	1.73	&	2.03	&	2.40	&	1.69	&	2.40	&	1.69	\\
2800	&	4.0	&	10.17	&	9.73	&	8.87	&	6.39	&	4.21	&	1.98	&	2.54	&	1.72	&	2.03	&	2.42	&	1.68	&	2.42	&	1.68	\\
2800	&	4.5	&	10.72	&	9.94	&	8.83	&	6.35	&	4.20	&	1.96	&	2.54	&	1.71	&	2.01	&	2.42	&	1.68	&	2.42	&	1.68	\\
2800	&	5.0	&	11.43	&	10.21	&	8.80	&	6.31	&	4.19	&	1.93	&	2.54	&	1.70	&	1.98	&	2.42	&	1.67	&	2.42	&	1.67	\\
2800	&	5.5	&	12.28	&	10.55	&	8.85	&	6.30	&	4.18	&	1.90	&	2.54	&	1.72	&	1.95	&	2.42	&	1.69	&	2.42	&	1.69	\\
2900	&	3.0	&	9.47	&	9.30	&	8.74	&	6.30	&	4.11	&	1.78	&	2.41	&	1.59	&	1.84	&	2.28	&	1.55	&	2.28	&	1.55	\\
2900	&	3.5	&	9.48	&	9.18	&	8.50	&	6.14	&	4.02	&	1.83	&	2.41	&	1.61	&	1.89	&	2.29	&	1.57	&	2.29	&	1.57	\\
2900	&	4.0	&	9.70	&	9.20	&	8.31	&	6.03	&	3.97	&	1.85	&	2.42	&	1.61	&	1.90	&	2.30	&	1.58	&	2.30	&	1.58	\\
2900	&	4.5	&	10.11	&	9.31	&	8.18	&	5.95	&	3.95	&	1.85	&	2.43	&	1.61	&	1.90	&	2.32	&	1.57	&	2.32	&	1.57	\\
2900	&	5.0	&	10.71	&	9.53	&	8.17	&	5.92	&	3.94	&	1.82	&	2.43	&	1.60	&	1.87	&	2.32	&	1.57	&	2.32	&	1.57	\\
2900	&	5.5	&	11.49	&	9.83	&	8.21	&	5.91	&	3.94	&	1.80	&	2.43	&	1.60	&	1.84	&	2.32	&	1.57	&	2.32	&	1.57	\\
3000	&	3.0	&	9.25	&	8.92	&	8.19	&	5.95	&	3.91	&	1.63	&	2.32	&	1.45	&	1.69	&	2.20	&	1.42	&	2.20	&	1.42	\\
3000	&	3.5	&	9.16	&	8.77	&	7.99	&	5.81	&	3.81	&	1.69	&	2.30	&	1.49	&	1.74	&	2.18	&	1.45	&	2.18	&	1.45	\\
3000	&	4.0	&	9.29	&	8.70	&	7.74	&	5.67	&	3.75	&	1.73	&	2.31	&	1.50	&	1.77	&	2.20	&	1.47	&	2.20	&	1.47	\\
3000	&	4.5	&	9.60	&	8.78	&	7.64	&	5.61	&	3.73	&	1.73	&	2.32	&	1.50	&	1.78	&	2.21	&	1.47	&	2.21	&	1.47	\\
3000	&	5.0	&	10.05	&	8.90	&	7.56	&	5.55	&	3.71	&	1.72	&	2.33	&	1.50	&	1.77	&	2.22	&	1.47	&	2.22	&	1.47	\\
3000	&	5.5	&	10.74	&	9.16	&	7.61	&	5.55	&	3.71	&	1.70	&	2.33	&	1.50	&	1.74	&	2.22	&	1.47	&	2.22	&	1.47	\\
\hline
	\end{tabular}
\end{table*}

\begin{table*}

	\caption{Complete set of {\sc ATLAS} synthetic fluxes for models of solar metalicity. The values in the filter columns correspond to $-2.5\rm{ log( {F_{\rm R1}} \slash {F_{\rm R1,Vega}}} )$, where R1 is the corresponding filter. Column 3-7 refer to the Johnson and columns 11-15 to the UKIDSS filters.}
	\label{TC4}
	
	\begin{tabular}{ccccccccccccccc}
		
	\hline
	
\bf{ T$_{\rm eff}$}  & \bf{log(g)} & \bf{U} & \bf{B} & \bf{V} & \bf{R} & \bf{I} & \bf{2MASS H} &\bf{2MASS J} & \bf{2MASS Ks} & \bf{H} & \bf{J} & \bf{K} & \bf{Y} & \bf{Z} \\
              	
\hline
3500	&	3.0	&	7.99	&	6.85	&	5.70	&	4.34	&	2.98	&	1.00	&	1.85	&	0.90	&	1.17	&	1.74	&	0.87	&	1.74	&	0.87	\\
3500	&	3.5	&	7.76	&	6.70	&	5.58	&	4.25	&	2.92	&	1.06	&	1.83	&	0.97	&	1.24	&	1.72	&	0.94	&	1.72	&	0.94	\\
3500	&	4.0	&	7.57	&	6.56	&	5.43	&	4.13	&	2.85	&	1.16	&	1.82	&	1.06	&	1.34	&	1.71	&	1.02	&	1.71	&	1.02	\\
3500	&	4.5	&	7.45	&	6.45	&	5.28	&	4.01	&	2.80	&	1.25	&	1.82	&	1.13	&	1.43	&	1.71	&	1.08	&	1.71	&	1.08	\\
3500	&	5.0	&	7.46	&	6.42	&	5.18	&	3.92	&	2.77	&	1.31	&	1.84	&	1.18	&	1.49	&	1.73	&	1.13	&	1.73	&	1.13	\\
4000	&	3.0	&	6.77	&	5.38	&	4.08	&	3.12	&	2.37	&	0.65	&	1.50	&	0.54	&	0.76	&	1.40	&	0.55	&	1.40	&	0.55	\\
4000	&	3.5	&	6.66	&	5.35	&	4.09	&	3.12	&	2.36	&	0.66	&	1.49	&	0.55	&	0.77	&	1.39	&	0.55	&	1.39	&	0.55	\\
4000	&	4.0	&	6.58	&	5.33	&	4.10	&	3.11	&	2.34	&	0.67	&	1.48	&	0.57	&	0.79	&	1.38	&	0.57	&	1.38	&	0.57	\\
4000	&	4.5	&	6.49	&	5.31	&	4.08	&	3.08	&	2.30	&	0.73	&	1.47	&	0.62	&	0.85	&	1.37	&	0.62	&	1.37	&	0.62	\\
4000	&	5.0	&	6.38	&	5.28	&	4.04	&	3.03	&	2.26	&	0.80	&	1.47	&	0.68	&	0.92	&	1.36	&	0.67	&	1.36	&	0.67	\\
\hline
	\end{tabular}
\end{table*}

\begin{table*}
 
    \caption{List of stars and associated broad-band photometric magnitudes used for comparison with synthetic colours of models. Data taken from \citet{koen2010}, tables 2 and 4}
    \label{TC5}
 
    \begin{tabular}{cccccc}
     
	\hline

HIP	&	J	&	K	&	V-R	&	Spec. Type	\\ \hline
439	&	5.344	&	4.535	&	0.973	&	M1.5	\\ \hline
523	&	8.517	&	7.616	&	1.068	&	M2.5	\\ \hline
1276	&	8.026	&	7.113	&	1.04	&	M2.5	\\ \hline
1463	&	7.672	&	6.771	&	0.995	&	M1.5	\\ \hline
1532	&	7.404	&	6.579	&	0.836	&	M0V	\\ \hline
1720	&	8.395	&	7.479	&	1.104	&	M3.0	\\ \hline
1734	&	7.749	&	6.815	&	1.009	&	M1.5	\\ \hline
1842	&	8.348	&	7.425	&	1.045	&	M2.5	\\ \hline
4569	&	7.832	&	6.926	&	1.138	&	M3V	\\ \hline
4845	&	7.456	&	6.585	&	0.862	&	M0V	\\ \hline
4927	&	7.7	&	6.796	&	1.072	&	M2	\\ \hline
5215	&	8.096	&	7.142	&	0.988	&	M2	\\ \hline
5410	&	8.532	&	7.607	&	1.084	&	M3	\\ \hline
5496	&	6.067	&	5.16	&	1.094	&	M2.5V	\\ \hline
5643	&	7.37	&	6.438	&	1.378	&	M4.5	\\ \hline
6005	&	7.899	&	6.964	&	1.02	&	M2.5V	\\ \hline
6008	&	7.84	&	6.918	&	0.942	&	M1	\\ \hline
6069	&	7.916	&	7.04	&	0.88	&	M0.5	\\ \hline
6097	&	8.427	&	7.513	&	1.008	&	M2	\\ \hline
6351	&	7.435	&	6.578	&	0.877	&	M0V	\\ \hline
6365	&	8.224	&	7.322	&	0.977	&	M1.0	\\ \hline
7170	&	8.015	&	7.121	&	0.976	&	M1.5V	\\ \hline
7646	&	8.264	&	7.322	&	1.019	&	M2.5V	\\ \hline
8051	&	7.429	&	6.543	&	1.036	&	M2	\\ \hline
8382	&	8.6	&	7.673	&	1.061	&	M2.5	\\ \hline
8691	&	8.445	&	7.627	&	1.005	&	M2	\\ \hline
9724	&	6.62	&	5.707	&	1.048	&	M2.5	\\ \hline
9749	&	8.073	&	7.174	&	0.899	&	M1+V	\\ \hline
9786	&	8.429	&	7.55	&	1.12	&	M2.5+V	\\ \hline
10279	&	6.921	&	6.089	&	0.95	&	M1.5	\\ \hline
10395	&	7.02	&	6.112	&	1.001	&	M2Vk	\\ \hline
10617	&	8.036	&	7.132	&	1.142	&	M3V	\\ \hline
10812	&	7.975	&	7.052	&	1.068	&	M2.5+V	\\ \hline
11439	&	7.831	&	6.929	&	0.959	&	M2V	\\ \hline
12097	&	7.285	&	6.377	&	1.017	&	M2	\\ \hline
12261	&	8.437	&	7.526	&	1.173	&	M3V	\\ \hline
12749	&	8.754	&	7.939	&	0.967	&	M1.5	\\ \hline
12781	&	6.852	&	5.972	&	1.073	&	M3	\\ \hline

\hline
	
    \end{tabular}
\end{table*}

\begin{table*}
 
    \contcaption{}
 
    \begin{tabular}{cccccc}
     
	\hline
12961	&	7.632	&	6.765	&	0.877	&	M0	\\ \hline
13218	&	7.427	&	6.536	&	1.003	&	M1.5	\\ \hline
13389	&	7.763	&	6.862	&	1.064	&	M2.5	\\ \hline
14165	&	8.284	&	7.327	&	1.046	&	M2.5V	\\ \hline
14555	&	7.297	&	6.391	&	0.926	&	M0V	\\ \hline
14731	&	8.408	&	7.485	&	1.025	&	M2	\\ \hline
15332	&	8.557	&	7.703	&	0.996	&	M2.5V	\\ \hline
15360	&	8.032	&	7.127	&	0.944	&	M1V	\\ \hline
15439	&	8.293	&	7.424	&	1.062	&	M2+V	\\ \hline
15844	&	7.203	&	6.276	&	0.984	&	M1	\\ \hline
15973	&	8.337	&	7.518	&	0.929	&	M0.5V	\\ \hline
16445	&	8.804	&	7.919	&	1.053	&	M2	\\ \hline
16536	&	7.88	&	6.993	&	1.079	&	M2.5V	\\ \hline
17743	&	8.005	&	7.12	&	0.958	&	M0.5	\\ \hline
18115	&	8.193	&	7.287	&	1.005	&	M2V	\\ \hline
19394	&	8.024	&	7.148	&	1.077	&	M3.5	\\ \hline
19948	&	7.585	&	6.681	&	1.003	&	M1.5+V	\\ \hline
21086	&	8.016	&	7.053	&	1.047	&	M2.5V	\\ \hline
21556	&	7.021	&	6.122	&	1.004	&	M1.5	\\ \hline
21932	&	6.563	&	5.635	&	1.02	&	M2	\\ \hline
22627	&	7.923	&	6.955	&	1.164	&	M3.5	\\ \hline
23512	&	7.885	&	6.98	&	1.141	&	M3.5	\\ \hline
24472	&	8.423	&	7.547	&	0.972	&	M0.5	\\ \hline
25578	&	8.413	&	7.568	&	1.157	&	M3.5	\\ \hline
26081	&	7.831	&	6.872	&	1.074	&	M2.5	\\ \hline
28035	&	7.252	&	6.314	&	1.04	&	M2.5V	\\ \hline
29295	&	5.062	&	4.162	&	0.961	&	M0.5	\\ \hline
30920	&	6.459	&	5.492	&	1.301	&	M4.5	\\ \hline
31126	&	7.596	&	6.682	&	0.938	&	M0V	\\ \hline
31300	&	7.995	&	7.061	&	1.058	&	M2.5	\\ \hline
31862	&	6.927	&	6.036	&	0.926	&	M0V	\\ \hline
31878	&	7.408	&	6.55	&	0.798	&	M1V	\\ \hline
33499	&	6.948	&	6.077	&	1.129	&	M3.0	\\ \hline
34104	&	7.389	&	6.425	&	1.116	&	M3.5	\\ \hline
34361	&	7.709	&	6.815	&	0.989	&	M2V	\\ \hline
35943	&	7.729	&	6.861	&	0.871	&	M0V	\\ \hline
36208	&	5.747	&	4.883	&	1.173	&	M3.5	\\ \hline
36349	&	6.681	&	5.753	&	0.974	&	M1V	\\ \hline
37217	&	7.959	&	7.067	&	1.096	&	M3	\\ \hline
39987	&	8.058	&	7.152	&	1.071	&	M3.0	\\ \hline

\hline
	
    \end{tabular}
\end{table*}

\begin{table*}
 
    \contcaption{}
 
    \begin{tabular}{cccccc}
     
	\hline
40239	&	6.719	&	5.849	&	0.874	&	M0V	\\ \hline
40501	&	6.708	&	5.808	&	1.02	&	M2	\\ \hline
41802	&	8.115	&	7.212	&	0.935	&	M2V	\\ \hline
42762	&	8.169	&	7.263	&	1.066	&	M2.5	\\ \hline
45908	&	6.495	&	5.589	&	0.948	&	M0.0	\\ \hline
46655	&	7.759	&	6.851	&	1.127	&	M3.5	\\ \hline
47103	&	7.39	&	6.51	&	1.048	&	M2.5V	\\ \hline
47425	&	6.964	&	6.07	&	1.077	&	M2.0	\\ \hline
47513	&	7.041	&	6.135	&	0.999	&	M1.5	\\ \hline
47619	&	8.384	&	7.5	&	1.051	&	M2.5V	\\ \hline
48336	&	7.038	&	6.167	&	0.938	&	M0.5	\\ \hline
48659	&	8.079	&	7.155	&	1.159	&	M3V	\\ \hline
48904	&	7.221	&	6.29	&	1.142	&	M3.5	\\ \hline
49091	&	7.668	&	6.719	&	1.077	&	M3.0	\\ \hline
49376	&	8.558	&	7.65	&	1.025	&	M2+V	\\ \hline
49969	&	7.103	&	6.206	&	1.048	&	M2.5	\\ \hline
49986	&	5.962	&	5.029	&	0.998	&	M1.5	\\ \hline
51317	&	6.233	&	5.349	&	1.019	&	M2	\\ \hline
52190	&	7.329	&	6.387	&	1.098	&	M2.5V	\\ \hline
52296	&	6.92	&	6.023	&	0.948	&	M0.5	\\ \hline
52596	&	7.906	&	7.008	&	1.009	&	M1.5V	\\ \hline
53767	&	6.412	&	5.503	&	1.068	&	M2.5	\\ \hline
55042	&	7.85	&	7.111	&	1.019	&	M3.5	\\ \hline
55625	&	8.041	&	7.148	&	0.965	&	M0.5	\\ \hline
56244	&	7.474	&	6.556	&	1.151	&	M3.5	\\ \hline
56284	&	8.369	&	7.433	&	0.963	&	M1.5V	\\ \hline
56466	&	8.142	&	7.277	&	0.955	&	M0	\\ \hline
56528	&	6.525	&	5.637	&	0.992	&	M1.5	\\ \hline
57459	&	8.008	&	7.076	&	1.061	&	M3	\\ \hline
57959	&	8.354	&	7.444	&	1.046	&	M2.5	\\ \hline
58688	&	7.598	&	6.696	&	0.914	&	M0V	\\ \hline
59406	&	7.974	&	7.075	&	1.09	&	M3	\\ \hline
60475	&	7.596	&	6.683	&	0.909	&	M0.5V	\\ \hline
60559	&	7.809	&	6.999	&	1.029	&	M2	\\ \hline
61495	&	7.748	&	6.85	&	1.003	&	M1.0	\\ \hline
61629	&	6.955	&	6.035	&	1.077	&	M2.0	\\ \hline
61706	&	7.645	&	6.691	&	1.108	&	M3	\\ \hline
61874	&	8.253	&	7.383	&	1.164	&	M3.0	\\ \hline
62452	&	7.29	&	6.377	&	1.16	&	M3.5	\\ \hline
63510	&	6.556	&	5.608	&	0.975	&	M0.5	\\ \hline
\hline
	
    \end{tabular}
\end{table*}

\begin{table*}
 
    \contcaption{}
 
    \begin{tabular}{cccccc}
     
	\hline
65520	&	7.748	&	6.856	&	0.996	&	M1	\\ \hline
65669	&	8.452	&	7.561	&	0.998	&	M1.5V	\\ \hline
65859	&	5.949	&	5.053	&	0.959	&	M0.5	\\ \hline
67164	&	7.835	&	6.925	&	1.147	&	M3.5	\\ \hline
67761	&	8.519	&	7.613	&	0.984	&	M2V	\\ \hline
67960	&	6.594	&	5.686	&	0.92	&	M0Vk	\\ \hline
68469	&	6.581	&	5.688	&	0.962	&	M1.5V	\\ \hline
69285	&	7.613	&	6.712	&	0.977	&	M2V	\\ \hline
69454	&	7	&	6.109	&	0.985	&	M2V	\\ \hline
70308	&	7.486	&	6.625	&	0.891	&	M1V	\\ \hline
70865	&	7.321	&	6.406	&	1.002	&	M2	\\ \hline
70956	&	6.715	&	5.825	&	0.875	&	M0.5-V	\\ \hline
70975	&	7.888	&	6.957	&	1.152	&	M3.5	\\ \hline
71253	&	6.957	&	5.985	&	1.224	&	M4	\\ \hline
72509	&	8.733	&	7.907	&	0.998	&	M1.5	\\ \hline
72511	&	8.48	&	7.619	&	0.97	&	M1	\\ \hline
72944	&	6.693	&	5.771	&	1.034	&	M2	\\ \hline
74190	&	7.792	&	6.873	&	1.068	&	M3	\\ \hline
74995	&	6.762	&	5.859	&	1.106	&	M3	\\ \hline
76074	&	5.705	&	4.779	&	1.046	&	M2.5	\\ \hline
76901	&	7.983	&	7.154	&	1.117	&	M3	\\ \hline
77349	&	7.637	&	6.754	&	1.066	&	M2.5V	\\ \hline
78353	&	7.275	&	6.335	&	0.99	&	M1	\\ \hline
79431	&	7.608	&	6.636	&	1.083	&	M3V	\\ \hline
80018	&	6.796	&	5.887	&	1.091	&	M2.0	\\ \hline
80229	&	8.537	&	7.67	&	0.979	&	M1.5	\\ \hline
80268	&	7.304	&	6.415	&	0.932	&	M0	\\ \hline
80612	&	7.779	&	6.866	&	0.959	&	M1V	\\ \hline
80817	&	8.487	&	7.608	&	1.086	&	M2.5V/M3V	\\ \hline
80824	&	6.009	&	5.102	&	1.158	&	M3.5	\\ \hline
82256	&	8.114	&	7.212	&	0.985	&	M0.5	\\ \hline
82283	&	7.715	&	6.806	&	0.974	&	M1.5V	\\ \hline
82817	&	5.279	&	4.406	&	1.086	&	M3V	\\ \hline
82926	&	7.386	&	6.477	&	1.111	&	M3V	\\ \hline
83405	&	7.893	&	6.999	&	0.94	&	M0	\\ \hline
83599	&	6.859	&	6	&	0.972	&	M2	\\ \hline
84051	&	6.933	&	6.045	&	0.961	&	M1-V	\\ \hline
84123	&	7.554	&	6.668	&	1.107	&	M3-V	\\ \hline
84212	&	8.688	&	7.883	&	0.922	&	M1V	\\ \hline
84277	&	8.181	&	7.275	&	1.098	&	M3.5	\\ \hline
\hline
	
    \end{tabular}
\end{table*}

\begin{table*}
 
    \contcaption{}
 
    \begin{tabular}{cccccc}
     
	\hline
84521	&	8.015	&	7.118	&	1.049	&	M2	\\ \hline
84652	&	7.663	&	6.769	&	0.928	&	M0	\\ \hline
85523	&	5.758	&	4.872	&	1.07	&	M2+V	\\ \hline
85647	&	6.787	&	5.88	&	0.91	&	M0.0	\\ \hline
85665	&	6.373	&	5.487	&	0.945	&	M0	\\ \hline
86057	&	6.689	&	5.785	&	1.026	&	M1.5V	\\ \hline
86214	&	6.642	&	5.641	&	1.217	&	M3.5	\\ \hline
86287	&	6.468	&	5.586	&	0.978	&	M1	\\ \hline
86707	&	7.542	&	6.651	&	0.976	&	M1	\\ \hline
86961	&	7.048	&	6.145	&	0.986	&	M2V	\\ \hline
86963	&	7.474	&	6.619	&	1.098	&	M2V	\\ \hline
87322	&	7.453	&	6.561	&	0.898	&	M0	\\ \hline
88574	&	6.237	&	5.358	&	0.975	&	M1	\\ \hline
91430	&	7.736	&	6.829	&	1.041	&	M2.5	\\ \hline
91608	&	7.495	&	6.575	&	0.975	&	M1	\\ \hline
92451	&	7.603	&	6.722	&	0.969	&	M3	\\ \hline
92573	&	7.237	&	6.347	&	0.93	&	M0	\\ \hline
92871	&	6.375	&	5.44	&	1.104	&	M3	\\ \hline
93101	&	6.329	&	5.424	&	0.931	&	M0.5	\\ \hline
93206	&	7.608	&	6.704	&	1.038	&	M2.0	\\ \hline
93873	&	7.372	&	6.522	&	1.029	&	M1.5	\\ \hline
93899	&	7.371	&	6.53	&	1.032	&	M2	\\ \hline
94349	&	7.182	&	6.333	&	1.127	&	M3.5	\\ \hline
94557	&	7.664	&	6.813	&	1.091	&	M3.5	\\ \hline
94739	&	6.482	&	5.583	&	0.924	&	M0V	\\ \hline
94761	&	5.591	&	4.663	&	1.039	&	M2.5	\\ \hline
96710	&	7.567	&	6.674	&	0.921	&	M1V	\\ \hline
97051	&	7.66	&	6.867	&	0.788	&	M0	\\ \hline
99150	&	8.275	&	7.416	&	1.21	&	M3.0	\\ \hline
99764	&	7.649	&	6.78	&	0.847	&	M0V	\\ \hline
100923	&	7.773	&	6.88	&	1.059	&	M3	\\ \hline
102235	&	7.645	&	6.764	&	0.943	&	M1.5	\\ \hline
102357	&	7.429	&	6.545	&	0.918	&	M0	\\ \hline
103039	&	7.137	&	6.212	&	1.207	&	M4V	\\ \hline
103388	&	7.856	&	6.924	&	1.059	&	M2.5	\\ \hline
103393	&	7.893	&	7.067	&	1.14	&	M4	\\ \hline
103441	&	8.512	&	7.659	&	1.058	&	M2	\\ \hline
103800	&	7.634	&	6.708	&	1.052	&	M3	\\ \hline
103910	&	8.786	&	7.883	&	1.17	&	M4	\\ \hline
104059	&	8.373	&	7.533	&	0.962	&	M1	\\ \hline
\hline
	
    \end{tabular}
\end{table*}

\begin{table*}
 
    \contcaption{}
 
    \begin{tabular}{cccccc}
     
	\hline
104137	&	8.651	&	7.72	&	1.038	&	M2.5	\\ \hline
104432	&	7.74	&	6.934	&	0.962	&	M1	\\ \hline
104644	&	8.55	&	7.701	&	1.04	&	M1	\\ \hline
105336	&	7.793	&	6.851	&	0.964	&	M1.5V	\\ \hline
105533	&	7.358	&	6.483	&	0.865	&	M0	\\ \hline
105932	&	8.079	&	7.213	&	0.962	&	M0.5	\\ \hline
106106	&	6.365	&	5.462	&	1.14	&	M3.5	\\ \hline
106255	&	7.376	&	6.402	&	1.275	&	M4	\\ \hline
106440	&	5.364	&	4.473	&	1.007	&	M1.5	\\ \hline
106803	&	7.551	&	6.633	&	0.951	&	M0.0	\\ \hline
107317	&	8.352	&	7.434	&	1.083	&	M3	\\ \hline
107705	&	6.576	&	5.663	&	0.942	&	M0.5	\\ \hline
107711	&	7.761	&	6.826	&	1.1	&	M2.5	\\ \hline
107772	&	8.007	&	7.144	&	0.863	&	M0	\\ \hline
108159	&	8.468	&	7.584	&	1.045	&	M2.5	\\ \hline
108380	&	7.801	&	6.869	&	0.975	&	M1.5	\\ \hline
108405	&	6.827	&	5.91	&	1.071	&	M2.5	\\ \hline
108569	&	6.694	&	5.795	&	0.953	&	M0.5	\\ \hline
108752	&	7.136	&	6.208	&	1.043	&	M2	\\ \hline
108782	&	6.257	&	5.355	&	0.933	&	M0	\\ \hline
108890	&	8.728	&	7.86	&	1.02	&	M1.5	\\ \hline
109084	&	7.281	&	6.413	&	0.906	&	M0	\\ \hline
109388	&	6.57	&	5.616	&	1.087	&	M3.5	\\ \hline
109555	&	6.793	&	5.845	&	1.028	&	M2	\\ \hline
110400	&	8.554	&	7.645	&	1.043	&	M1.0	\\ \hline
110534	&	7.681	&	6.776	&	0.96	&	M1-V	\\ \hline
110951	&	7.891	&	6.999	&	0.919	&	M1V	\\ \hline
110980	&	7.674	&	6.796	&	0.922	&	M1V	\\ \hline
111313	&	7.265	&	6.371	&	0.974	&	M1	\\ \hline
111391	&	7.808	&	6.905	&	1.03	&	M2+V	\\ \hline
111766	&	7.358	&	6.445	&	1.179	&	M3.5V	\\ \hline
111932	&	8.758	&	7.919	&	0.929	&	M0V	\\ \hline
112120	&	8.139	&	7.201	&	1.05	&	M2.5	\\ \hline
112312	&	7.886	&	6.943	&	1.188	&	M3	\\ \hline
112388	&	8.982	&	8.097	&	0.953	&	M1V	\\ \hline
112774	&	7.021	&	6.146	&	0.914	&	M0.5-V	\\ \hline
113020	&	5.993	&	5.044	&	1.182	&	M4	\\ \hline
113201	&	8.42	&	7.401	&	0.981	&	M0.5	\\ \hline
113229	&	6.722	&	5.829	&	1.065	&	M3-V	\\ \hline
113244	&	8.199	&	7.289	&	0.945	&	M1	\\ \hline
\hline
	
    \end{tabular}
\end{table*}

\begin{table*}
 
    \contcaption{}
 
    \begin{tabular}{cccccc}
     
	\hline
113602	&	8.343	&	7.445	&	0.993	&	M1	\\ \hline
113850	&	7.742	&	6.851	&	0.932	&	M0.0	\\ \hline
114233	&	7.924	&	7.07	&	0.947	&	M0	\\ \hline
114252	&	7.997	&	7.113	&	0.908	&	M0	\\ \hline
114411	&	7.939	&	7.032	&	1.017	&	M2V	\\ \hline
114719	&	7.444	&	6.497	&	0.95	&	M0.5V	\\ \hline
114954	&	8.141	&	7.278	&	0.887	&	M0V	\\ \hline
115332	&	7.469	&	6.542	&	1.196	&	M4	\\ \hline
116003	&	7.292	&	6.37	&	1.102	&	M3	\\ \hline
116317	&	7.7	&	6.861	&	1.019	&	M2.5	\\ \hline
116645	&	8.441	&	7.488	&	1.044	&	M2.0	\\ \hline
117473	&	5.887	&	5.068	&	0.95	&	M1	\\ \hline
117966	&	7.681	&	6.743	&	1.031	&	M2.5V	\\ \hline
118200	&	7.964	&	7.064	&	1.085	&	M3	\\ \hline

	
	\hline
	
    \end{tabular}

\end{table*}

\end{document}